\documentclass[12pt, a4paper]{article}
\pdfoutput=1
\usepackage{jheppub}
\usepackage{enumerate}
\usepackage{epsfig}
\usepackage{float}
\usepackage[caption=false]{subfig}
\usepackage{multirow}
\usepackage{slashed}
\usepackage{wasysym} 
\usepackage{mathrsfs} 
\usepackage{amsfonts}
\usepackage{amsbsy}
\usepackage{amscd}

\usepackage{amsmath}
\usepackage{amssymb}
\usepackage{slashed}
\usepackage{subfiles}
\usepackage{import} 
\usepackage{blindtext}
\def \beq{\begin{equation}}
\def \eeq{\end{equation}}
\def \bea{\begin{eqnarray}}
\def \eea{\end{eqnarray}}
\def \ba{\begin{array}}
\def \ea{\end{array}}
\usepackage{color}
\definecolor{myred}{rgb}{0.6,0,0} %usage:  {\textcolor{myred}{Hello World}}
\definecolor{myblue}{rgb}{0,0.2,0.4}
\definecolor{mygreen}{rgb}{0,0.9,0.1}
\definecolor{hc}{rgb}{.9,0.1,0.7}
\definecolor{hcout}{rgb}{.9,0.7,0.9}
\definecolor{Orange}{rgb}{1.,0.65,0.}
\numberwithin{equation}{section}
\numberwithin{figure}{section}
\numberwithin{table}{section}

\title{Natural SUSY at LHC with Right-Sneutrino LSP}
\author[a]{Arindam Chatterjee,} 
\author[b]{Juhi Dutta,}
\author[b]{and Santosh Kumar Rai}
\affiliation[a]{Indian Statistical Institute, 203 B.T. Road, Kolkata-700108, India}
\affiliation[b]{Regional Centre for Accelerator-based Particle Physics, \newline Harish-Chandra Research Institute, HBNI, Chhatnag Road, Jhusi, Allahabad-211019, India.}
\emailAdd{arindam.chatterjee@gmail.com}
\emailAdd{juhidutta@hri.res.in}
\emailAdd{skrai@hri.res.in}

\abstract{
We study an extension of the minimal supersymmetric standard model 
(MSSM) with additional right-handed singlet neutrino superfields. While such 
an extension incorporates a mechanism for the neutrino mass, it also opens 
up the possibility of having the right-sneutrinos ($\widetilde{\nu}$) as the 
lightest supersymmetric particle (LSP). In this work, we focus on the the viability 
of rather small ($\lesssim 500$ GeV) higgsino mass  parameter ($\mu$), an important 
ingredient for ``naturalness", in the presence of such a LSP. For simplicity, 
we assume that the bino and wino mass parameters are much heavier; thus we only 
consider (almost) pure and compressed higgsino-like states, with small 
$\mathcal{O}(10^{-2})$ gaugino admixture which nevertheless still affect the decay 
of the low-lying higgsino-like states, thus significantly affecting the proposed 
signatures at colliders. Considering only prompt decays of the higgino-like states, especially 
the lightest chargino, we discuss the importance of leptonic channels consisting of 
up to two leptons with large missing transverse energy to probe this scenario 
at the Large Hadron Collider (LHC). In addition we also comment on the dark matter 
predictions for the studied scenario.}

\preprint{HRI-RECAPP-2017-13}


\begin{document}

\maketitle

\section{Introduction}
\label{intro}
The TeV scale limits from LHC searches on the masses of strongly interacting 
supersymmetric particles set a dismal tone for naturalness concerns, a prime 
motivation for invoking Supersymmetry (SUSY) in particle physics studies. 
While several studies in the literature attempt to quantify ``naturalness" in 
a supersymmetric scenario, the interpretation and the measure of naturalness are 
often debated \cite{Barbieri:1987fn,Ellis:1986yg,Feng:2013pwa,Giudice:2013nak,
Baer:2012cf,Mustafayev:2014lqa}. Nevertheless, in minimal supersymmetric 
extensions of the standard model (MSSM), a small value of the higgsino mass parameter 
$\mu$ and possibly with light \cite{Barbieri:1987fn,Ellis:1986yg,Feng:2013pwa,Giudice:2013nak} or a rather light stop squarks and gluinos 
($\lesssim 1.5$ TeV) \cite{Baer:2012up,Baer:2012cf,Baer:2013ava,Mustafayev:2014lqa}
remain desirable in ``natural" scenarios at the electro-weak (EW) scale. However, even 
with not-so-light strong sector \cite{Baer:2015rja, Baer2017}, ``natural" scenarios 
without much fine-tuning is possible impressing the fact that low $|\mu|$ is of more 
essence to the ``natural'' scenarios at EW scale. 

While the constraints on stop squarks and gluinos are rather stringent 
due to their large production cross-section at the LHC, the weakly interacting sector 
with rather light electroweakinos in general, and higgsinos in particular, 
remain viable \cite{MoriondATLAS,MoriondCMS}. There have been several analyses 
on light electroweakinos, 
assuming a  simplified spectra with one or more specific decay channels 
\cite{Choudhury:2012tc,Belanger:2013pna, Choudhury:2013jpa,Boehm:2013gst,
Chakraborti:2014gea,diCortona:2014yua,Drees:2015aeo,
Chakraborti:2015mra,Badziak:2015qca,Cao:2015efs,Beneke:2016jpw,Baer:2015tva,
Abdughani:2017dqs,Buckley2017}. 
Further, the constraints on the mass of the light higgsino-like states 
have been studied in detail because of their importance in a ``natural" 
supersymmetric scenario \cite{Baer:2011ec,Han:2013usa,Han:2014kaa, 
Drees:2015aeo,Barducci:2015ffa,Baer:2016usl,Fukuda:2017jmk,Mahbubani:2017gjh}. 
However, note that these analyses assume the lightest neutralino as 
the lightest supersymmetric particle (LSP).    
In scenarios with conserved R-parity, the search strategies, and therefore 
the limits of various sparticle masses, depend on the nature of the LSP. 
This is because in such scenarios the LSP appears at the end of the decay 
chain of each sparticle, therefore dictating the possible search channels. 
This warrants investigation of supersymmetric scenarios with different 
types of LSP. While within the paradigm of the MSSM, the lightest neutralino 
is the LSP, and most supersymmetric searches are based on the same assumption.  
There have been studies with gravitino LSP, discussing implications on cosmology 
and signatures at the LHC \cite{Ambrosanio:2000ik,Ellis:2004bx,
Roszkowski:2004jd,Ellis:2006vu,Kribs:2008hq,Bailly:2009pe,Feng:2010ij,
Figy:2010hu, deAquino:2012ru,Barnard:2012au,Bobrovskyi:2012dc,
Roszkowski:2012nq,Cyburt:2013fda,DHondt:2013cwd,Heisig:2013sva,
Covi:2014fba,Maltoni:2015twa,Arvey:2015nra,Kim:2017pvm,Dutta:2017jpe}. 
In other simple extensions, axion and/or axino as the LSP 
\cite{Baer:2010kd,Baer:2011hx,Bae:2013hma,Bae:2014efa, Bae:2015rra} 
and  right-sneutrino LSP have also been considered in minimal extensions 
of the MSSM \cite{ArkaniHamed:2000bq,Gopalakrishna:2006kr,Asaka:2005cn,
Asaka:2006fs,Page:2007sh,Kumar:2009sf, Kadota:2009fg,Belanger:2010cd,
Dumont:2012ee, Kakizaki:2015nua, Chatterjee:2014bva, Cerna-Velazco:2017cmn}. 
While the former sets out to resolve the strong CP-problem, the 
latter provides a weak-scale solution to the neutrino mass generation 
issue, an important aspect missing in the MSSM. 

In this work we consider a similar extension to the MSSM with 
three generations of right-neutrino superfields. This scenario, which provides 
a weak-scale solution to the neutrino mass generation issue, has been widely 
studied in supersymmetric extensions. While the left-sneutrinos have been 
ruled out as a Dark Matter (DM) candidate long ago, thanks to the stringent limit 
from direct detection experiments \cite{Falk:1994es}, right-sneutrinos continue 
to be widely studied as a candidate for DM in simple extensions of the MSSM 
\cite{ArkaniHamed:2000bq,Gopalakrishna:2006kr, Asaka:2005cn,Asaka:2006fs, 
Arina:2007tm, Page:2007sh, Kadota:2009fg,Belanger:2010cd,Dumont:2012ee,
Kakizaki:2015nua}. In its simplest incarnation as ours, the right-sneutrinos at 
EW scale remain very weakly interacting, thanks to the small Yukawa coupling 
$\mathcal{O}(10^{-6}-10^{-7})$ determining their coupling strength to other 
particles. However, as in the case of charged sfermions, a rather large 
value of the  corresponding tri-linear soft supersymmtry breaking parameter 
can induce significant left-admixture in a dominantly right-sneutrino and 
therefore can substantially increase the interaction strengths \cite{Arina:2007tm,Belanger:2010cd,Dumont:2012ee}. 
In both of these scenarios, DM aspects as well as search strategies at LHC 
have been studied for certain choices of the SUSY spectra 
\cite{deGouvea:2006wd, Gupta:2007ui, Choudhury:2008gb,Kumar:2009sf,
Guo:2013asa,Arina2014,Arina:2015uea,Banerjee:2016uyt}. 

We note that in the light of ``naturalness", it becomes equally important 
to investigate the supersymmetric spectrum in such a scenario. In particular 
we focus on a minimalistic spectrum, motivated by ``naturalness" at the 
EW scale, with light higgsino-like states and a right-sneutrino LSP. However, 
analysing collider signatures from the third generation squarks and gluinos 
will be beyond the scope of the present work and will be addressed in a 
subsequent extension. For the present case, the strongly interacting 
sparticles have been assumed to be very heavy adhering to the ``naturalness'' scheme 
proposed in Ref.~\cite{Baer:2015rja, Baer2017}. 
Further, we will also assume the gaugino mass parameters to be large 
enough ($\gtrsim \mathcal{O}(1)$ TeV). Thus the light electroweakinos are 
higgsino-dominated states. Note that the presence of a mixed right-sneutrino as the 
LSP can lead to a very different signature from the compressed higgsino-like states, 
mostly due to the leptonic decay of the light chargino. Although leptonic channels 
provide a cleaner environment for new 
physics searches at a hadron machine such as the LHC, one expects that the level of 
compression in the mass spectra of the electroweakinos would also play a major role 
in determining the efficacy of the leptonic channels. We investigate the prospects of 
discovery of such channels at the 13 TeV run of LHC. We focus on the following apsects 
in our study:
\begin{itemize}
\item We consider a right-sneutrino LSP along with a compressed electroweakino sector  sitting above the LSP, where the lighter states are almost Higgsino-like with a very small admixture of gauginos. 
\item We give a detailed account of how the decay of the light electroweakinos depend on the
various supersymmetric parameters that govern the mixing, mass splitting and, in which region of the parameter space 
the decays are prompt. We also highlight how even the smallest gaugino admixture plays a significant role 
in their decays. 
\item  We comment on the DM predictions for a thermal as well as non-thermal nature of the 
right-sneutrino DM candidate in regions of parameter space of our interest.
\item We then look at possible leptonic signals that arise from such a spectrum and 
analyze the signal at LHC.  
\end{itemize}

%As the gaugino mass parameters have been assumed to be large, the production 
%cross-section for gaugino-like neutralinos and charginos at the LHC would hardly be significant. However, the gaugino parameters 
%are still found to be relevant through their mixing with the light higgsino-like 
%states. At tree-level the mixing with the gaugino-like states, even though very 
%small $\left(\mathcal{O}(10^{-2})\right)$ affects the hierarchy as well as the 
%mass difference of the three light higgsino-like states. In addition, these parameters 
%significantly affect the decay properties of the higgsino-like states which in turn 
%determines the branching fractions of the different decay modes of the light 
%higgsino-like states. We also note that depending on the neutrino Yukawa coupling 
%and the amount of left-right mixing in the sneutrino sector in general, the collider 
%signatures for the electroweakinos strongly rely on the relative sign and 
%magnitude of gaugino mass parameters. 

The article is organized as follows. In Section \ref{model} we discuss the model and the 
underlying particle spectrum of interest in detail. In the following Section \ref{survey} 
we focus on identifying the parameter space satisfying relevant constraints as well as 
implications on neutrino sector and a sneutrino as DM. In Section \ref{lhc} we discuss 
the possible signatures at LHC and present our analysis for a few representative points 
in the model parameter space. We finally conclude in Section \ref{sec:conclusion}.

\section{The Model}
\label{model}
We consider an extension to the Minimal Supersymmetric Standard Model (MSSM) by
introducing a right-chiral neutrino superfield for each generation. This 
extension addresses the important issue of neutrino mass generation 
which is otherwise absent in the MSSM. In particular, we adopt a 
phenomenological approach for ``TeV type-I seesaw mechanism". The 
superpotential, suppressing the generation indices, is given by 
\cite{ArkaniHamed:2000bq,Grossman:1997is,Arina:2007tm}:
\begin{align*}
 \mathcal{W} \supset \mathcal{W_{MSSM}} + y_\nu \hat{L} \hat{H_u} \hat{N^c} + 
 \frac{1}{2} M_R \hat{N^c} \hat{N^c}
\end{align*}
where $y_\nu$ is the neutrino Yukawa coupling, $\hat{L}$ is the left-chiral 
lepton doublet superfield, $\hat{H_u}$ is the Higgs up-type chiral superfield 
and $\hat{N}$ is the right-chiral neutrino superfield. Besides the usual MSSM
superpotential terms denoted by $\mathcal{W_{MSSM}}$, we now have an added 
Yukawa interaction term involving the left-chiral superfield $\hat{L}$ coupled 
to the up-type Higgs superfield $\hat{H_u}$, and $\hat{N}$. SM neutrinos 
obtain a Dirac mass $m_D$ after electroweak symmetry breaking once the neutral 
Higgs field obtains a vacuum expectation value ($\textit{vev}$) $v_u$, such 
that $m_D = y_\nu v_u$. The third term $\frac{1}{2} M_R \hat{N^c} \hat{N^c}$ 
is a lepton-number violating ($ \slashed{L} $) term ($\triangle L = 2 $). 

In addition to the MSSM contributions, the soft-supersymmetry breaking 
scalar potential receives additional contributions as follows:
\begin{align*}
 \mathcal{V}^{soft} \supset \mathcal{V}^{soft}_{MSSM} + 
 m^2_{R} |\widetilde{N}|^2 + \frac{1}{2} B_M \widetilde{N}^c \widetilde{N}^c + 
 (T_\nu \widetilde{L}.H_u \widetilde{N}^c + \text{ h.c.})
\end{align*}
where $m^2_{R}$ is the soft-supersymmetry breaking mass parameter for the 
sneutrino, $B_M$ is the soft mass-squared parameter corresponding to the 
lepton-number violating term and $T_\nu$ is the soft-supersymmetry breaking 
L-R mixing term in the sneutrino sector. We have suppressed the generation 
indices both for the superpotential as well as for the soft 
supersymmetry-breaking terms so far. 

Note that a small $\mu$-parameter is critical to ensure the 
absence of any fine-tuning at the EW scale ($\Delta_{\rm EW}$) 
\cite{Baer:2012up,Baer:2012cf,Baer:2013ava,Mustafayev:2014lqa}. 
Fine-tuning arises if there is any large cancellation involved at 
the EW scale in the right hand side of 
the following relation \cite{Barbieri:1987fn,Ellis:1986yg} :  
\begin{equation}
\dfrac{M_Z^2}{2}= \dfrac{m^2_{H_d}+ \Sigma_d -(m^2_{H_u}+\Sigma_u)\tan \beta^2}{\tan\beta^2 -1} -\mu^2,\
\end{equation}
where $m^2_{H_u},~ m^2_{H_d}$ denote the soft-supersymmetry breaking terms 
for the up-type and the down-type Higgses at the supersymmetry breaking mass scale 
(which is assumed to be the geometric mean of the stop masses in the present 
context) and $\tan\beta$ denotes the ratio of the respective $vev$s while $\Sigma_u \,\, {\rm and} \,\,
\Sigma_d$ denote the radiative corrections. 
Note that, since we are not considering any specific high-scale framework in the 
present context, we are only concerned about the EW fine-tuning. Typically 
$\Delta_{\rm EW} \lesssim 30$ is achieved with $|\mu| \lesssim 300$ GeV 
\cite{Baer:2012up,Baer:2012cf,Baer:2013ava,Mustafayev:2014lqa}. The assurance of EW
naturalness is the prime motivation in exploring small $\mu$ scenarios. However it is quite
possible that obtaining such a spectrum from a high-scale theory may require larger 
fine-tuning among the high-scale parameters and the corresponding running involved, 
especially considering that $m_{H_u}$ evolves significantly to ensure radiative 
EW symmetry breaking. Therefore, $\Delta_{\rm EW}$ can be interpreted as a lower 
bound on fine-tuning measure \cite{Baer:2012up,Baer:2012cf,Baer:2013ava,
Mustafayev:2014lqa}. Note that, stop squarks and gluinos contribute to the 
radiative corrections to $m_{H_u}$ at one and two--loop levels respectively. 
It has  been argued \cite{Baer:2015rja, Baer2017} that an EW fine-tuning of less 
than about 30 can be achieved with $\mu \lesssim 300$ GeV and with stop squarks 
and (gluinos) as heavy as about  3 TeV (4 TeV). It is, therefore, important to 
probe possible scenarios with low $\Delta_{\rm EW}$ and therefore with low $|\mu|$. 

%%%%%%
\subsection{The (s)neutrino sector}
\label{sec:snu}
In presence of the soft-supersymmetry-breaking terms $B_M$, a split is generated between 
the CP-even and the CP-odd part of right-type sneutrino fields. In terms of CP eigenstates 
we can write: $\widetilde{\nu}_L = \dfrac{\widetilde{\nu}^e_L + i\widetilde{\nu}^o_L }{\sqrt{2}}$; $\widetilde{\nu}_R = \dfrac{\widetilde{\nu}^e_R + i\widetilde{\nu}^o_R }{\sqrt{2}}$,
 where superscripts $e$, $o$ denote ``even" and ``odd" respectively.
The sneutrino ($\widetilde{\nu}$) mass-squared matrices in the basis $\widetilde{\nu}^{e} = \{ \widetilde{\nu}^e_L, \widetilde{\nu}^e_R\}^T$ and 
$\widetilde{\nu}^{o} = \{ \widetilde{\nu}^o_L, \widetilde{\nu}^o_R\}^T $ 
are given by, 
\begin{center}
$\mathcal{M}^{j~2} =$
$ \begin{pmatrix}
   m_{LL}^2 & \text{  }  &     m_{LR}^{j2} \\
     &  \\
    m_{LR}^{j~2} & \text{  }  &  m_{RR}^{j~2}\\ 
  \end{pmatrix}$,
 \label{eq:numass}
\end{center}
where, 
\begin{eqnarray}
   m_{LL}^2 &=& m^{2}_L + \frac{1}{2}m^2_Z \cos{2\beta} + m^{2}_D, \nonumber \\ 
   m_{LR}^{j ~2} & = &   (T_\nu\pm y_{\nu} M_R) v \sin \beta - \mu m_D \cot{\beta}, \nonumber\, \\ 
   m_{RR}^{j~2} &=& m^{2}_R + m^{2}_D  + M^{2}_R \pm B_M,
   \label{eq:numass1}
\end{eqnarray}
with $j \in \{e,o\}$ and the `+' and the `-' signs correspond to $j=e$ and $j=o$ respectively, and $v= \sqrt{v_u^2+v_d^2} = 174$ GeV, where $v_u, ~v_d$ denotes the $vevs$ of the up-type and the down-type CP-even neutral Higgs bosons. Further, we have assumed $T_{\nu}$ to be real and with no additional CP-violating parameters in the sneutrino sector. The physical masses 
and the mass eigenstates can be obtained by diagonalizing these matrices. The eigenvalues
are given by : 
\begin{equation}
m^{j~2}_{1,2}=\dfrac{1}{2} \left(m_{LL}^2 + m_{RR}^{j~2} \pm \sqrt{(m_{LL}^2-m_{RR}^{j~2})^2 +4 m_{LR}^{j~4}}\right).
\label{eq:numass2}
\end{equation}
The corresponding mass eigenstates are give by, 
\begin{eqnarray}
\widetilde{\nu}^{j}_{1} & = & \cos\varphi^{j} \widetilde{\nu}^{j}_L 
- \sin\varphi^{j} \widetilde{\nu}^{j}_R \nonumber \\ 
\widetilde{\nu}^{j}_{2} & = & \sin\varphi^{j} \widetilde{\nu}^{j}_L 
+ \cos\varphi^{j} \widetilde{\nu}^{j}_R.
\end{eqnarray}
The mixing angle $\theta =  \frac{\pi}{2}- \varphi $ is given by, 
\begin{equation}
\sin 2 \theta^{j} = \dfrac{(T_{\nu}\pm y_{\nu} M_R)v \sin \beta - \mu m_D \cot \beta}{m_{2}^{j 2}-m_{1}^{j2}},
\label{eq:numix}
\end{equation}
where $j$ denotes CP-even ($e$) or CP-odd ($o$) states. 
 
The off-diagonal term involving $T_\nu$ is typically proportional to the coupling $y_\nu$, 
ensuring that the left-right (L-R) mixing is small. However, the above assumption relies on 
the mechanism of supersymmetry-breaking and may be relaxed. The  phenomenological 
choice of a large $T_{\nu} \sim \mathcal{O} (1) {\rm GeV}$ leads to increased mixing 
between the left and right components of the sneutrino flavor eigenstates in the 
sneutrino mass eigenstates \cite{Arina:2007tm,Belanger:2010cd,Dumont:2012ee}. Further, 
if the denominator in eq. (\ref{eq:numix}) is suitably small, it can also lead to 
enhanced mixing.

As for the neutrinos, at tree-level with $M_R \gg 1$ eV, their masses are given by 
$m_{\nu} \simeq \dfrac{y_{\nu}^2v_u^2}{M_R}$, as in the case of Type-I see-saw mechanism 
\cite{MINKOWSKI1977421,Yanagida:1979as, Mohapatra}. Thus, with $M_R \sim \mathcal{O}(100)$ 
GeV, neutrino masses of $\mathcal{O} (0.1)$ eV requires $y_{\nu} \sim 10^{-6}-10^{-7}$.
Although we have ignored the flavor indices in the above discussion of the sneutrino 
sector, the neutrino oscillation experiments indicate that these will play an important 
role in the neutrino sector. We will assume that the leptonic Yukawa couplings are 
flavor diagonal, and that the only source of flavor mixing arises from $y_{\nu}$ 
\cite{Casas:2001sr}; see also \cite{Drewes:2013gca,Rasmussen:2016njh}.
\begin{figure}[H]
\begin{center}
 \includegraphics[width=3in]{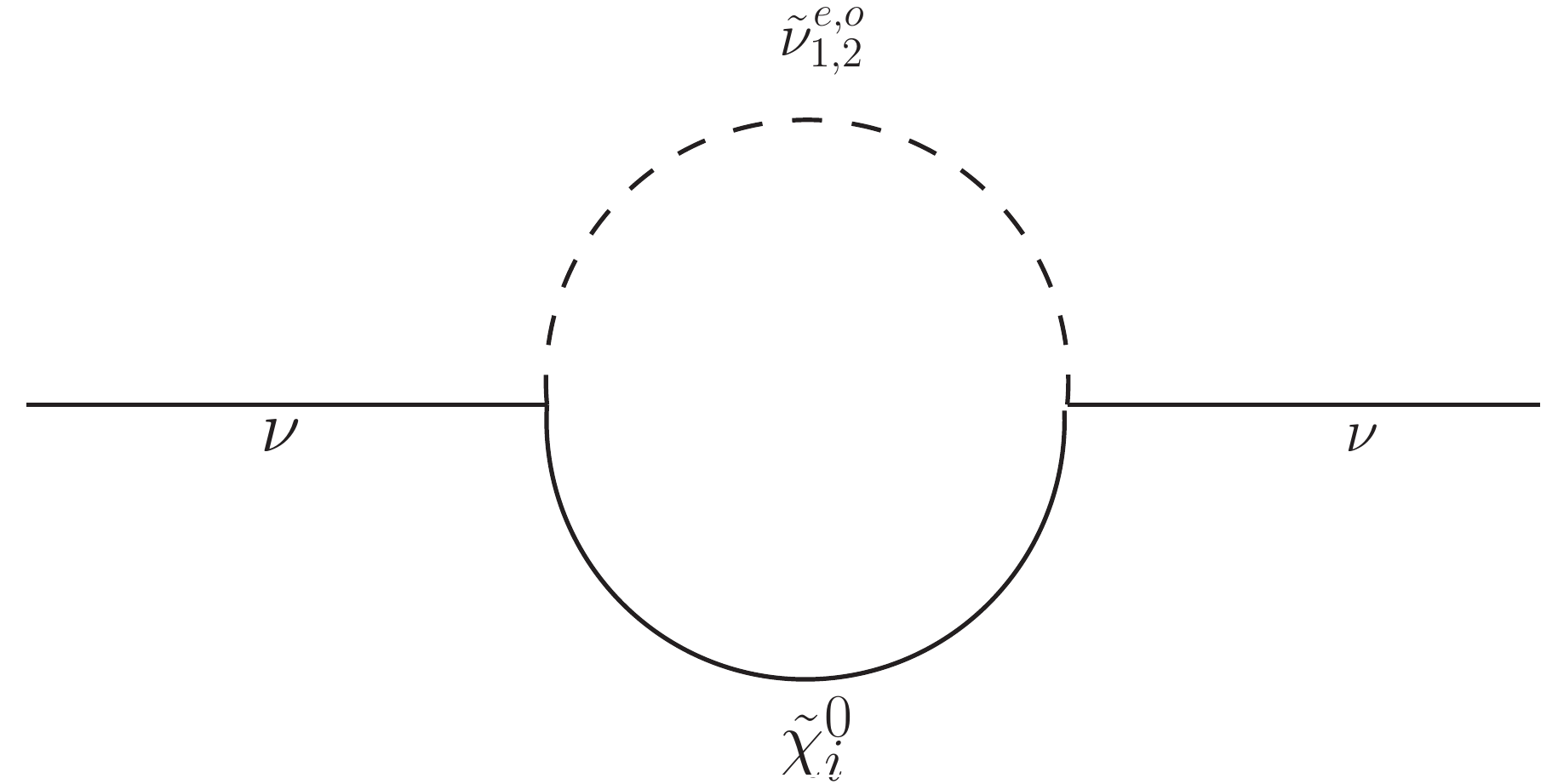}
  \caption{Schematic diagram showing the leading one-loop contribution to the light neutrino mass. } 
\label{fig:nuloop}
\end{center}
\end{figure}
Further, at one-loop, flavor diagonal $B_M$ can also contribute to the 
neutrino mass matrix \cite{Grossman:1997is,Dedes:2007ef} which can be quite significant in 
the presence of large $T_{\nu}$ in particular.\footnote{Note that flavor off-diagonal 
terms in $B_M$ can lead to flavor mixing in the neutrino sector via higher order effects 
which we avoid in our discussions for simplicity.} 
The dominant contribution to the 
Majorana mass of the active neutrino arises from the sneutrino-gaugino loop 
as shown in fig. \ref{fig:nuloop}. The contributions from the loop are proportional to 
the mass splitting between the CP-even and the CP-odd left-sneutrino state which makes it  
significant in the presence of a rather large $T_{\nu}$ which is responsible for 
left-right mixing in the sneutrino sector (see eq. \ref{eq:numix}). These additional 
contributions to the neutrino mass give significant constraints in the $\{T_{\nu}, B_M\}$ 
parameter space. %We shall discuss this in more detail in Section \ref{survey}.
%Also note that, as discussed in ref \cite{Kumar:2009sf}, if one 
%considers Dirac gauginos instead (i.e.in an extension of the MSSM), the large 
%radiative contribution to the neutrino mass does not appear at one-loop.} 

Finally, some comments on the scenario with $M_R = 0$ and $B_M = 0$ are in order. With 
$M_R = 0$ (and $B_M = 0$), only Dirac mass terms would be present for neutrinos, which 
is given by  $y_{\nu} v_u$. The oscillation data for neutrinos can only be satisfied by 
assuming $y_{\nu} $ (and/or $T_{\nu}$, at one-loop order) to be flavor off-diagonal. 
In addition, $\mathcal{O}(0.1)$ eV neutrino mass, then, requires a very small $y_{\nu}\simeq 10^{-11}$. 

In the sneutrino sector, the relevant mass eigenstates may be obtained simply by 
substituting $M_R=0=B_M$ in equations (\ref{eq:numass}, \ref{eq:numass1},  
\ref{eq:numass2}). Since the mass matrices for both CP-even and the CP-odd sneutrinos 
are identical in this scenario, any splitting between the corresponding mass 
eigenstates would be absent. Consequently there will be only two complex-scalar mass 
eigenstates $\tilde{\nu}_1, \tilde{\nu}_2$. Also, there will be no large one-loop 
contribution to the Majorana neutrino mass, relaxing the constraint on large $T_{\nu}$ 
significantly. 
 
\subsection{The Electroweakino sector}
\label{sec:delm}
The other relevant sector for our study is the chargino-neutralino sector, in 
particular the higgsino-like states. This sector resembles the chargino-neutralino 
sector of the MSSM. The tree-level mass term for the charginos, in the gauge 
eigen-basis, can be written as \cite{Drees:2004jm}
\beq
-\mathcal{L}^{\rm c}_{\rm mass}  =  \psi^{-T} M^c \psi^{+} + h.c.
\eeq
where,
\beq 
\psi^+ = (\widetilde{W}^{+},~~\tilde{h}^{+}_{2})^{T},~~ 
\psi^- = (\widetilde{W}^{-},~~\tilde{h}^{-}_{1})^{T}
\eeq
are column vectors whose components are Weyl spinors. The mass matrix $M^{\rm c}$ 
is given by
\beq \label{mc}
M^{\rm c}= \left( \begin{array}{cc}
M_{2} & \sqrt{2} M_W\sin \beta \\
\sqrt{2} M_W \cos \beta & \mu \\
\end{array} \right).
\eeq
In the above equation, $M_2$ is the supersymmetry breaking $SU(2)$ gaugino (wino) 
mass parameter, $\mu$ is the supersymmetric higgsino mass parameter, $M_W$ is the mass of the 
$W$ boson, and $\tan\beta$ is the ratio of $vevs$ as described before. The non-symmetric 
$M^{\rm c}$ can be diagonalized with a bi-unitary transformation using the
unitary matrices $U $ and $V$ to obtain the diagonal mass matrix,
\beq \label{mcd}
M^{\rm c}_{D} = U^*M^{\rm c}V^{-1} = {\rm Diagonal} (m_{\tilde{\chi}_{1}^{+}}
~m_{\tilde{\chi}_{2}^{+}}) . 
\eeq
The eigenstates are ordered in mass such that $m_{\tilde \chi_{1}^{+}} 
\leq m_{\tilde \chi_{2}^{+}}$. 
%\footnote{Note that scenarios with $m_{\tilde \chi_1^+} \lesssim 100$ GeV are excluded by chargino searches at LEP \cite{PDG}.} 
The left-- and right--handed components of the corresponding Dirac mass eigenstates, 
the charginos $\tilde \chi^+_i$ with $i \in \{1,2\} $, are 

\beq \label{ec}
P_L \tilde \chi^{+}_i = V_{ij} \psi^+_j,~~ P_R \tilde \chi^+_i = U^*_{ij}
  \overline{\psi^-_j}\,,
\eeq
where $P_L$ and $P_R$ are the usual projectors, $\overline{\psi^-_j} = \psi^{- 
\dagger}_j$, and summation over $j$ is implied. 

For the electrically neutral neutralino states, in the gauge eigenbasis, $\psi^{0} = \left( \begin{array}{cccc} \tilde{B^{0}}, & 
\widetilde W^3, & \tilde h_1^0, & \tilde h_2^0 \end{array}\right)^{T}$, the 
tree level mass term is given by \cite{Drees:2004jm} 
\beq
-\mathcal{L}^{\rm n}_{\rm mass}  =  \frac{1}{2} \psi^{0T} M^n
\psi^{0} + h.c.  
\eeq
The neutralino mass matrix $M^{\rm n}$ can be written as
\beq \label{mn}
M^{\rm n} =  \left( \begin{array}{cccc}
M_{1} & 0 & -M_{Z}s_W c_\beta & M_{Z}s_W s_\beta\\
0 &  M_{2} & M_{Z}c_W c_\beta & -M_{Z}c_W s_\beta \\
-M_{Z}s_W c_\beta & M_{Z}c_W c_\beta & 0 & -\mu \\ 
M_{Z}s_W s_\beta & -M_{Z}c_W s_\beta & -\mu & 0 \end{array} \right)\,.
\eeq
In the above mass matrix $s_W, s_\beta, c_W$ and $c_\beta$ stand for $\sin \theta_W, 
\sin \beta, \cos \theta_W$ and $\cos \beta$ respectively while $\theta_W$ is the 
weak mixing angle. $M_Z$ is the mass of the $Z$ boson, and $M_1$ is the 
supersymmetry breaking $U(1)_Y$ gaugino (bino) mass parameter. $M^{\rm n}$ can be 
diagonalized by a unitary matrix $N$ to obtain the masses of the neutralinos 
as follows, 
\beq \label{mnd}
M^{\rm n}_{D}=N^* M^{\rm n} N^{-1}= {\rm  Diagonal} (m_{\tilde{\chi}^{0}_{1}} ~
m_{\tilde{\chi}^{0}_{2}} ~m_{\tilde{\chi}^{0}_{3}} ~m_{\tilde{\chi}^{0}_{4}})
\eeq
Again, without loss of generality, we order the eigenvalues such that
$ m_{\tilde{\chi}^{0}_{1}} \leq m_{\tilde{\chi}^{0}_{2}} \leq m_{\tilde{\chi}^{0}_{3}} \leq m_{\tilde{\chi}^{0}_{4}}. $

The left--handed components of the corresponding mass eigenstates, described 
by four--component Majorana neutralinos $\tilde \chi_i^0$ with $i \in \{1,2,3,4\}$, 
may be obtained as,
\beq
P_L \tilde \chi^{0}_i = N_{ij}  \psi_j^{0},
\eeq
where summation over $j$ is again implied; the right--handed components of the 
neutralinos are determined by the Majorana condition $\tilde \chi_i^c = \tilde 
\chi_i$, where the superscript $c$ stands for charge conjugation. 

Since the gaugino mass parameters do not affect ``naturalness", for 
simplicity we have assumed $M_1, ~M_2 \gg |\mu|$. In this simple scenario there are 
only three low-lying higgsino-like states, $\tilde\chi^0_1$, $\tilde\chi^0_2$  and  $\tilde\chi^{\pm}
_1$. The EW symmetry breaking induces mixing between the gaugino and the 
higgsino-like states, via the terms proportional to $M_Z, ~ M_W$ in the mass 
matrices above. 
The contributions of the right-chiral neutrino superfields to the chargino and neutralino mass matrices 
are negligible, thanks to  the smallness of $y_{\nu}$ ($\simeq 10^{-6}$). Thus lightest neutralino and 
charginos are expected to be nearly the same as in the MSSM. 
Following \cite{Drees:1996pk} (see also \cite{Giudice:1995np}), in the 
limit $M_1, ~M_2 \gg |\mu|$, we give the analytical expression for the masses below, 
\bea
m_{\tilde\chi^{\pm}_1} & = & |\mu |\left(1- \frac{M_W^2 \sin2\beta}{\mu  M_2}\right) + 
\mathcal{O}(M_2^{-2})+ {\rm rad. corr.} \nonumber\\
m_{\tilde\chi^{0}_{a,s}} & = & \pm \mu - \frac{M_Z^2}{2}(1\pm \sin2\beta)\left(\frac{\sin
\theta_W^2}{M_1}+\frac{\cos \theta_W^2}{M_2}\right) + {\rm rad. corr.} 
\label{eq:mhiggsino}
\eea
 where the subscripts $s$ ($a$) denote symmetric (anti-symmetric) 
states respectively, and the sign of the eigenvalues have been retained. For 
the symmetric state  $N_{i3}, ~N_{i4}$ share the same sign, while for the 
anti-symmetric state there is a relative sign between these two terms.   
Although the leading contribution to the mass eigenvalues are given by $|\mu|$ (which 
receives different radiative corrections in $M^n$ and $M^c$), $M_1,~ M_2 $ and 
$\tan \beta$ affects the mass splitting between the three light higgsino-like 
states due to non-negligible gaugino-higgsino mixing. The radiative corrections, 
mostly from the third generation (s)quarks, contribute differently for 
$m_{\tilde\chi^{\pm}_1}$ and $m_{\tilde\chi^{0}_{1,2}}$ and have been estimated in 
\cite{Pierce:1993gj,Pierce:1994ew,Lahanas:1993ib,Drees:1996pk}. As we are interested 
in a spectrum where the lighter chargino and the neutralinos play a major role and the knowledge 
of their mass differences would become crucial, it is necessary to explore what role the 
relevant SUSY parameters have in contributing to the masses of the higgsino dominated states.
It is quite evident from our choice of large $M_1$ and $M_2$ that the three states according to 
eq. \ref{eq:mhiggsino} would be closely spaced. 

We now look at how the variation of the the above 
gaugino parameters affect the shift in mass of $m_{\tilde\chi^{\pm}_1}$ and 
$m_{\tilde\chi^{0}_{1,2}}$.
\begin{figure}[ht!]
\label{fig:delm}
\includegraphics[width=3.2in]{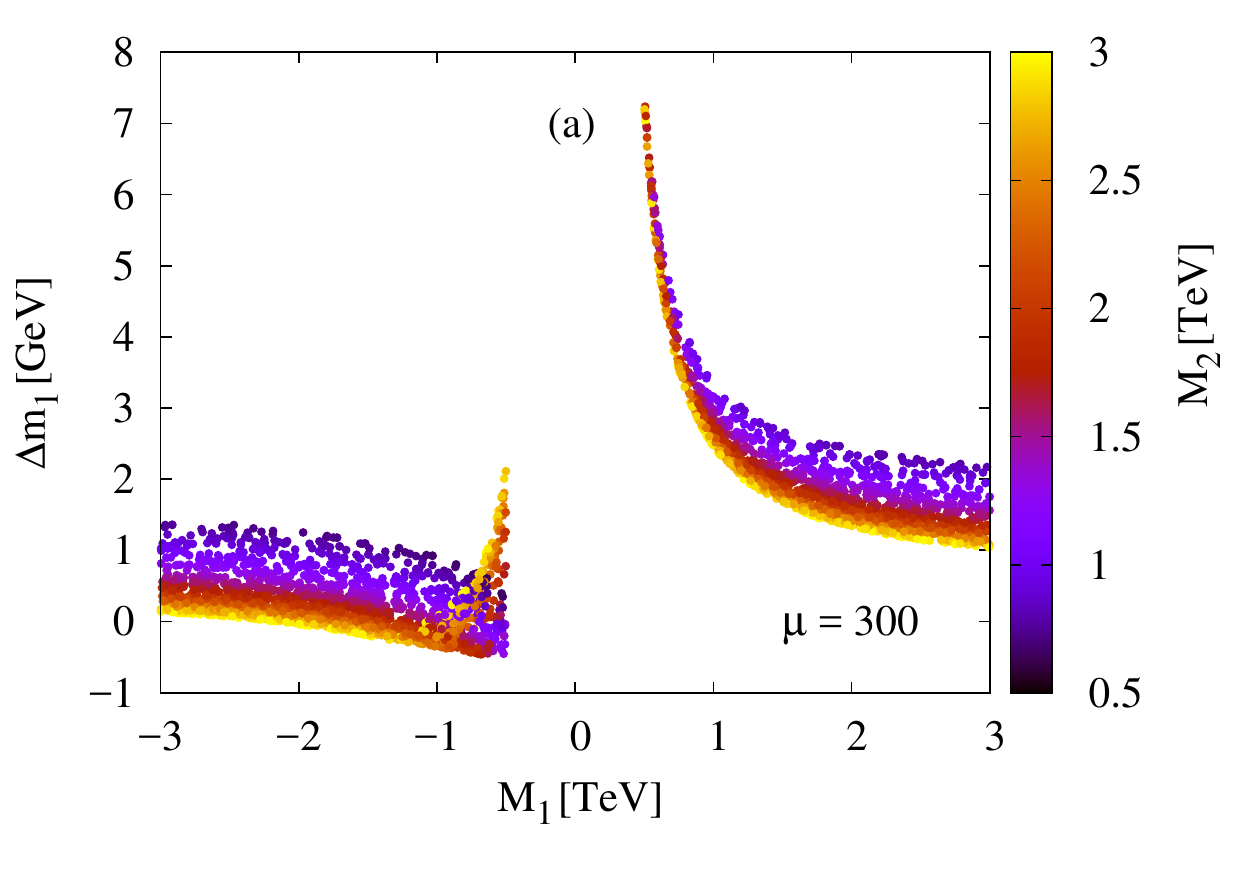} 
\includegraphics[width=3.2in]{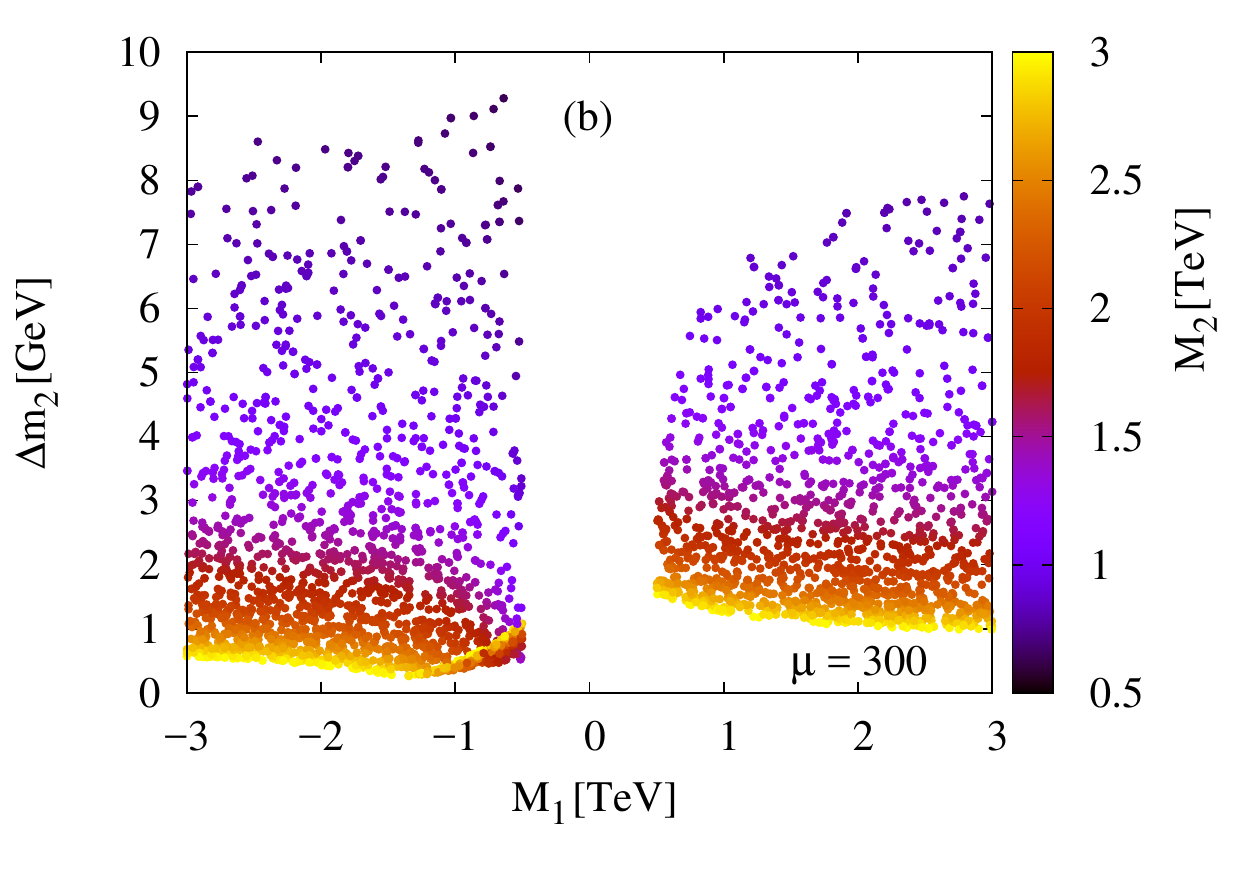}
\includegraphics[width=3.2in]{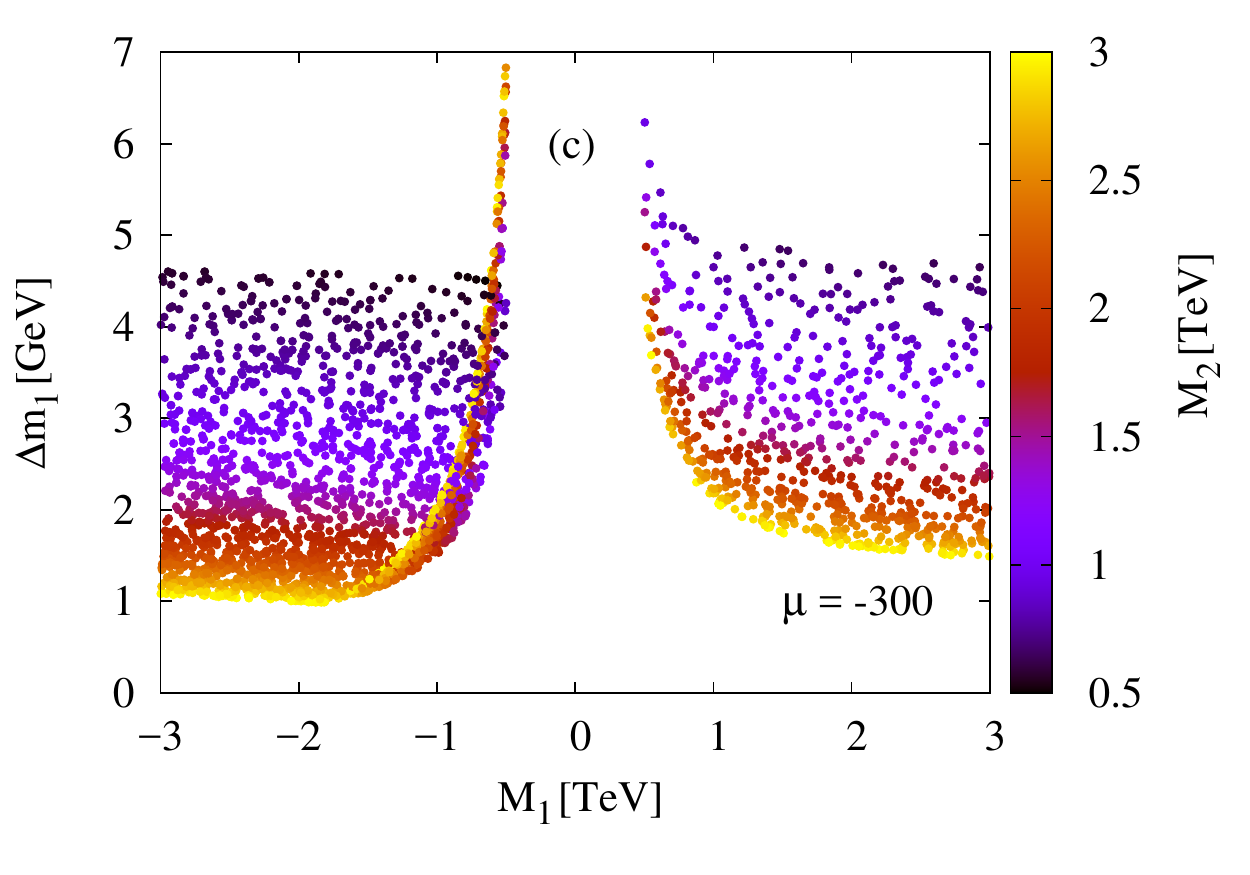} 
\includegraphics[width=3.2in]{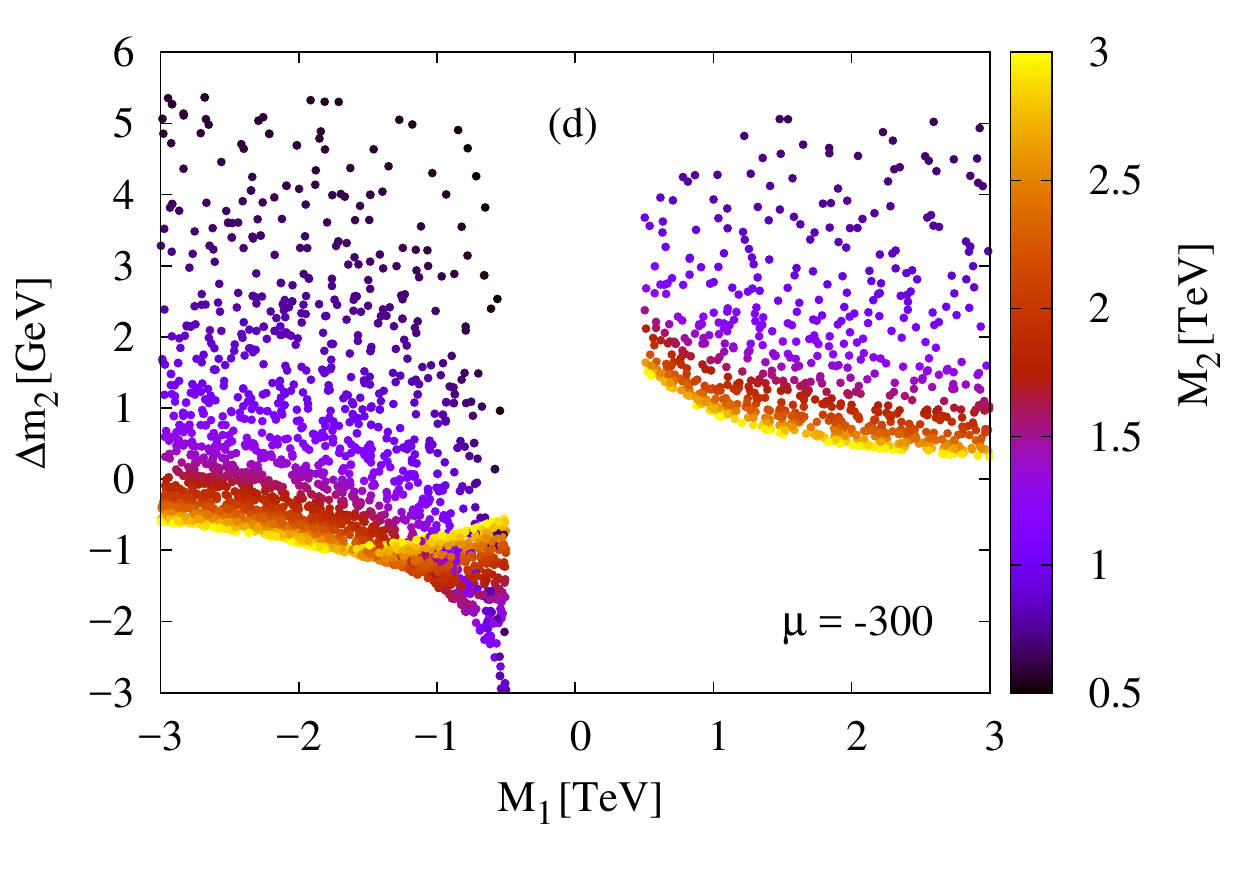}
\caption{The left (right
) panel shows the variation of the mass difference 
$\Delta m_1 = m_{\widetilde{\chi}^{\pm}_1}-m_{\widetilde{\chi}^0_1}~(\Delta m_2= 
m_{\widetilde{\chi}^0_2}-m_{\widetilde{\chi}^{\pm}_1})$
between $\tilde{\chi}_1^{\pm}$ and  $\tilde{\chi}^0_1$  [$\tilde{\chi}_2^0$] 
for $\tan\beta =5 $ with respect to $M_1$, with $M_2$ on the palette.} 
\end{figure}
Assuming $\mu=300$ GeV, $\tan\beta= 5$, in fig. \ref{fig:delm} we show the variation 
of the mass differences $\Delta m_1 = m_{\widetilde{\chi}^{\pm}_1}-m_{\widetilde{\chi}^0_1}$ and $ \Delta m_2= 
m_{\widetilde{\chi}^0_2}-m_{\widetilde{\chi}^{\pm}_1} $ as a function of the gaugino mass 
parameters. $M_1$ and $M_2$ have been varied from 500 GeV to 3 TeV. Further, 
we have set $T_t = 2.9 \,\,{\rm TeV}, ~M_{Q_3} = 
1.3 \,\, {\rm TeV}, ~M_{U_3} = 
2 \,\,{\rm TeV}$ and $ M_3 = 2 \,\,{\rm TeV}$. 
We have used \texttt{SARAH} \cite{Staub:2008uz,Staub:2013tta} to generate model 
files for \texttt{SPheno} \cite{Porod:2003um,Porod:2011nf}, and have used the 
same to estimate the masses. Since \texttt{SLHA} \cite{Skands:2003cj} convention 
has been followed, the input parameters, as shown in the figures above, are 
interpreted as $\overline{\rm DR}$ parameters at $\sim 1.6$ TeV. Note that the 
same model and spectrum generators have been used for all subsequent figures. 
The following features are noteworthy from fig. \ref{fig:delm} :\footnote{{Although our numerical analysis, as shown in figure \ref{fig:delm}, 
includes radiative corrections, the generic features also appear at the tree-level} 
for $|\mu|= 300$ GeV, $M_1,~ M_2 \gg |\mu|$ and $\tan\beta = 5$.We have checked 
this using a \texttt{Mathematica} code.} 
\begin{itemize} 
\item For $\mu>0$; $M_1, M_2 \gg \mu$: Here $\tilde{\chi}_2^0$ is the 
heaviest higgsino-like state while $\tilde{\chi}_1^{\pm}$ remains between 
the two neutralinos.  For a fixed $M_1 \gg |\mu|$, the mass difference $\Delta m_1$
increases as $M_2$ decreases. 
This feature can be  simply understood from eq. \ref{eq:mhiggsino}. A similar conclusion 
also holds for $\Delta m_2$. Further, as shown in panels (a) and (b) of 
fig. \ref{fig:delm}, the variation in $\Delta m_2$ is larger compared to 
$\Delta m_1$ in this case.   

\item For $\mu>0$;$M_1< 0$:  We find that negative $M_1$ can lead to 
negative $\Delta m_1$, since the lightest chargino can become lighter than this state 
for a wide range of $M_2$ \cite{Kribs:2008hq,Barducci:2015ffa,Han:2014kaa}. 
As shown in fig. \ref{fig:delm}(a), such a scenario occurs for large $M_2$ values 
($\gtrsim 2$ TeV) with  $|M_1| \lesssim 1$ TeV. Further, 
for  $|M_1| \ll M_2$, as $|M_1|$ decreases one observes 
an upward kink in the $\Delta m_1$ and $\Delta m_2$ plots as shown in 
figs. \ref{fig:delm}(a) and \ref{fig:delm}(b) which can be attributed to the  change in 
nature of the lightest neutralino state from anti-symmetric to the symmetric 
state.
\item For $\mu<0$; $M_1>0$:  As shown in figs. \ref{fig:delm}(c) and \ref{fig:delm}(d), similar 
to the $\mu > 0$ case, $\Delta m_i$ smoothly increases with decreasing $M_2$ in this region as well. 
\item For $\mu< 0$; $M_1<0$:  In fig. \ref{fig:delm}(c) we again see (due to the change in nature of LSP) 
a  sharp rise of $\Delta m_1$ for large $M_2 \gtrsim 2$ TeV and $|M_1| \lesssim 1.5$ TeV. Note that in this case the $\tilde{\chi}_1^{\pm}$ can be the heaviest 
higgsino-like state in a substantial region of the parameter space for $M_2 \gtrsim 2$ TeV, 
as shown in fig. \ref{fig:delm}(d).   
\end{itemize}
 
%We will further discuss the implications of the variation of $\Delta m_i$ among these 
%light higgsino-like states with rather heavy gaugino mass parameters in section 
%\ref{survey}.

%%%%%%%%%%%%%
%\subsection{The spectrum} 
\subsection{Compressed Higgsino spectrum and its decay properties}
\label{sec:decay}
As we have already emphasized, the focus of this work is on  higgsino-like NLSPs in 
a scenario with a right-sneutrino LSP where the choice of small $|\mu|$ is motivated by 
the ``naturalness" criteria \cite{Baer:2013ava,Mustafayev:2014lqa,Baer:2015rja}. 
Thus we will restrict our discussions to scenarios where the higgsino mass 
parameter $|\mu| \lesssim 500$ GeV. The gaugino mass parameters have been 
assumed to be heavy for simplicity; thus the light higgsino-like states are 
quite compressed in mass (fig. \ref{fig:delm}). 
\begin{figure}[H]
\begin{center}
\includegraphics[height=2.0in,width=2.0in]{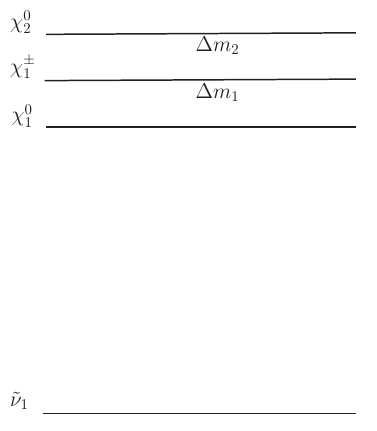}
\includegraphics[height=2.05in,width=2.1in]{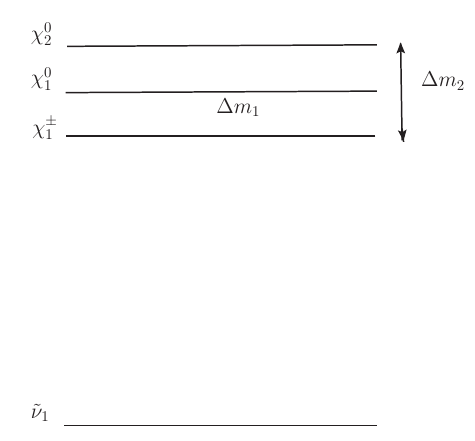} 
 \caption{Schematic description of the mass spectrum with $|M_1|, M_2 \gg |\mu|$, 
  and $m_{\tilde{\nu}_1} < |\mu|$. Here $|m_{\tilde{\chi}^0_2}| -  m_{\tilde{\chi}^{\pm}_1}  = \Delta m_2, 
  ~  m_{\tilde{\chi}^{\pm}_1}-|m_{\tilde{\chi}^{0}_1}|  = \Delta m_1$.} 
 \label{fig:spec}
\end{center}
\end{figure}

Note that since the gaugino mass parameters are much heavier, the gaugino fraction 
in the higgsino--like states are small ($\mathcal{O}(10^{-2})$). However, $M_1$ 
and $M_2$ play significant role in determining $\Delta m_1$ and $\Delta m_2$ and 
also the hierarchy between the higgsino--like states. While for most parameter space 
the spectra shown in the left panel of fig. \ref{fig:spec} is realized, for $M_1 < 0 $ (i.e. 
sign($M_1 M_2$) = -1), it is possible to achieve the chargino as the lightest 
higgsino-like state which leads to a spectra as shown in the right panel of fig. \ref{fig:spec}. 
Further, with $\mu, M_1 < 0$ one can also have the chargino as the heaviest of the 
three higgsino-like state. However, as we will discuss subsequently in section 
\ref{lhc}, this does not contribute to any new signature. 
%and we will therefore mostly focus on the spectra  shown in the figure.  
Fig. \ref{fig:spec} schematically shows  the mass hierarchies of our interest. 
%
%We consider the LSP to 
%be dominantly right-sneutrino with a small left admixture 
%$(\lesssim \mathcal{O}(10^{-3})$. This is to ensure prompt decays of 
%the lightest chargino into LSP via the gaugino fraction, typically 
%$\mathcal{O}(10^{-2})$.
For the electroweakinos which are dominantly higgsino-like, their production rates and subsequent decay 
properties would have serious implications on search strategies at accelerator machines like LHC. This in 
turn would play an important role in constraining the higgsino mass parameter $\mu$
in the natural SUSY framework. 

We now try to briefly motivate the compositions of the 
LSP as well  as the higgsino-like states of our interest and their decay properties. 
In the presence of $\tilde{\chi}_1^0$ as the lightest higgsino--like state, the 
decay modes available to the chargino are 
$\tilde{\chi}_1^{\pm} \rightarrow l \, \tilde{\nu}^j_k$ and 
$\tilde{\chi}_1^{\pm} \rightarrow \tilde{\chi}_1^0 W^{\pm *}$, where $j,~k$  
corresponds to a particular lighter sneutrino species. The partial width 
to the 3-body decay modes, mostly from the off-shell $W$ boson mediated 
processes, are suppressed by the small mass difference while small $y_{\nu} 
(\lesssim 10^{-6})$  suppresses the 2-body decay mode. In such a scenario, 
the gaugino fraction 
in $\tilde{\chi}^{\pm}_1$, can contribute to the 2-body mode significantly 
in the presence of small left-right mixing $(\sim \mathcal{O}(10^{-5}))$ 
in the sneutrino sector. 

\begin{figure}
\begin{center}
\includegraphics[width=3.5in]{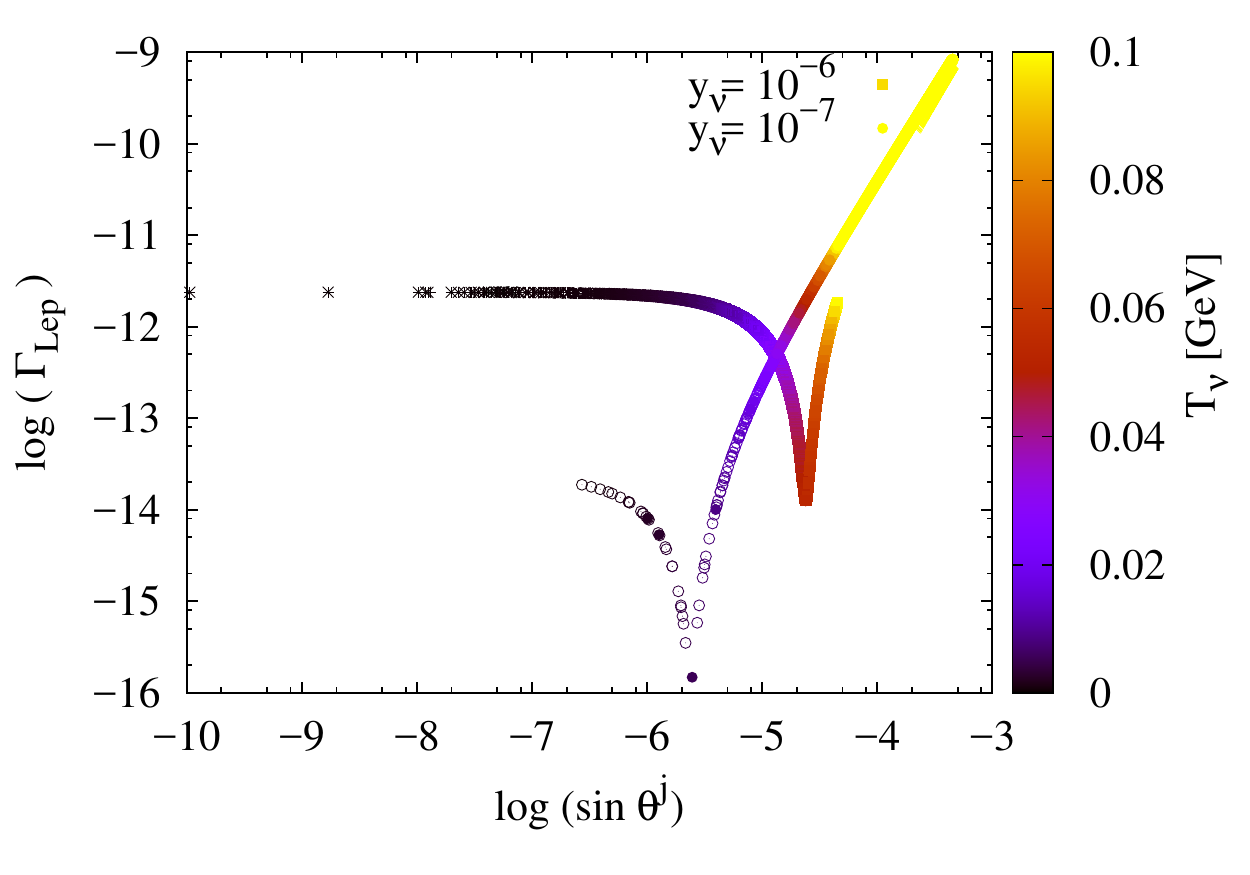}
\caption{Variation of the partial decay width of $\widetilde{\chi}^{\pm}_{1} 
\rightarrow l \, \widetilde{\nu}$  versus sin($\theta^j$) in 
logarithmic scale for $M_1 = -1.5$ TeV, $M_2 = 1.8 $ TeV and gaugino 
fraction $\sim \mathcal{O} (10^{-2})$. Further, $M_R$ = 100 GeV, $M_{L_{1/2}} 
= 600$ GeV, $m^{soft}_{\tilde\nu}=100$ GeV, $\mu=$ 300 GeV and $\tan\beta=$ 5. 
The colored palette corresponds to  $T_{\nu}$, the soft left-right mixing 
parameter in the sneutrino sector. The plot shows the required $T_{\nu}$ and 
mixing angle sin($\theta^j$) for prompt decay of the chargino. We focus on 
the values of $T_{\nu}$ in our study ensuring prompt decays of the 
chargino.} 
\label{fig:promptcha}
\end{center}
\end{figure}

We illustrate the decay properies of $\tilde{\chi}^{\pm}_1$ based on the composition of 
the LSP in fig. \ref{fig:promptcha}.\footnote {The particular choice of gaugino mass 
parameters correspond to $\Delta m_1 \lesssim 1$ GeV, and the partial width in the 
corresponding hadronic channel is quite small ($\simeq 10^{-16}$ GeV). Thus, the 
leptonic partial width resembles the total width of $\tilde{\chi}_1^{\pm}$.}
As shown in fig. \ref{fig:promptcha}, for small $T_{\nu}$ and therefore for 
small left admixture in the sneutrino sector, $y_{\nu}$ dominates the decay 
of $\tilde{\chi}^{\pm}_1$. As $T_{\nu}$ increases past $\mathcal{O}(10^{-2})$, 
the gaugino fraction plays a crucial role, which explains the rise of the 
partial width in the 2-body leptonic decay mode. With $y_{\nu} \sim 10^{-6}$ 
prompt decay of the lightest chargino to the sneutrino and lepton is always 
ensured. However, for $y_{\nu}\sim 10^{-7}$ prompt decay of the chargino 
in the leptonic channel is not viable in the absence of adequate left 
admixture; $T_{\nu} \gtrsim \mathcal{O}(10^{-2})$ GeV is required to ensure 
prompt decay in the leptonic channel. The dip in fig. \ref{fig:promptcha} 
appears as a consequence of possible cancellation between the gaugino and 
the higgsino contributions to the vertex factor (e.g. $\propto (g_2 V_{11} \sin\theta^j - y_{\nu} V_{12} \cos \theta^j)$, $g_2$ is the $SU(2)$ gauge coupling). 
It is of our interest to study the scenario where the 2-body decay mode into 
$l \,\tilde{\nu}$ competes with the 3-body decay mode. Since the present work 
focuses on prompt decays, we ensure small left admixture with 
$T_{\nu} \gtrsim \mathcal{O}(10^{-2})$ GeV in the dominantly right-sneutrino 
LSP to ensure prompt decay of $\tilde{\chi}^{\pm}_1$ in the 2-body leptonic 
decay mode. The mass splitting $\Delta m_1 \gtrsim 1$ GeV has been considered 
to ensure a competing 3-body mode. 

Since we have assumed a compressed higgsino spectrum, together with a mostly
right-sneutrino LSP, the light higgsino states include $\tilde{\chi}_1^0,
~\tilde{\chi}_2^0,~\tilde{\chi}_1^{\pm}$ and at least one generation of CP-odd 
and/or CP-even sneutrino LSP as described in section \ref{model}. 
 In fig. \ref{fig:delm} we showed that for a fixed $|\mu|$, the hierarchy and 
the mass differences between the higgsino--like states are affected significantly 
by the choice of the gaugino mass parameters $M_1,~ M_2$, and sign($\mu$). 
In a similar compressed scenario within the MSSM, the higgsinos $\tilde{\chi}_2^0$ 
and $\tilde{\chi}_1^{\pm}$ decay into soft leptons or jets \cite{Djouadi:2001fa}
and $\tilde{\chi}_1^0$, producing $\slashed{E}_T$. 
%These mostly receive 
%contributions from the $Z$ and $W$ boson mediated processes respectively. 
Scenarios with compressed higgsinos in MSSM have been studied in the light of 
recent LHC data \cite{Baer:2011ec,Han:2014kaa, Barducci:2015ffa,Baer:2016usl,
Fukuda:2017jmk,Mahbubani:2017gjh}. For smaller mass differences, $130 ~{\rm MeV} 
\lesssim \Delta m_1 \lesssim 2$ GeV, the effective two-body process 
$\tilde{\chi}_1^{\pm} \rightarrow \pi^{\pm} \tilde{\chi}_1^0$ 
\cite{Chen:1995yu,Chen:1996ap,Chen:1999yf} can dominate the hadronic branching 
fraction. Further, when $\tilde{\chi}_2^0$ is also almost degenerate with $\tilde{\chi}_1^{\pm}$, 
for an even smaller mass difference $\Delta m_2$, $\tilde{\chi}_2^{0} \rightarrow \gamma 
\tilde{\chi}_1^0 $ can become significant \cite{Haber:1988px, Ambrosanio:1995az,
Ambrosanio:1996gz,Baer:2002kv}. Note that while the three-body decay 
modes (soft leptons/ jets and $\tilde{\chi}_1^0$) suffer from phase space 
suppression $\left((\Delta m)^5\right)$, the two-body mode ($ \gamma \tilde{\chi}_1^0$) 
is also suppressed by a loop factor.  

In addition to the above decay channels of the compressed higgsino-like states, 
the present scenario with a sneutrino LSP offers additional decay channels to the lighter 
sneutrinos. While a $\tilde{\chi}_1^{0} \rightarrow  \nu \, \tilde{\nu} $  would lead to 
missing transverse energy (as in the case for MSSM) without altering the signal topology if the neutralino 
was the LSP,  $\tilde{\chi}_1^{\pm} \rightarrow  l \, \tilde{\nu} $ would have a significant impact on the 
search strategies. For a pure right-sneutrino LSP this decay is driven by $y_{\nu}$.  In the 
presence of large $T_{\nu}$ and therefore a large left-right mixing in the sneutrino LSP, a 
gaugino fraction of $\gtrsim \mathcal{O}(10^{-2})$ in the higgsino-like chargino 
begins to play a prominent role as the decay is driven by a coupling proportional 
to $g \delta \epsilon$ where $\delta$ represents the gaugino admixture and $\epsilon$ 
represents the L-R mixing in the sneutrino sector.  The presence of multiple flavors of 
degenerate sneutrinos would lead to similar decay probabilities into 
each flavor and would invariably increase the branching to the two-body leptonic 
mode when taken together. 

In the present context, as has been emphasized, only prompt decays into the 
leptonic channels such as $\tilde{\chi}^{\pm}_1 \rightarrow l \, \tilde{\nu}$ and 
$\tilde{\chi}^0_i \rightarrow \tilde{\chi}^{\pm}_1 j_s j'_s$, where $j_s,~j'_s$ denote 
soft-jets or soft-leptons can give us a signal with one or more hard charged leptons in the 
final state. Since the latter consists of $\tilde{\chi}_1^{\pm}$ in the cascade, 
it can also lead to leptonic final states. These branching fractions would be affected by 
any other available decay channels and therefore it is important to study the different regions 
of parameter space for all possible decay modes of the light electroweakinos. As shown in 
fig. \ref{fig:delm}, while in most of the parameter space $\tilde{\chi}^{0}_1$ is the lightest 
higgsino-like state, and  $\tilde{\chi}_1^{\pm}$ is placed in between the two neutralinos 
(i.e. $m_{\tilde{\chi}_1^{0}} < m_{\tilde{\chi}_1^{\pm}} < m_{\tilde{\chi}_2^{0}}$), it 
is also possible to have $\tilde{\chi}_1^{\pm}$ as the lightest or the heaviest 
higgsino-like state. The important competing modes for 
$\tilde{\chi}^{\pm}_1$ and $\tilde{\chi}^{0}_2$ where 
$m_{\tilde{\chi}_1^{0}} < m_{\tilde{\chi}_1^{\pm}} < m_{\tilde{\chi}_2^{0}}$ include  
\begin{align*}
(a) \,\, \tilde{\chi}^{\pm}_1 \rightarrow \tilde{\chi}^{0}_1 j_s j_s^\prime/ \pi^{\pm}, && 
(b) \,\, \tilde{\chi}^{0}_2 \rightarrow  \tilde{\chi}^{0}_1 j_s j_s/ \gamma, && (c) \,\, \tilde{\chi}^{0}_2 \rightarrow 
\tilde{\chi}^{\pm}_1 j_s j_s^\prime /\pi^{\mp} 
\end{align*}
where $(c)$ is usually small. However, if 
$\tilde{\chi}_1^{\pm}$ is the lightest higgsino-like state, decay modes $(b)$ and $(c)$, 
together with  $\tilde{\chi}^{0}_1 \rightarrow \tilde{\chi}^{\pm}_1 j_s j_s^\prime/\pi^{\mp}$ can 
be present. Similarly, when $\tilde{\chi}_1^{\pm}$ is the heaviest higgsino-like state, decay modes 
$(a), \,\, (b)$ and   $\tilde{\chi}^{\pm}_1 \rightarrow \tilde{\chi}^{0}_2 j_s j_s^\prime/ 
\pi^{\pm}$ can be present, although the latter would be sub-dominant. 

\begin{figure}[ht]
\includegraphics[width=3.1in]{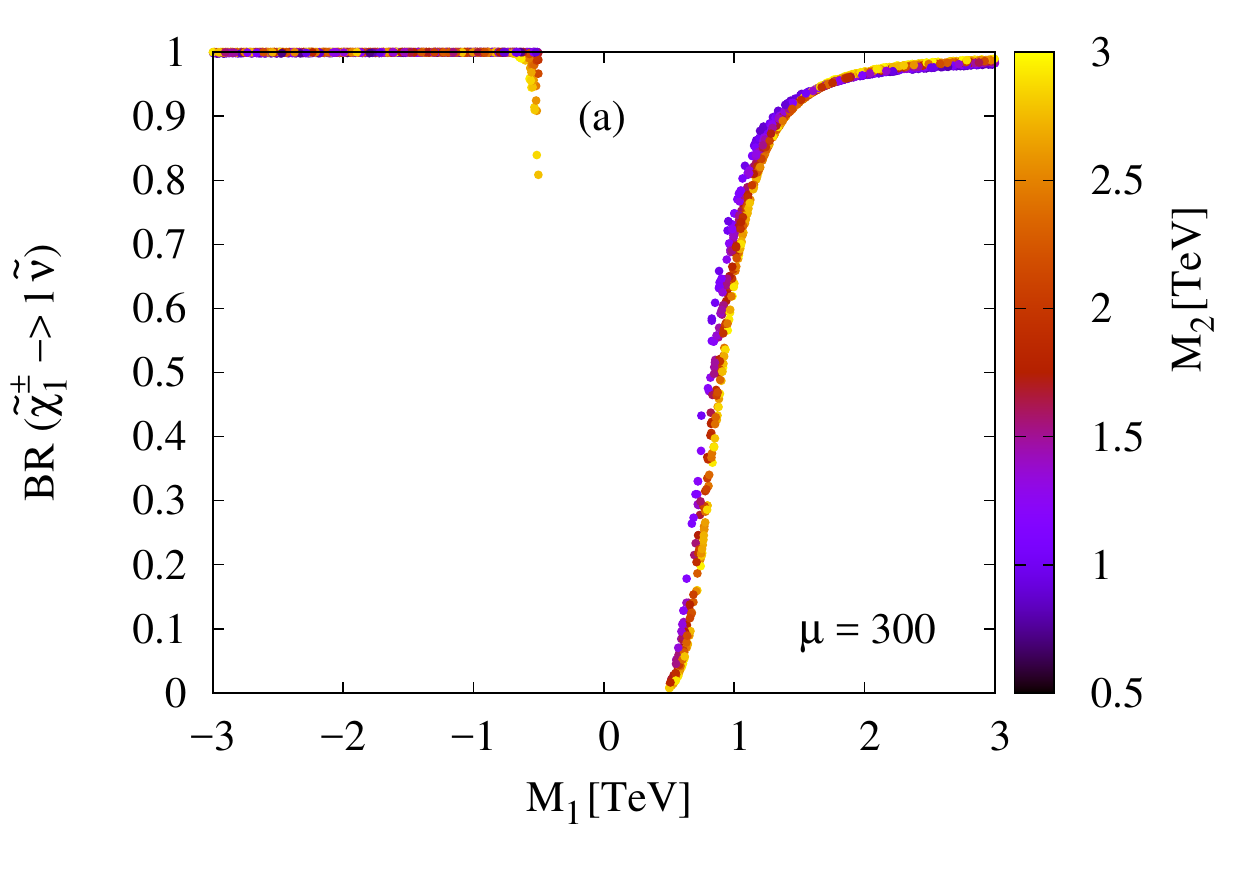}
\includegraphics[width=3.1in]{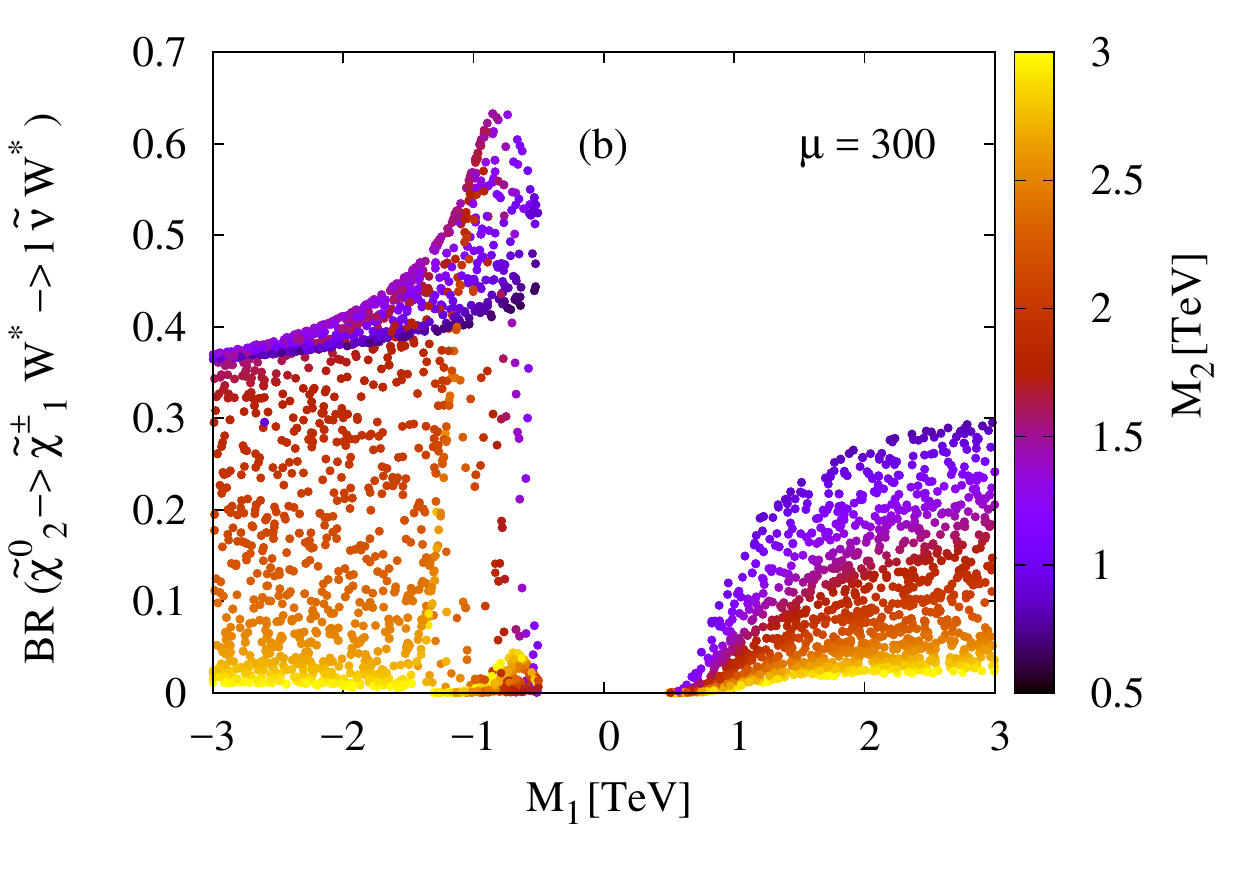}
  \caption{Variation of the leptonic branching ratios of 
$\widetilde{\chi}^{\pm}_1 \rightarrow l \widetilde{\nu} $ and 
$\widetilde{\chi}^{0}_2 \rightarrow l \widetilde{\nu} W^*$ 
against the bino soft mass parameter, $M_1$ for the Higgsino mass parameter, 
$\mu =300$ GeV. The wino soft mass parameter $M_2$ is shown in the palette.}
\label{fig:brpmu}
\end{figure}

\begin{figure}[ht]
\includegraphics[width=3.1in]{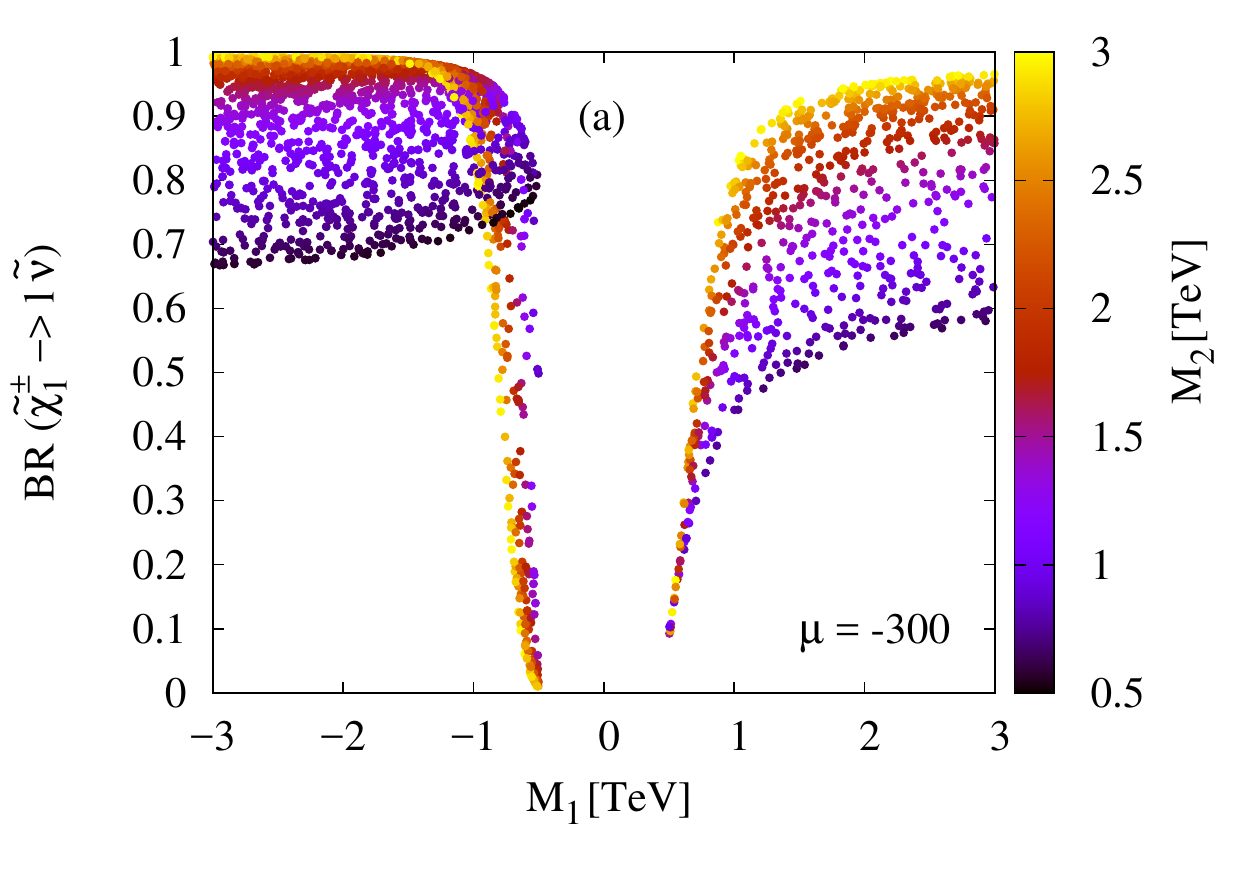}
\includegraphics[width=3.1in]{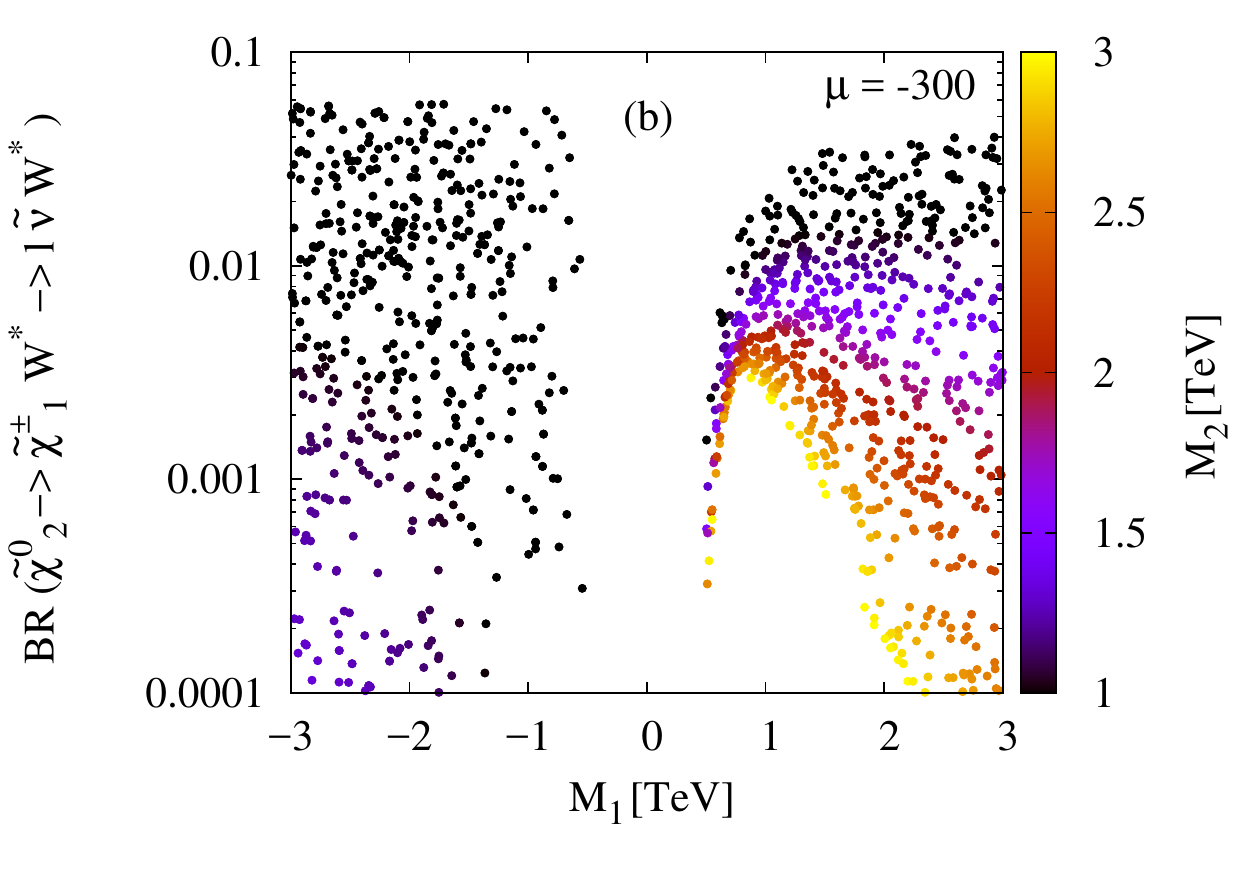}
  \caption{Variation of the leptonic branching ratios of $\widetilde{\chi}^{\pm}_1 \rightarrow l \widetilde{\nu}$ and $\widetilde{\chi}^{0}_2  \rightarrow l \widetilde{\nu} W^*$ against the bino soft mass parameter, $M_1$ for the Higgsino mass 
parameter, $\mu= -300$ GeV. The wino mass parameter $M_2$ is indicated in the 
palette.}
\label{fig:brnmu}
\end{figure}

\begin{table}[t!]
\begin{center}
 \small
 \begin{tabular}{|c|c|c|c|c|c|}
\hline
Parameters & $|M_1|$ (GeV) &$|M_2|$ (GeV) & $|\mu|$ (GeV)&$\tan\beta$ & $T_{\nu} (GeV)$  \\
\hline
Values & (500-3000) &(500-3000)  & 300   & 5 & 0.5\\
  \hline
 \end{tabular} 
\caption{Relevant input parameters for the parameter-space scan have been 
presented. Other parameters kept at fixed values include :   
$M_R$ = 100 GeV, $B_M  = 10^{-3}\,\, {\rm GeV}^2$, $M_3 = 2$ TeV, $M_{Q_3} = 
1.3$ TeV, $M_{U_3} = 
2$ TeV, $T_t = 2.9$ TeV, $M_{L_{1/2}} = 600 $ GeV, $m^{soft}_{\tilde\nu}=100$ GeV, 
$M_A=2.5$ TeV,  and $y_{\nu}=10^{-7}$. }
\label{tab:parbr}
\end{center}
\end{table}
In figures \ref{fig:brpmu} ($\mu >0$) and \ref{fig:brnmu} ($\mu < 0$) we show the 
variation of branching fraction in the leptonic decay 
channels $\tilde{\chi}^{\pm}_1 \rightarrow l \, \tilde{\nu}_i$ and 
$\tilde{\chi}^0_i \rightarrow l \, \tilde{\nu}_i \, W^{*}$. 
The relevant parameters for the scan can be found in table \ref{tab:parbr}.

Since the sneutrino masses and mixing matrices do not change in the scan, the two body 
partial decay widths $\Gamma(\tilde{\chi}_1^{\pm} \rightarrow l \, \tilde{\nu}_i)$ 
and $\Gamma(\tilde{\chi}_i^{0} \rightarrow \nu \tilde{\nu}_j)$ are only affected 
by the variation of the gaugino-admixture in the higgsino-like states. 
However, the choice of gaugino mass parameters do affect the mass splittings 
$\Delta m_1$ and $\Delta m_2$ through mixing and can even alter the hierarchy. These 
alterations in the spectrum mostly affect the 3-body decay modes described above
which has a significant effect on the branching ratio. 

As shown in fig. \ref{fig:delm}(a), for $sgn(\mu) = +$ (i.e. $\mu = 300$ GeV) 
and for  $M_1<0$, $\Delta m_1$ is almost entirely $\lesssim 1$ GeV. 
With large $M_2$ and $|M_1| \lesssim 2$ TeV, $\tilde{\chi}_1^{\pm}$ can become the 
lightest higgsino-like state making its leptonic branching probability close to 
100\% as shown in fig. \ref{fig:brpmu}(a). However, for small $|M_1|$, and 
large $M_2$, where $\Delta m_1$ increases, this branching is somewhat reduced to 
about 0.8 and the 3-body decays start becoming relevant.  
%In addition a large $M_2$ also leads to a small gaugino fraction 
%in $\tilde{\chi}_1^{\pm}$ thus reducing the two-body leptonic partial width. 
For $M_1>0$ region the branching ratio increases  as $M_1$ increases. This can be 
attributed to the consistent decrease in  $\Delta m_1$ (fig. \ref{fig:delm}(a)) and therefore 
of the three-body partial decay width. 

Fig. \ref{fig:brpmu}(b) shows the variation of Br$(\tilde{\chi}_2^0 \rightarrow 
\tilde{\chi}_1^{\pm} W^{\mp *})$ as a function of $M_1$ and $M_2$. For  
$M_1 <0 $, generally the branching grows for larger $\Delta m_2$ (fig. \ref{fig:delm}(b)) 
and decreases for smaller $M_2$ as the mass splitting goes down. 
%Note that for most regions of the parameter space the negative eigenvalue (symmetric) state is 
%$\tilde{\chi}_2^0$ and the nature of the state changes for large  $ M_2 $ values 
%as $|M_1|$ approaches $\mu$. 
It is again worth pointing out here that for large $M_2$ and
with $|M_1| \lesssim 2$ TeV,  $\Delta m_1 < 0$ and $\tilde{\chi}^{\pm}_1$ becomes the lightest state.
Thus in this region the three-body mode into $\tilde{\chi}_1^0$ is 
more phase-space suppressed compared to the decay mode into 
$\tilde{\chi}^{\pm}_1$.\footnote{Note that, because of $\Delta m_i \lesssim 1.5$ 
GeV, decay modes involving $\pi^{\pm}$ can dominate the hadronic branching fractions 
in this region. While we have estimated the same to be significant using routines 
used in \texttt{SPheno-v4} \cite{Porod:2003um,Porod:2011nf}, see also 
ref. \cite{Chen:1995yu,Chen:1996ap,Chen:1999yf}, the presence of large $T_{\nu}$ 
typically ensures that the two-body decay mode shares rather large 
branching fraction in these regions. In the plot we have only included 
three-body partial widths. A similar strategy has been adopted for regions with 
small $\Delta m_2$ as well.}  
Further,  as $|M_1|$ approaches $\mu$, the symmetric state, which mixes well with 
the bino, acquires larger bino fraction and there can be a cancellation in the 
vertex factor $\propto g_2 (N_{22} - \tan \theta_W N_{21})$ for the two-body 
decay width into sneutrino. This can reduce the corresponding width and 
then increase again as $|M_1|$ decreases. Thus the branching ratio for the 
three-body decay shows a discontinuous behavior in such regions. For positive 
$M_1$, the branching ratio shows similar pattern as $\Delta m_2$ variation, 
as expected. 
%The negative eigenvalue (symmetric) state is  $\tilde{\chi}_2^0$ 
%in this case, which have smaller gaugino fraction compared to the anti-symmetric state. 
Larger $\Delta m_1$ in this region implies that the three-body decay 
$(\tilde{\chi}_2^0 \rightarrow \tilde{\chi}_1^0 j_s j_s)$ can be larger, and consequently 
Br$(\tilde{\chi}_2^0 \rightarrow \tilde{\chi}_1^{\pm} j_s j'_s)$ is rather 
small. 
  
For $\mu = -300$ GeV, there are marked differences in the decay probabilities as 
the $\tilde{\chi}_1^{\pm}$ can become the heaviest when $M_1<0$, for large regions 
of the parameter space in contrast to what was observed for $\mu>0$. 
Figure \ref{fig:brnmu}(a) shows the branching ratio of 
$\tilde{\chi}_1^{\pm}\rightarrow l \, \tilde{\nu}_j$ which decreases as $M_2$ increases. Although, for large $M_2$, 
the gaugino fraction in $\tilde{\chi}_1^{\pm}$ would be small, thus possibly 
reducing the partial width in this two-body decay mode; smaller $\Delta m_1$ 
in this region ensures that the competing three-body mode decreases even more. 
Therefore, the branching ratio in the two-body mode is enhanced. This holds true 
for almost the entire range of $M_1$. The feature in the negative $M_1$ region, 
as $|M_1|$ approaches $|\mu|$, where the branching ratio rises faster for larger 
$M_2$ values, corresponds to a similar fall in $\Delta m_1$ (see fig. ~\ref{fig:delm}(c)). 
%As has been explained in section \ref{model}, in this region as  $|M_1|$ 
%decreases and approaches  $|\mu|$ the symmetric (anti-symmetric) state becomes 
%$\tilde{\chi}_2^0$ ($\tilde{\chi}_1^0$).

In figure \ref{fig:brnmu}(b) we show the variation of  
Br$(\tilde{\chi}_2^0 \rightarrow \tilde{\chi}_1^{\pm} j_s j_s^\prime)$ with $M_1,~ M_2$. 
For negative $M_1$, this branching ratio increases with decreasing $M_2$, since
the corresponding mass difference $\Delta m_2$ also increases 
(see figure \ref{fig:delm}). The larger $M_2$ values are not shown for $M_1 <0$, 
since $\tilde{\chi}_1^{\pm}$ becomes the heaviest higgsino-like state in this 
region. Thus, $\Delta m_2 <0$ as shown in see fig. \ref{fig:delm}(d), and this 
decay mode does not contribute. For $M_1 > 0$ smaller $M_2$ values correspond 
to larger branching fractions, since $\Delta m_2$ becomes larger, increasing 
the partial width. However, for large $M_2$ values, the partial width decreases 
rapidly as $\Delta m_2$ decreases. 

Note that $T_{\nu} = 0.5$ GeV has been used in the figure. For smaller 
values of $T_{\nu}$ the leptonic branching ratio of $\tilde{\chi}_1^{\pm}$ would 
generally be reduced when it is not the lightest higgsino-like state. However, 
the generic features described above would remain similar. Note that, $y_{\nu} 
\sim 10^{-6}$ can lead to prompt decay even in the absence of large 
left-admixture, as induced by large $T_{\nu}$. Therefore, even for small 
$T_{\nu} \lesssim \mathcal{O}(10^{-2})$, for certain choice of the gaugino mass 
parameters, the leptonic branching can be competing, and thus would be relevant 
to probe such scenario at collider.

\section{Survey of the relevant parameter space }
\label{survey}
We now consider the model parameter space in light of various constraints. 
%The allowed decay channels of the 
%light higgsino-like states and the corresponding branching ratios will be discussed 
%in more detail later. Needless to emphasize that these would play an important role in deciding about 
%the possible signatures at LHC. 

\subsection{General  constraints }
\label{const}

We implement the following general constraints on the parameter-space$:$
\begin{itemize}
 
 \item The lightest CP-even Higgs mass $m_h$ has been constrained 
 within the range : $ 122 \leq m_{h} \text{ (GeV)} \leq 128$   
 \cite{Aad:2015zhl,Aad:2012tfa,Chatrchyan:2012xdj}. While the 
 experimental uncertainty is only about 0.25 GeV, the present range of 
 $\pm 3$ GeV is dominated by uncertainty in the theoretical estimation 
 of the Higgs mass, see e.g. \cite{Carena:2013ytb} and references 
 there.\footnote{Note that, besides the MSSM contributions, rather large 
 $T_{\nu}$ can induce additional contributions to the Higgs mass 
 \cite{Belanger:2010cd}.  Our numerical estimation takes 
 this effect into account.}
 
 \item The lightest chargino satisfies the LEP lower bound: 
 $ m_{\widetilde{\chi}^{\pm}_{1}} \geq 103.5$  GeV \cite{LEPSUSYWG}. 
 The LHC bounds, which depend on the decay channels of the chargino, 
 will be considered only for prompt channels in more detail in section \ref{lhc}. 

\item  The light sneutrino(s) (with small left-sneutrino admixture) can contribute to 
the non-standard decay channels of (invisible) Higgs and /or $Z$ boson. The latter 
requires the presence of both CP-even and CP-odd sneutrinos below $\simeq 45$ GeV.  
Constraints from the invisible Higgs decay ($\simeq 20\%$) \cite{Khachatryan:2016whc} 
and the Z boson invisible width ($\simeq 2$ MeV) \cite{PDG} can impose significant 
constraints on the parameter space where these are kinematically allowed.
 
\item We further impose $B_s \rightarrow \mu^+ \mu^-$ \cite{Aaij:2012nna} and
$b \rightarrow s \gamma$ constraints \cite{Amhis:2012bh}.  
\end{itemize}
 
\subsubsection{Implication from neutrino mass}
Recent analyses by \texttt{PLANCK} \cite{Ade:2015xua} imposes the following 
constraint on the neutrino masses : $ \sum m_{\nu}^i \lesssim 0.7 \,\, {\rm eV}.$ 
\begin{figure}[H]
\begin{center}
 \includegraphics[width=3.5in]{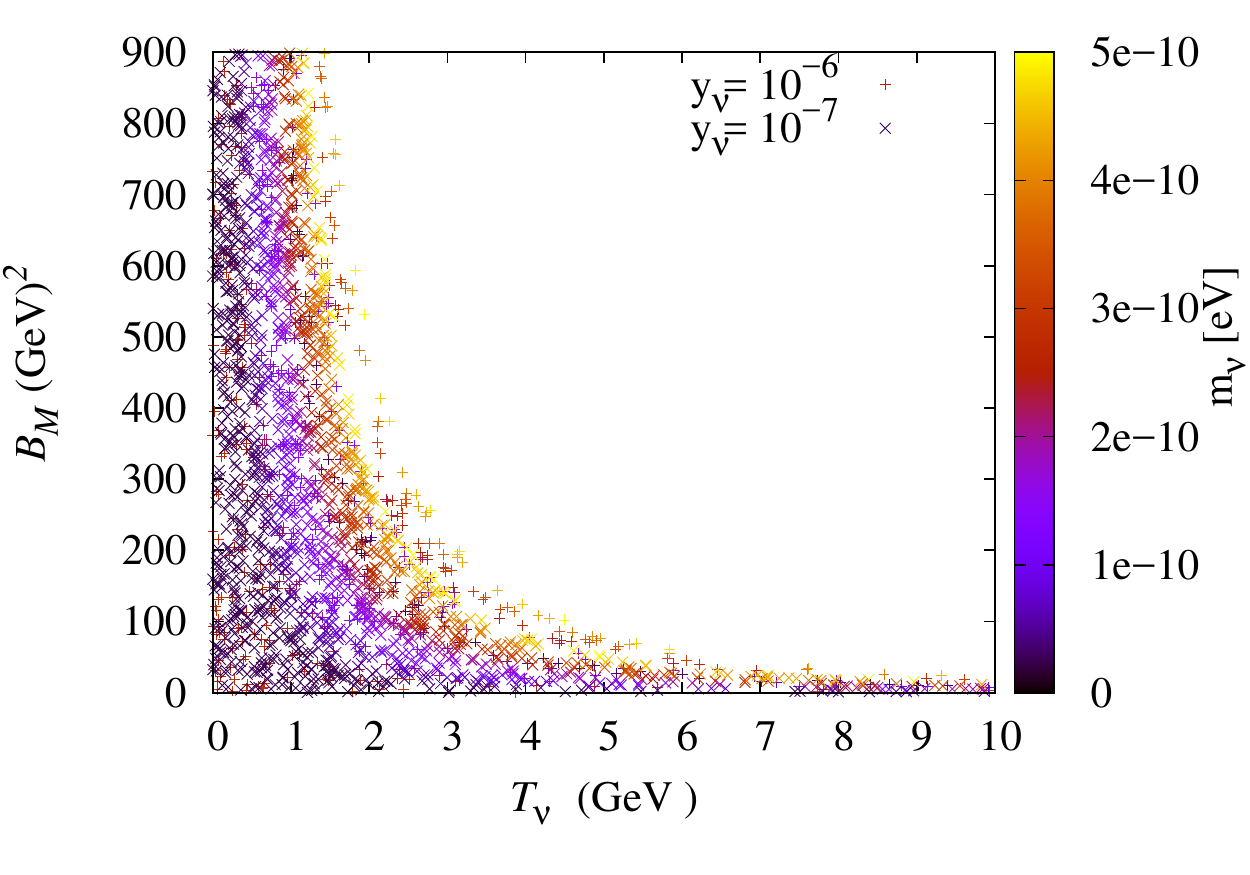}
\caption{Allowed regions of $B_M$ and $T_\nu$ plane for $M_1 = 1.5 $ TeV, 
$M_2 = 1.8 $ TeV and gaugino fraction $\sim \mathcal{O} (10^{-2})$. The colored palette denotes the 
mass of the heaviest neutrino.}
 \label{fig:bmrtnu}
 \end{center}
 \end{figure}
In the present scenario, the neutrinos can get a tree-level mass, as 
is usual in the Type-I see-saw scenario. For $y_{\nu} \sim 10^{-6}$, 
and $M_R \sim 100$ GeV, the active neutrino mass is of $\mathcal{O}(0.1)$ 
eV. Further, as discussed in section \ref{sec:snu}, a non-zero Majorana 
mass term $M_R$, and the corresponding soft-supersymmetry breaking 
term $B_M$ introduce a splitting between the CP-even and CP-odd mass 
eigenstates of right-sneutrinos. In the presence of sizable left-right 
mixing, significant contribution to the Majorana neutrino mass can be 
generated at one-loop level in such a scenario, the details  depend on 
the gaugino mass parameters \cite{Grossman:1997is, Dedes:2007ef}. Thus, 
regions of large $B_M$, in the presence of large left-right mixing in the 
sneutrino sector (induced by a large $T_{\nu}$) can be significantly 
constrained from the above mentioned bound on (active) neutrino mass. In 
fig. \ref{fig:bmrtnu} we show the allowed region in the $T_{\nu}-B_M$ 
plane. We consider $y_{\nu} \in \{10^{-6},10^{-7}\}$ while the other parameters 
are fixed as follows: $\mu = 300 $ GeV, $M_3 = 2$ TeV, $M_{Q_3} = 1.5$ TeV,
$T_t = 2.9$ TeV, $M_{L_{1/2}} = 600 $ GeV, $m^{soft}_{\tilde\nu}=100$ GeV and 
$M_A=2.5$ TeV. While in the former case the tree-level and radiative 
contributions to the neutrino mass can be comparable (with each being 
$\mathcal{O}(0.1)$ eV), the radiative corrections often dominate for
the latter. As shown in the figure, clearly larger $T_{\nu}$ values are 
consistent with neutrino mass for smaller $B_M$. 
 
\subsubsection{Implications for Dark Matter}
 \label{section_scan2}
 \begin{figure}[H]
  \begin{center}
  \includegraphics[width=6.0in]{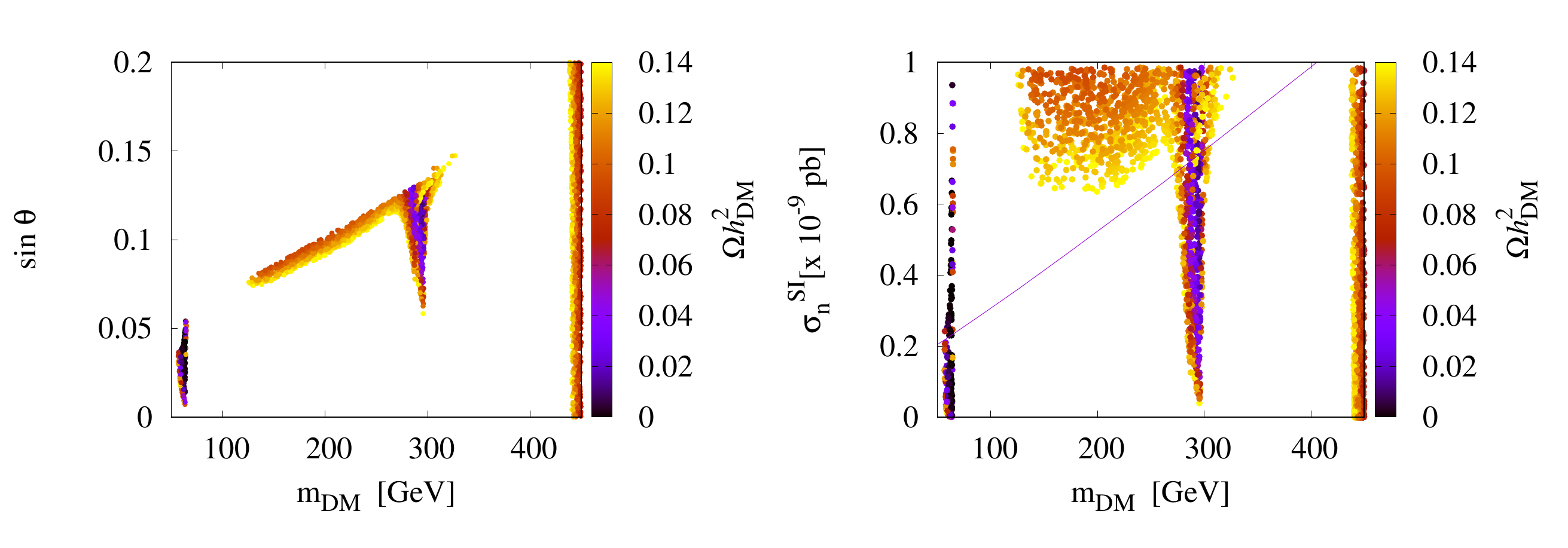} 
  \caption{In the left panel, the dependence of the relic abundance of $\tilde{\nu}_1$ has been shown 
  on its mass and left-fraction. The right panel shows 
  the allowed region respecting the direct detection constraint from \texttt{XENON-1T}.}
\label{fig:dm}
 \end{center}
 \end{figure}

Within the paradigm of standard model of cosmology the relic abundance is 
constrained as $0.092 \leq \Omega h^2 \leq 0.12$ \cite{Ade:2015xua}. 
Stringent constraints from {\sl direct} search constraints require the 
DM-nucleon (neutron) interaction to be less than about $10^{-9}$ pb, which 
varies with the mass of DM, see e.g. \texttt{LUX} \cite{Akerib:2016vxi}, 
\texttt{PANDA-II} \cite{Tan:2016zwf} and \texttt{Xenon-1T} \cite{Aprile:2017iyp}. 
 
In figure \ref{fig:dm}, with $y_{\nu} \in \{10^{-6},10^{-7}\},~ \tan \beta 
\in\{5,10\},~\mu= -450 \,\, {\rm GeV},~m_A = 600 \,\, {\rm GeV}$ and all other 
relevant parameters are fixed as before, we scan over the set of parameters
 $\{T_{\nu}, \,\, m_{\tilde{\nu}},\,\, B_M\}$ (first generation only). We plot 
the left-admixture ($\sin \theta^j$) in the LSP required to obtain the  
thermal relic abundance and direct detection cross section against its mass 
in the left and the right panel respectively. We have used \texttt{micrOMEGAs-3.5.5} \cite{Belanger:2013oya} to compute the thermal relic abundance and direct detection 
cross-sections. 

With $T_{\nu} \gtrsim \mathcal{O}(10^{-3})$, which is the region of interest to allow 
left-right mixing in the sneutrino sector, the right-sneutrino LSP thermalizes with 
the (MS)SM particles via its interaction with left-sneutrino and Higgs bosons. 
The important annihilation processes involve s-channel processes mediated by Higgs
bosons, as well as four-point vertices leading to $hh,~W^{\pm} W^{\mp}, ~ZZ, 
~t\bar{t}$ final states. However, large left-right mixing induces large direct 
detection cross-section. In fig.\ref{fig:dm} we have only shown parameter regions 
with a mass difference of at least 1 MeV between the CP-even and the CP-odd states 
to prevent the $Z$ boson exchange contribution to the direct detection\cite{Hall:1997ah, ArkaniHamed:2000bq, TuckerSmith:2001hy}.\footnote{We have checked that 
with 1 MeV mass splitting and a left-admixture 
of $\mathcal{O}(10^-2)$, as is relevant for thermal relic, the heavier of the 
CP-even and the CP-odd state has a decay width of $\sim 10^{-20}$ GeV, mostly 
into the LSP and soft leptons/quarks via off-shell $Z$ boson. This corresponds 
to a lifetime of $\lesssim 10^{-3}$ s. Thus it would decay well before the onset 
of Big Bang Nucleosynthesis (BBN) and is consistent with constraints from the same.} 
There are t-channel contributions mediated by Higgs bosons, mostly from the 
D-term, as well as the tri-linear term $T_{\nu}$, and is proportional to 
the left-right mixing ($\sin\theta$) in the sneutrino sector. Note that, we have 
only shown points with spin-independent direct detection cross-section less than $10^{-9}$ pb.

As shown in the left panel of fig.\ref{fig:dm}, $\sin\theta$ of $\mathcal{O}(0.1)$ 
is required to achieve  the right thermal relic abundance. The right relic abundance 
is achieved soon after the dominant annihilation channels into the gauge bosons 
(and also Higgs boson) final states are open (i.e. $m_{\rm DM} \gtrsim  130$ GeV),  
while at the Higgs resonances ($m_h = 125 \,\, {\rm GeV}$ and $m_A \sim 600 \,\, 
{\rm GeV}$) a lower admixture can be adequate. 
%\footnote{As discussed in section \ref{subsec:numass}, with small enough $B_M$ 
%such a sizable mixing can still be consistent with the observed neutrino mass. 
%We have allowed for up to $\mathcal{O}(10\%)$ tuning 
%among the the relevant parameters (e.g. between the gaugino mass parameters), 
%so that the radiative corrections do not lead to large neutrino mass 
%\cite{Ma:2011zm, Dedes:2007ef}.} 
Further, co-annihilation with the low-lying 
higgsino-like states ($|\mu| \sim 450 
\,\, {\rm GeV}$), when the LSP mass is close to 450 GeV, can also be effective. 
As shown in the right panel of the same figure, for $m_{\rm DM} \lesssim 450$ GeV, 
most parameter space giving rise to the right thermal relic abundance is tightly 
constrained from direct searches (spin-independent cross-sections) 
from Xenon-1T \cite{Aprile:2017iyp} (similar constraints also arise from \texttt{LUX} \cite{Akerib:2016vxi}, \texttt{PANDA-II} \cite{Tan:2016zwf}), the exceptions being 
the resonant annihilation and co-annihilation regions. 
%As will be discussed 
%in subsection \ref{subsec:benchmarks}, we have considered a benchmark scenario, 
%where the right-sneutrino DM achieves the correct thermal relic abundance.

Note that for very small $T_{\nu}$ and $y_{\nu} \lesssim 10^{-6}$,  the effective interaction strength of 
right-sneutrinos may be smaller than the Hubble parameter at $T \simeq m_{\rm DM}$. In such 
a scenario, non-thermal production, especially from the decay of a 
thermal NLSP, can  possibly generate the relic abundance
\cite{Asaka:2005cn, Asaka:2006fs, Gopalakrishna:2006kr, Page:2007sh}.
Further, non-thermal productions can also be important in certain non-
standard cosmological scenarios, e.g. early matter domination or low 
reheat temperature, see e.g. \cite{Baer:2014eja, Acharya:2008bk}. In addition, 
large thermal relic abundance can be diluted if substantial entropy 
production takes place after the freeze-out of the DM. For such regions of parameter space our 
right-sneutrino LSP is likely to be a non-thermal DM candidate. 
\section{Signatures at LHC }
\label{lhc}
We now focus on the LHC signal of the higgsino-like electroweakinos in the presence 
of a right-sneutrino LSP. As discussed, the various decay 
modes available to $\widetilde{\chi}^{\pm}_1$, $\widetilde{\chi}^{0}_2$ and 
$\widetilde{\chi}^{0}_1$ in presence of a right-sneutrino LSP depend not only on the mixing among the 
various sparticle components but also crucially on the mass splittings. The LHC signals would 
then reflect upon the above dependencies on the parameter space. We therefore look at all possible signals for 
different regions of $\Delta m_{1/2}$ and $T_{\nu}$. While there are regions of $\Delta m_{1/2}$ 
where the chargino decays non-promptly to pions that lead to the chargino traveling in the 
detector for some length and then decay into a soft pion and neutralino. In such cases, since both 
decay products are invisible, the relevant search channel at LHC is 
the \textit{disappearing tracks} \cite{Aaboud:2017iio,Mahbubani:2017gjh,Fukuda:2017jmk}. 
%Since both the two body and three body decays may compete when the roles of the mass splitting 
%$\Delta m_{1/2}$ and $T_{\nu}$ are evenly matched for the 3-body and 2-body decays 
%respectively, alternate signals to the leptonic channels worth looking at would 
%be mono-jet signals \cite{ATLAS-CONF-2017-060} or photon(s) and/or associated with 
%hard ISR jets and missing energy since the decay products from the three body 
%decays are soft. 
Our focus however, is primarily on the prompt decay of the chargino to hard 
leptons (small $\Delta m_{1/2}$ and large $T_{\nu}$) which would be clean signals 
to observe at LHC. 
 
The following production channels are of interest to us: 
\begin{equation}
 p\text{ }p \rightarrow \widetilde{\chi}^{\pm}_1 \,\, \widetilde{\chi}^{0}_2, \text{ } \widetilde{\chi}^{\pm}_1 \,\, 
 \widetilde{\chi}^{0}_1, \text{ }\widetilde{\chi}^{+}_1 \,\, \widetilde{\chi}^{-}_1,\text{ } 
 \widetilde{\chi}^{0}_1 \,\, \widetilde{\chi}^{0}_2, \text{ }\widetilde{\chi}^{0}_1 \,\, 
 \widetilde{\chi}^{0}_1, \text{ }\widetilde{\chi}^{0}_2 \,\, \widetilde{\chi}^{0}_2, \text{ }\widetilde{l} \,\, \widetilde{l}, 
 \text{ } \widetilde{l} \,\, \widetilde{l}^*,  \text{ } \widetilde{l} \,\, \widetilde{\nu},  \text{ } \widetilde{\nu} \,\, \widetilde{\nu}
 \label{eq:process}
\end{equation}

where the sleptons and sneutrinos are heavier than the electroweakinos here. The 
LSP pair production is excluded in the above list. 
\begin{figure}[ht]
\begin{center}
 \includegraphics[width=3.8in,height=3.4in] {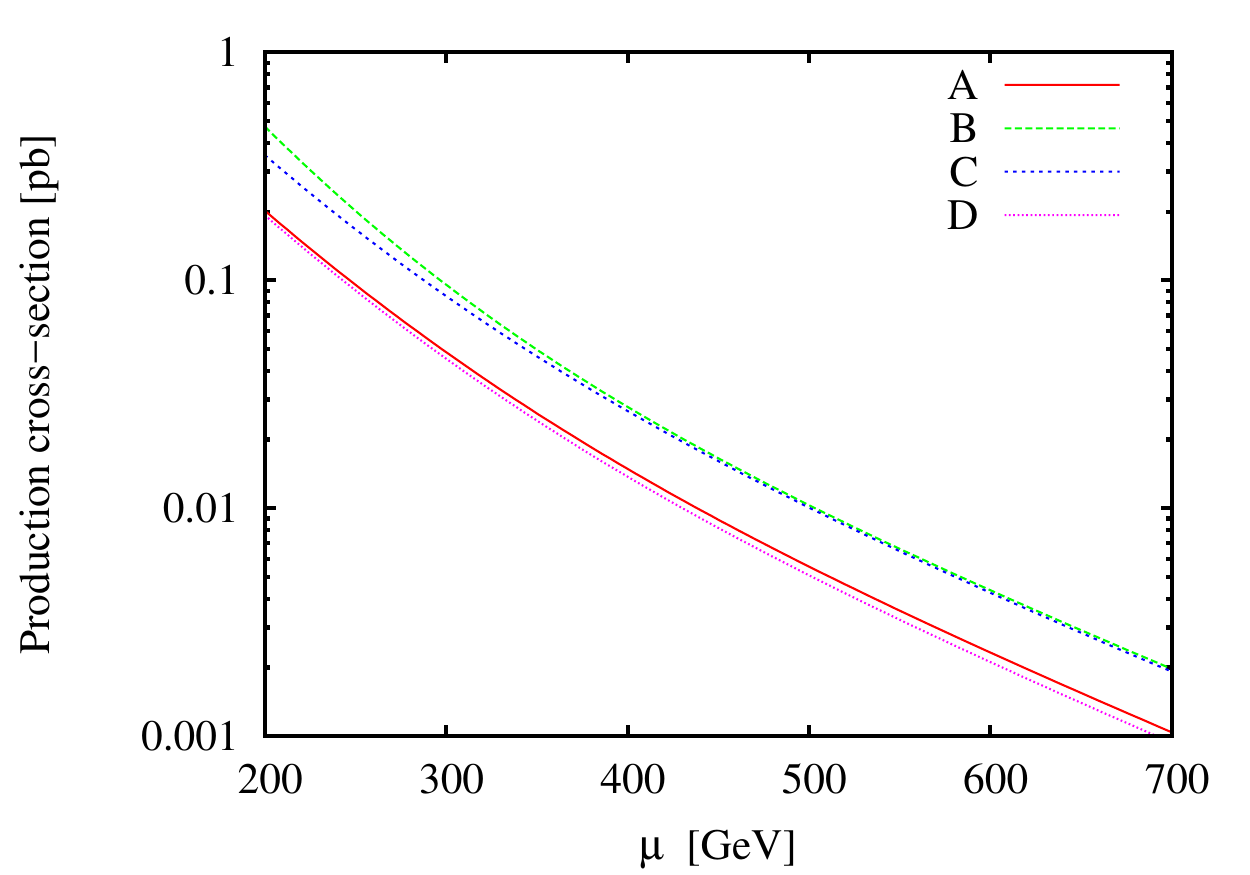}
 \caption{LO cross-sections of the different production channels at $\sqrt{s}=13$ TeV. Here, A = $\widetilde{\chi}^{+}_1
 \widetilde{\chi}^{-}_1$, B = $\widetilde{\chi}^{\pm}_1
 \widetilde{\chi}^{0}_1$, C = $\widetilde{\chi}^{\pm}_1
 \widetilde{\chi}^{0}_2$ and D = $\widetilde{\chi}^{0}_1
 \widetilde{\chi}^{0}_2$. $\widetilde{t}_1$ and $\widetilde{b}_1$
 are of mass $\sim 1.4$ TeV. The NLO cross-sections can be estimated by using a K
factor $\sim$ 1.25.}
 \label{fig:sigma}
 \end{center}
\end{figure}
The processes as given in eq. \ref{eq:process} are in decreasing order of 
production cross-sections as obtained from \texttt{Prospino} 
\cite{Beenakker:1996ed,Plehn:2004rp,Spira:2002rd}.
The associated chargino neutralino pair, i.e $\widetilde{\chi}^{\pm}_1 \,\, 
\widetilde{\chi}^{0}_{1/2}$ production has the largest cross-section followed 
by the chargino pair production, $\widetilde{\chi}^{+}_1 \,\, \widetilde{\chi}^{-}_1$. 
In the pure higgsino limit, the pair production cross-section of $\widetilde{\chi}^{0}_1 
\,\, \widetilde{\chi}^{0}_1$ and $\widetilde{\chi}^{0}_2 \,\, \widetilde{\chi}^{0}_2$ 
are negligible compared to the other processes. Since the strong sector is kept 
decoupled and the compressed higgsino sector leads to soft jets and leptons, the 
only source of hard jets are from initial-state radiations (ISR). The suppressed 
jet multiplicity in the signal could prove to be a potent tool for suppressing 
SM leptonic backgrounds coming from the strongly produced $t \bar{t}$ and single top 
subprocesses which would give multiple hard jets in the final state in association 
with the charged leptons. Therefore we shall focus on the following leptonic 
signals with low hadronic activity: 
%to probe such a compressed higgsino sector in presence of $\widetilde{\nu}$ LSP:
 \begin{itemize}
  \item Mono-lepton + $\leq 1$ jet + $\slashed{E}_T$ 
 \item Di-lepton + $0$ jet + $\slashed{E}_T$
 \end{itemize}
The mono-lepton signals would come from the pair production of 
$\widetilde{\chi}^{+}_1 \widetilde{\chi}^{-}_1$, ($\widetilde{\chi}^{\pm}_1 
\rightarrow l \,\, \widetilde{\nu}$ and $\widetilde{\chi}^{\pm}_1 \rightarrow \widetilde{\chi}^0_1 \,\, W^{*\pm}$), and associated pair production, $\widetilde{\chi}^{\pm}_1 \,\, \widetilde{\chi}^{0}_i$ with ${i=1,2}$ 
($\widetilde{\chi}^{\pm}_1 \rightarrow l  \,\, \widetilde{\nu}$ and 
$\widetilde{\chi}^0_i \rightarrow \nu \widetilde{\nu}$). A smaller contribution 
also comes from the production of $\widetilde{\chi}^{0}_1 \,\, \widetilde{\chi}^{0}_2$  
($\widetilde{\chi}^{0}_{2/1} \rightarrow \widetilde{\chi}^{\pm}_1 \,\, W^{*\mp}$ and 
$\widetilde{\chi}^{0}_{1/2} \rightarrow \nu \widetilde{\nu}$) leading to missing energy. 
Among the di-lepton signals we look into both opposite sign leptons and same sign 
lepton signal with missing energy. Opposite sign leptons arise from the pair produced 
$\widetilde{\chi}^{\pm}_1 \,\, \widetilde{\chi}^{\pm}_1$, with 
the chargino decaying leptonically as $\widetilde{\chi}^{\pm}_1 \rightarrow l  \,\, 
\widetilde{\nu}$. In regions of the parameter space where $\widetilde{\chi}^{0}_{2/1}$ 
is heavier than $\widetilde{\chi}^{\pm}_1$, $\widetilde{\chi}^{\pm}_1
\widetilde{\chi}^{0}_i$, ${i=1,2}$ process may contribute to the di-lepton state 
via $\widetilde{\chi}^{0}_i \rightarrow \widetilde{\chi}^{\pm}_1 \,\, W^{*\mp}$ 
followed by $\widetilde{\chi}^{\pm}_1 \rightarrow l  \,\, \widetilde{\nu}$. In 
such cases, there could be either opposite sign di-lepton signal or same-sign 
di-lepton signal owing to the Majorana nature of the neutralinos 
($\widetilde{\chi}^{0}_i$). A similar contribution to both channels come 
from $\widetilde{\chi}^{0}_1\widetilde{\chi}^{0}_2$ with $\widetilde{\chi}^{0}_{1/2} 
\rightarrow \widetilde{\chi}^{\pm}_1 \,\, W^{*\mp}$. Also there are sub-leading 
contributions from slepton pair productions which can become relevant if 
light sleptons are also present in the spectrum.
 
It is worth pointing out that in very particular regions of the parameter space 
$\widetilde{\chi}^{\pm}_1$ is the NLSP and therefore always decays to a hard 
lepton and sneutrino LSP. In such cases, signal rates for the di-lepton channel 
would be most interesting and dominant rates for same-sign di-lepton would be a 
particularly clean channel which will be important to probe high values of $\mu$ very 
effectively. This is very particular of the parameter region when $M_1$ is negative 
and one has a sneutrino LSP. 
\subsection{Constraints on electroweakino sector from LHC}   
Before setting up our analysis on the above signals we must consider the role of existing LHC studies that 
may be relevant for constraining the parameter space of our interest. 
LHC has already looked for direct production of lightest $\widetilde{\chi}^{\pm}_1$, 
$\widetilde{\chi}^{0}_2$ and $\widetilde{\chi}^{0}_1$ in both run 1 and run 2 
searches at 7, 8 and 13 TeV respectively albeit assuming simplified models. Search results have been reinterpreted both in terms of 
non-prompt as well as prompt decays of the Higgsinos. Since the focus of our study is on 
prompt decays of $\widetilde{\chi}^{\pm}_{1}$, we consider the prompt search results only. 

The LHC results have been reinterpreted assuming simplified models with and without intermediate left 
and right sleptons with $\widetilde{\chi}^{0}_1$ LSP contributing to $\slashed{E}_T$ (see Table 
\ref{tableLHC}). 
\begin{itemize}
\item Assuming 100$\%$ leptonic branching of the sparticles and an uncompressed spectra, 
CMS has ruled out degenerate wino-like $m_{\widetilde{\chi}^{\pm}_1}$, $m_{\widetilde{\chi}^{0}_2}
< 1.2$ TeV
for a bino-like $m_{\widetilde{\chi}^{0}_1} <$ 600 GeV from 
same-sign di-lepton, three lepton and four lepton searches with at most 1 jet 
\cite{CMS-PAS-SUS-16-039}. The limits vary slightly depending on the choice of slepton masses.
%for slepton masses halfway between the $\widetilde{\chi}^{\pm}_1$ and $\widetilde{\chi}^{0}_1$. 
%For slepton masses closer to the $\widetilde{\chi}^{\pm}_1$ or very close to the LSP, the limits 
%are quite similar and reduce to $m_{\widetilde{\chi}^{\pm}_1}$, $m_{\widetilde{\chi}^{0}_2}$ $\sim$ 1 TeV
%for $m_{\widetilde{\chi}^{0}_1} <$ 400 GeV.
For the nearly compressed higgsino sector and assuming 
mass degeneracy of the lightest chargino ($\widetilde{\chi}^{\pm}_1$) and 
next-to-lightest neutralino ($\widetilde{\chi}^{0}_2$) the alternate channels 
probed by LHC are soft  opposite sign di-leptons searches \cite{CMS-PAS-SUS-16-048}. The mass limits on the compressed higgsino sector relax 
to $\sim 230 \text{ } (170)$ GeV for $m_{\widetilde{\chi}^{0}_1}$ 
$\sim $ 210 (162) GeV. 

\item  ATLAS has also extensively looked for compressed higgsinos in opposite 
sign dilepton and trilepton final states excluding $m_{\widetilde{\chi}^{0}_2} \sim 150$ GeV 
for splittings as low as
$\Delta m(\widetilde{\chi}^{0}_2-\widetilde{\chi}^{0}_1) = 3$ GeV
while the limit further improves by 20 GeV for degenerate $\widetilde{\chi}^{0}_2$ and 
$\widetilde{\chi}^{\pm}_1$  \cite{Aaboud:2017leg}. For the uncompressed case, 
searches with low hadronic activity \cite{ATLAS-CONF-2017-039} look for di-lepton 
and trilepton signal with upto  0 or 1 jet. The di-lepton search with no jet from chargino pair 
production, with charginos decaying to the neutralino LSP via intermediate sleptons 
sets a limit on the $m_{\widetilde{\chi}^{\pm}_1} > $ 720 GeV for 
$m_{\widetilde{\chi}^{0}_1}<$ 200 GeV. The trilepton 
channel excludes mass degenerate $m_{\widetilde{\chi}^{0}_2}$, 
$m_{\widetilde{\chi}^{\pm}_1} \sim$ 1.15 TeV for $m_{\widetilde{\chi}^{0}_1}$ up to 
580 GeV. 
\end{itemize}
%For $\widetilde{\chi}^{0}_{2}  \widetilde{\chi}^{\pm}_1$ decaying via 
%$(Z \rightarrow l^+l^-)$ and $(W \rightarrow l \nu)$ bosons respectively, the limits reduce to 
%$m_{\widetilde{\chi}^{0}_2}$, $m_{\widetilde{\chi}^{\pm}_1} \sim$ 360 GeV
%for $m_{\widetilde{\chi}^{0}_1} \sim$ 100 GeV \cite{ATLAS-CONF-2017-039}. 
% It is important to note, that limits from channels 
%associated with multiple jets usually have higher SM background events leading 
%to slightly weaker limits. For example in the ATLAS di-lepton search with jets 
%reinterpreted from $\widetilde{\chi}^{\pm}_1 \,\, \widetilde{\chi}^{0}_2$ 
%associated production, with the sparticles decaying via $(W \rightarrow j j )$ 
%and $(Z \rightarrow l^+l^-)$ and the LSP, the limits on the mass degenerate 
%$m_{\widetilde{\chi}^{0}_2}, \,\, m_{\widetilde{\chi}^{\pm}_1} \sim 580$ GeV for 
%$ m_{\widetilde{\chi}^{0}_1} < 150$ GeV, being 140 GeV lighter than the limits 
%from di-lepton + 0 jet search \cite{ATLAS-CONF-2017-039}.
%
 \begin{table}[ht]
 \begin{center}
 \scriptsize
  \begin{tabular}{|c|c|c|c|c|}
  \hline
  \textbf{A} &\multicolumn{4}{|c|}{$\sqrt{s}=8$ TeV}\\
  \hline
   S.No. & Final State & Luminosity (in fb$^{-1}$) & ATLAS & CMS \\
   \hline
   1 & 3 leptons $+$ $\slashed{E}_T$ &20.3 &\cite{Aad:2014nua} & -\\
   2 & Stop search with 2 leptons $+$ $\slashed{E}_T$ & 20.3&\cite{Aad:2014qaa} &-\\
   3& Stop search with Z boson and b jets $+$ $\slashed{E}_T$ & 20.3& \cite{Aad:2014mha}& -\\
  4&  2 same-sign leptons or 3 leptons $+$ $\slashed{E}_T$ &20.3 & \cite{Aad:2014pda}& -\\
   5& 1 lepton $+$ (b) jets $+$ $\slashed{E}_T$ &20.3 & \cite{Aad:2014kra}&- \\
   6& 2 leptons $+$ jets $+$ $\slashed{E}_T$ &20.3 & \cite{Aad:2015wqa}& -\\
  7&  Monojet $+$ $\slashed{E}_T$ &20.3,19.5 & \cite{Aad2015} & \cite{Khachatryan2015}\\
  8&  2 leptons $+$ 2 b jets $+$ $\slashed{E}_T$ & 20.3& \cite{Aad:2015pfx} &- \\
  9& 1 lepton $+$ $\geq 4 $ jets $+$ $\slashed{E}_T$ & 20.5&\cite{ATLAS:2012tna} &- \\
  10& 3 leptons $+$ $\slashed{E}_T$ &20.3 &\cite{ATLAS:2013rla} &- \\
  11& 2 leptons $+$ $\slashed{E}_T$ & 20.3&\cite{TheATLAScollaboration:2013hha} &- \\
   12&0$-$1 lepton $+$ $\geq 3 $ b jets $+$ $\slashed{E}_T$&20.1&\cite{TheATLAScollaboration:2013tha} & -\\
  13& 2 leptons $+$ jets $+$ $\slashed{E}_T$ &20.3,19.5 & \cite{s}& \cite{Khachatryan:2015lwa}\\
  14& 1 lepton $+$ $\geq 3 $ jets $+$ $\geq 1$ b jet$+$ $\slashed{E}_T$ (DM $+$ 2 top) & 19.7&- &\cite{Khachatryan:2015nua} \\
  15& Opposite sign leptons $+$ 3 b tags & 19.5&- &\cite{CMS-PAS-SUS-13-016} \\
   \hline
   \bf{B}&\multicolumn{4}{|c|}{$\sqrt{s}=13$ TeV} \\
   \hline
   S.No. & Final State & Luminosity (in fb$^{-1}$) & ATLAS & CMS \\
   \hline
   1& 2 same-sign or 3 leptons $+$ jets $+$ $\slashed{E}_T$ & 3.2 & \cite{Aad:2016tuk} & -\\
   2& Mono jet $+$ $\slashed{E}_T$ &3.2 &\cite{Aaboud:2016tnv} & - \\
   3& 1 lepton $+$ jets $+$ $\slashed{E}_T$ & 3.3 & \cite{Aad:2016qqk}&  -\\
   4& 0$-$1 lepton $+$ $3 $ b jets $+$ $\slashed{E}_T$ & 3.3&\cite{Aad:2016eki} &  -\\
   5& 1 lepton $+$ (b) jets $+$ $\slashed{E}_T$ & 3.2&\cite{Aaboud:2016lwz} & - \\
   6& 2 leptons(Z) $+$ jets $+$ $\slashed{E}_T$ &3.2 &\cite{TheATLAScollaboration:2015nxu} & - \\
   7& 1 lepton $+$ jets  &3.2 &\cite{TheATLAScollaboration:2016gxs} & \\
   8& 2 leptons $+$ jets $+$ $\slashed{E}_T$ &13.3 &\cite{ATLAS:2016xcm} &  -\\
   
  9&  1 lepton $+$ (b) jets $+$ $\slashed{E}_T$ &13.2 &\cite{ATLAS-CONF-2016-050} & - \\
    10&   2 leptons $+$ jets $+$ $\slashed{E}_T$ &2.2&  - &\cite{CMS:2015bsf} \\
   \hline
  \end{tabular}
 \caption{List of LHC analyses at s = 8,13 TeV implemented in the public software
CheckMATE. All the benchmarks considered in our study pass these analyses, without
showing any excess above the observed number of events at 95$\%$ CL.}
\label{checkmate}
\end{center}
 \end{table}
In Table~\ref{checkmate}, we list all relevant searches implemented in the public reinterpretation software \texttt{CheckMATE}. 
The ones, not implemented in \texttt{CheckMATE} have been recast in 
\texttt{Madanalysis-v5} as shown in table~\ref{tableLHC} and benchmarks have been 
chosen to pass all the relevant searches.
Alternate results from LHC which constrain 
the compressed higgsino sector is the 
monojet $+$ $\slashed{E}_T$ channel \cite{Aaboud:2017phn}. However no limits on the electroweak sector are yet placed from it.
We ensure that our chosen benchmarks pass all of these discussed analyses.  
\begin{table}[ht]
\small
\begin{tabular}{|c|c|c|c|}
\hline
Final State& ATLAS & CMS & Madanalysis$-$v5\\
\hline

1 lepton + $\slashed{E}_T$& \cite{Aaboud:2017efa} & \cite{Khachatryan:2016jww,CMS-PAS-SUS-16-052}&  Yes\\
2 leptons + $\slashed{E}_T$   &\cite{ATLAS-CONF-2017-039,Aaboud:2017leg} & \cite{201757,CMS-PAS-SUS-17-009,CMS-PAS-SUS-17-002}&  Yes\\
2 same-sign leptons + $\slashed{E}_T$  & - & \cite{Sirunyan:2017lae}&  Yes\\
3 or more leptons + $\slashed{E}_T$  & \cite{ATLAS-CONF-2017-039,Aaboud:2017dmy,ATLAS-CONF-2016-075} &\cite{Sirunyan:2017lae,Sirunyan:2018ubx,Khachatryan2017}&  Not relevant for this study \\
\hline 
\end{tabular}
\caption{Leptonic searches at $\sqrt{s} = 13$ TeV LHC with few jets (i,e, $N_j \leq 2 $), as relevant for this study.}
\label{tableLHC} 
\end{table}

\subsection{Impact of additional related searches at LHC}

For \textit{monolepton} signal, multiple searches in both ATLAS and CMS look for single lepton final states with multiple (b) jets (refer Table~\ref{checkmate} B.5) thus focussing on the production of 
coloured sparticles. However, there are not many dedicated SUSY search results for exploring low hadronic jet activity and low $\slashed{E}_T$.
One of the closest analysis is \cite{CMS-PAS-SUS-16-052}, however employing a large cut on $\slashed{E}_T > 300$  GeV.
Such a large cut on missing energy depletes the SUSY signal for low higgsino masses, ie., $\mu \sim 300-500$ GeV.
Both ATLAS and CMS have also looked into resonant searches for heavy gause bosons and considered monoleptonic channels with missing energy and placed limits on mass of heavy gauge bosons 
\cite{Aaboud:2017efa,Khachatryan:2016jww}. Since no reinterpretation exists on SUSY models in run 1 or run 2 so far, we reinterpret and take into account these 
limits in the context of our study and impose any constraints that apply on 
our parameter space. Few studies on soft leptons \cite{CMS-PAS-SUS-16-052,20169}, involve high cuts on missing transverse energy ($\slashed{E}_T$) and hadronic energy ($H_T$) even though requiring 
upto two light jets and atmost one b jet. It is important to point out that since the signals from the higgsino sector here are devoid of 
sources of b jets, and only ISR jets are present, signal
efficiency reduce substantially with cuts on large values of hadronic energy and hard b jet requirements. Further, owing to a light $\mu$ parameter, and consequently a low-lying higgsino states, a large missing energy cut of $\sim \mu$ reduces signals significantly. Hence, these searches would require large luminosity to probe the compressed higgsino sector. 
Reducing the hard cuts on $\slashed{E}_T$, b-veto and number of jets coupled with hadronic energy would allow better sensitivity to such signals and we have attempted to 
give an estimate of the results for the run 2 of LHC.  

For \textit{di-lepton} final states, both ATLAS and CMS have looked at stop searches or gluino searches giving rise to opposite sign or same sign leptons 
accompanied with multiple jets and b jets along with missing energy \cite{Beauchesne:2017yhh,Sirunyan:2017qaj}. 
As argued above, these searches weakly constrain the scenario we are interested in this study. 
However there are some searches specifically for soft leptons studied in 8 TeV \cite{20169} 
against which we check our benchmarks using \texttt{Madanalysis-v5}.  For \textit{opposite sign dileptons}, 
important searches from LHC which constrain our scenario are from ATLAS \cite{ATLAS-CONF-2017-039} and CMS
\cite{CMS-PAS-SUS-17-009}. Among other kinematic variables such as high lepton pt cuts, 
both studies focus on using large cuts on $M_{T_2}$ ($\ge 90$ GeV). 
%From Figure, we see that the benchmarks \textbf{BP2-b} and \textbf{BP3} which have a smaller mass gap $\Delta M$ between the higgsinos and LSP, and hence
%the distribution of $M_{T_2}$  starts falling close to 90-100 GeV as discussed in Section ~\ref{subsec:benchmarks}. Thus, necessarily requiring this 
%removes signal events and therefore not sensitive to such scenarios. 
%For other spectra, ie., \textbf{BP2-a} and \textbf{BP1}, with larger mass gaps, the searches are relevant 
%as they focus on large mass gaps. 
We implement these analyses in {\tt Madanalysis-v5} and choose benchmarks such that 
they are not excluded by current data.  
For \textit{same sign dilepton}, the most constraining limit comes from CMS \cite{Sirunyan:2017lae} with atmost 1 jet. Other searches usually focus on the strong sector thus requiring large number of jets.
%%%%
\subsection{Benchmarks}
\label{subsec:benchmarks}
In the context of natural supersymmetry with degenerate first and second 
generation sneutrino as LSP, we look into regions of parameter space allowed 
by neutrino physics constraints, LHC data and direct detection cross-section 
constraints. We select five representative points of the parameter space and 
analyze their signal at the current run of LHC. We check the viability of the 
chosen benchmarks for multi-leptonic signatures by testing the signal strengths 
against existing experimental searches implemented in the public software 
\texttt{CheckMATE} \cite{Drees:2013wra,Dercks:2016npn}. 
Amongst the searches implemented in \texttt{CheckMATE}, mono-jet along with 
missing energy search \cite{Aaboud:2016tnv} provides the most stringent constraint. 
Among the other 13 TeV searches as listed in Table~\ref{tableLHC}, same-sign di-lepton 
and opposite-sign di-lepton searches also impose a stringent constraint on the 
current scenario. The allowed same-sign di-lepton branching is restricted 
to $4\%$ or lower for $\mu=300$ GeV for uncompressed scenarios. 
A higher value of $\mu$ and hence a lower production cross-section allows a larger 
same-sign di-lepton branching thereby allowing us to probe a wider range of the 
parameter space.

We choose parameters with $|\mu|$ in the range 300-500 GeV, $M_{1/2} \sim 2$ TeV and 
$\tan \beta \sim 5-10 $ GeV as listed in Table~\ref{bp}. The 
choice of the benchmarks ensure prompt decay of the chargino to a hard lepton and 
LSP, i.e., $\Gamma > 10^{-13}$ GeV. The gaugino mass parameters $M_1$ and $M_2$ are 
large such that the spectrum consists of two light higgsino-like neutralinos 
$\widetilde{\chi}^{0}_1, \,\, \widetilde{\chi}^{0}_2$ and a nearly degenerate 
light higgsino-like $\widetilde{\chi}^{\pm}_1$ within $\mathcal{O}$(2-4) GeV. 
However there is considerable amount of freedom in choosing the relative sign among
the soft parameters $M_1$, $M_2$ and $\mu$. Both the first and second 
generation squarks as well as gluino soft mass parameter are set to $\sim$ 2 TeV. 
The stops are also kept heavy to ensure the light CP-even Higgs mass and signal 
strengths to be within the allowed experimental values. Both the first two 
generation left and right sleptons are kept above the higgsino sector and 
when possible within the reach of LHC, in the range 360$-$600 GeV, in the 
different benchmark points studied. Following our discussion in section 
\ref{sec:delm} on the $M_1-M_2$ dependence of the masses, the benchmarks 
represent points in the following regions of parameter space:
\begin{itemize}
 \item \textbf{Region A}: $M_1 >$ 0, $M_2 >$ 0 and $\mu >0$, with 
 $\widetilde{\chi}^0_{1}$ as NLSP (\textbf{BP1}).
 \item \textbf{Region B}: $M_1 >$ 0, $M_2 >$ 0 and $\mu < 0$ , with 
 $\widetilde{\chi}^0_{1}$ as NLSP (\textbf{BP2-a} and \textbf{BP2$-$b}).
 \item \textbf{Region C}: $M_1 <$ 0, $M_2 >$ 0 and $\mu > 0$, with 
 $\widetilde{\chi}^{\pm}_{1}$ as NLSP (\textbf{BP3} and \textbf{BP4}).
\end{itemize}
 \begin{table}
 \begin{center}
\small
 \begin{tabular}{|c|c|c|c|c|c|}
  \hline
  Parameters & \bf{BP1} & \bf{BP2$-$a} & \bf{BP2$-$b} & \bf{BP3} & \bf{BP4}\\
   \hline
   $\mu$ & 300& -500& -300&300 & 400\\
   $\tan \beta$ & 5 &5 &10  &5  &6.1\\
   $M_1$ &1500 &1500 &2000 &  -860  & -1150 \\
   $M_2$ &1800 &1800 &1000&  2500  & 2500\\
   $M_3$ & 2000&2000 &2000 &  2000 & 2000\\
   $M_A$ &2500 &803.2 &2500 &  2500 &2500  \\
   $T_t$ &2900 &2900 & -2500&  2950 & 2750\\
   $M_R$ & 100&100 &100 & 100  &100 \\
   $BM_R$ (GeV$^2$) &$ 10^{-3}$ &10.7&143 &$ 10^{-3}$  &$ 10^{-3}$ \\
   $m_{\widetilde{\nu}}$ &100 &404 & 245 &245 & 316.2\\
   $Y_{\nu} (\times 10^{-7})$ & 1  &  10 &1  & 1   &1 \\
   $T_{\nu}$ &0.02 &140&0.8 & 4.0 & 0.06 \\
   \hline
    $m_{\widetilde{\chi}^{\pm}_1}$ &303.6  &510.9 & 307.5&305.4   &407.2 \\
   $m_{\widetilde{\chi}^0_1}$ &301.7 &508.4&303.5  &305.5 &407.3 \\
    $m_{\widetilde{\chi}^0_2}$ &305.8 & 512.2 &  311.5&305.8 &407.5 \\
  %  $m_{\widetilde{\chi}^{\pm}_2}$ &1806.8 &1018.8 &1018.8  &2477.3&\\
   $m_{\widetilde{t}_1}$ &1034.6 & 1528.3&1024.7 & 1514.8   &1523.5\\
 %  $m_{\widetilde{t}_2}$ & 2011.2&2008.0 & 2008.0 &2018.2&\\ 
   $m_{\widetilde{b}_1}$ &1064.3 &1568.6& 1057.8 &1552.1 &  1555.2\\
  % $m_{\widetilde{b}_2}$ &2050.2 &2049.9 & 2049.9 &2065&\\ 
   $m_{\widetilde{l}_L}$ &380.3  &627.3&617.5 &617.9  & 618.7\\
   $m_{\widetilde{l}_R}$ &364.5 & 611.8&606.7 &608.5 &610.6\\   
   $m_{\widetilde{\nu}_L}$ & 372.4& 624.5&611.7&612.9 & 613.5\\
   $m_{\widetilde{\nu}_R}$ &141.4 &412.2 &264.1 &264.6  &331.7\\  
     $m_h$ &124.6 & 124.1& 126.1 &124.5 &124.7 \\
     
  \hline
   $\Delta{m_{CP}}$ (MeV) &0.004 & 25.7 & 900 &0.004 & 0.003\\
   $\Delta{m_1}$ & 1.9 &2.5 & 4.0&-0.1 & -0.1\\
   $\Delta{m_2}$ &2.2 &3.8& 4.0&0.4  &0.2\\
  % $\Delta{m_3}$ &4.1 &3.5& 8&0.3 \\

   $\Delta M $ &162.2 &96.2 &43 &40.9 &75.5 \\
   $\Omega h^2$ & &0.11 & &  &\\
   $\sigma_{SI}$ (pb) & &1.4$\times 10^{-10}$ & &  &\\
   \hline
   $\sin \theta^{j} (\times 10^{-2})$ & 0.002&10.9 & 0.046& 0.224 &0.004\\
   $BR(\widetilde{\chi}^{\pm}_1 \rightarrow l \text{ } \widetilde{\nu})$ & 0.13&1.00 &0.34 & 1.0 &1.0\\
   $BR(\widetilde{\chi}^{0}_2 \rightarrow W^{\mp *} \, \widetilde{\chi}^{\pm}_1)$ & 0.12&0.0&0.09 &0.0001 &  0.10 \\
   $BR(\widetilde{\chi}^{0}_2 \rightarrow W^{\mp *} \, \widetilde{\chi}^{\pm}_1 \rightarrow l \widetilde{\nu} \, W^{*\mp} $) & 0.015&0.0 &0.031 &0.0001  &0.10\\
   \hline
\end{tabular}
\caption{Low energy input parameters and sparticle masses for the benchmarks 
used in the current study. All soft mass parameters and mass differences are in 
GeV. Mass differences amongst the different higgsino sector sparticles, $\Delta m_1$ 
and $\Delta m_2$, are as defined in Section 2.2. Additionally, $\Delta M = 
m_{\widetilde{\chi}^{\pm}_1} - m_{\widetilde{\nu}}$ represents the mass gap 
between the chargino and the sneutrino LSP and $\theta^j$ represents the 
mixing angle between the lightest left and right sneutrinos.}
\label{bp}
\end{center}
\end{table}

\textbf{BP1} represents a point in the $M_1M_2>$ 0 and $\mu>0$ plane with 
$M_1 = 1.5 $ TeV, $M_2 = 1.8 $ TeV, $\tan \beta = 5$ and $\Delta m_{1/2} \sim 2$ 
GeV. The LSP mass is $\sim$ 140 GeV and therefore there is a large mass gap 
between the higgsinos and the LSP, $\Delta M( = m_{\widetilde{\chi}^{\pm}_1} - 
m_{\widetilde{\nu}_{LSP}}) \sim 162$ GeV. The first two generation sleptons are 
of masses $\sim$ 360 GeV to facilitate left-right mixing in the sneutrino sector. 
The mixing in the left-right sneutrino is $\mathcal{O} (10^{-5})$, such that for 
\textbf{BP1} the three body decay of $\widetilde{\chi}^{\pm}_1$, i.e.
BR($\widetilde{\chi}^{\pm}_1 \rightarrow \widetilde{\chi}^{0}_1 \,\, {W^{\pm}}^*$) 
dominates ($\sim 88\%$) over the two-body decay, BR($\widetilde{\chi}^{\pm}_1 
\rightarrow l  \,\, \widetilde{\nu}$) ($\sim 12\%$). For a heavier slepton mass, 
a larger $T_{\nu}$ value is required for a similar left-right mixing angle and 
vice versa. Thus, we can fix the leptonic branching of the chargino either by 
lowering the left slepton mass or increasing $T_{\nu}$, and hence the left-right 
mixing in the sneutrino sector. Since the softer decay products from the three body 
decay produced from the off-shell $W$ pass undetected owing to the compression 
in the electroweakino sector, the two body leptonic decay is of interest, although 
subdominant. $\widetilde{\chi}^{0}_1$ and $\widetilde{\chi}^{0}_2$ dominantly 
decay to a $\nu \,\, \widetilde{\nu}$ pair contributing to missing energy signal. 
The dominant signals to look for in this case are mono-lepton $+$ $\slashed{E}_T$ 
and to a lesser extent opposite-sign and same-sign di-lepton events owing to the 
small leptonic branching of chargino. 

Further, we choose a benchmark $\bf{BP2-a}$, consistent with current data and similar to $\bf{BP1}$, but 
with an increased left-right mixing in the sneutrino sector and satisfying thermal relic density in presence of heavy higgs resonance, 
 of mass $\sim 824$ GeV.
It represents a point in the $M_1M_2> 0 $ 
and $\mu<0$ plane with $M_1 = 1.5 $ TeV and $M_2 =  1.8 $ TeV with the chargino decaying completely to the lepton and sneutrino mode. 
%Note that for $\mu<0$ a larger region of parameter space exists.
Here $\Delta m_1 = 2.5 $ GeV, $\Delta m_2 = 3.8 $ GeV. We focus only on signals from the electroweakino sector and choose to keep the 
first and second generation left and right sleptons $\sim 600 $ GeV such that 
their production cross-sections are negligible at 13 TeV LHC, thus reducing any 
additional contributions to the leptonic final states. Owing to the large left-right mixing in the sneutrino sector, 
the higgsinos decay entirely to the sneutrino final state. Thus, the dominant signals from this scenario are monolepton and opposite sign dileptons along with missing energy.

We choose another spectrum \textbf{BP2-b} similar to \textbf{BP2-a} but for a lighter $
\mu $  = 300 GeV and LSP mass $ \sim 264$ GeV such that 
$\Delta M \sim 40$ GeV. Thus, we choose a nearly compressed spectrum \textbf{BP2-b} 
where the leptons are much softer as compared to those of \textbf{BP2-a} in order 
to study the prospects of such a spectrum in presence of a $\widetilde{\nu}$ 
LSP. The dominant signals to look for are 
mono-lepton, opposite-sign di-lepton and same-sign di-lepton along with missing 
transverse energy.

\begin{table}[t!]
\begin{tabular}{|c|c|c|c|c|c|c|}
\cline{3-7}
 \multicolumn{2}{}{}& \multicolumn{5}{|c|}{Luminosity (in $fb^{-1}$) for $3\sigma$ excess}\\
 \hline
Analyses & Reference & \textbf{BP1} & \textbf{BP2-a} & \textbf{BP2-b} & \textbf{BP3} & \textbf{BP4}\\ 
%$l^{\pm}$ + $\slashed{E}_T$ & \cite{CMS-PAS-SUS-16-052} & & & &&\\ 
% & \cite{Aaboud:2017efa} & & & &&\\ 
 %\hline
 \hline
$l^{\pm}l^{\mp}$ (SF)+ 0 jet  + $\slashed{E}_T$& \cite{ATLAS-CONF-2017-039} &13397  & 812 &- &- & 958   \\ %ATLAS
$l^{\pm}l^{\mp}$ (DF)+ 0 jet  + $\slashed{E}_T$& \cite{ATLAS-CONF-2017-039} & 2191& 162  &- & -& 104  \\ %ATLAS
$l^{\pm}l^{\mp}$ + 0 jet  + $\slashed{E}_T$& \cite{CMS-PAS-SUS-17-009} & - & 2223 & -&- & 385 \\ %ATLAS
 \hline
$l^{\pm}l^{\pm}$ + 0 jet + $\slashed{E}_T$ &  \cite{Sirunyan:2017lae}& -&- & 1997   &- & 2726  \\
$l^{\pm}l^{\pm}$ + 1 jet + $\slashed{E}_T$ &  \cite{Sirunyan:2017lae}&- &- & 4039 &- &4901  \\
 \hline
\end{tabular}
\caption{Forecast for luminosity for 3$\sigma$ excess using present experimental searches using 36 $fb^{-1}$ of data at LHC. The blank spaces
indicate that the benchmark is not sensitive to the final state analysis. We do not show the forecast from current monoleptonic searches as it gives much weaker sensitivity to 
our scenario.}
\label{LHCforecast}
\end{table}
%We find that a limit of 4$\%$ for $\mu=300 $GeV for large values of $\Delta M \ge 40 $ GeV. 
\textbf{BP3} and \textbf{BP4} represent spectra with $M_1 M_2 < 0$ and $\mu>0$ with $M_1 = -860 (-1150)$ 
GeV, $M_2 = 2.5 $ TeV and $\tan \beta = 5$ such that the NLSP is the $\widetilde{\chi}^{\pm}_1$. For \textbf{BP3}
we also choose a large left-right 
sneutrino mixing ($\mathcal{O}(10^{-3}$) while the LSP mass is 264 GeV. This 
leads to a tightly compressed electroweakino sector with $\widetilde{\chi}^{\pm}_1$ 
as the NLSP as discussed in section \ref{sec:delm}. Hence the only allowed decay 
of the chargino is the two body leptonic decay to the LSP with BR$(\widetilde{\chi}^{\pm}_1 \rightarrow l \widetilde{\nu}) = 100\%$. Thus this region of parameter space 
favors the di-lepton channel with missing energy from chargino pair production. 
However the di-lepton channel suffers from a huge SM background and is much difficult 
to observe. Again the larger cross section for chargino-neutralino production only 
contributes to the mono-lepton channel as the decay of the heavier neutralinos 
to the chargino is rather suppressed for \textbf{BP3} in order to respect the 
bounds from existing same-sign di-lepton searches. Hence for this particular benchmark 
the dominant signal to look for is mono-lepton $+$ $\slashed{E}_T$ and, to 
a lesser extent, opposite sign di-lepton $+$ $\slashed{E}_T$. However, other 
choices of benchmark points in this region of parameter space would allow 
same-sign di-lepton signal along with missing energy making it very interesting and 
clean mode for discovery. This can be the preferred channel for much larger $\mu$. We demonstrate a single benchmark, \textbf{BP4},
with $\mu=400$ GeV for this purpose. 

Note that \textbf{BP2-a} is the only benchmark shown with the correct relic density
($\Omega h^2 =0.11$) suggesting that the LSP in this case is a thermal DM candidate. 
While the other benchmarks are assumed to have non-thermal DM we could have made 
them thermal by adjusting the mass of one of the heavy Higgs to achieve resonant 
annihilations and satisfy the relic density criterion. However, from the collider point of 
view the relic value will not affect the signals at LHC for any of the benchmark points and in 
the process neither differentiate a thermal relic from a non-thermal one. 

Before we propose our analysis for observing the signal at LHC we use the 
existing analyses and forecast the integrated luminosity that would be required to observe 
a 3$\sigma$ excess at LHC for each of the benchmark points. We summarize our observations in table~\ref{LHCforecast}. For the above estimates, we have used the SM background events from the given references for respective analyses as shown in the 
table while we have computed the signal events in \texttt{Madanalysis-v5}. 
%%%
\subsection{Collider Analyses}
%%%%%
\subsubsection*{Simulation set-up and Analyses}

Our focus in this study is on leptonic channels with up to one ISR jet ($p_T > $ 40 GeV). 
We consider no extra partons at the matrix element level while generating the parton-level 
events for the signal using \texttt{MadGraph-v5} \cite{Alwall:2011uj,Alwall:2014hca,
Sjostrand:2006za}. Following the event generation at parton-level, showering and 
hadronisation of the events are performed using \texttt{Pythia-v8} 
\cite{Sjostrand:2007gs, Sjostrand:2014zea}. Subsequently detector simulation is 
performed using \texttt{Delphes-v3} \cite{deFavereau:2013fsa,Selvaggi:2014mya,Mertens:2015kba}. 
Default dynamic factorization and renormalization scales of \texttt{MadGraph-v5} have 
been used with \texttt{CTEQ6L} \cite{Pumplin:2002vw} as the parton distribution functions 
(PDF). Jets are reconstructed using {\tt Fastjet} \cite{Cacciari:2011ma} with a minimum 
$p_{T}$ of 20 GeV in a cone of $\Delta R = 0.4$ using the {\sl anti}-$k_t$ algorithm 
\cite{Cacciari:2008gp}. The charged leptons ($e,\mu$) are reconstructed in a cone 
of $\Delta R = 0.2$ with the maximum amount of energy deposit from other objects
in the cone limited to 10$\%$ of the $p_{T}$ of the lepton. Photons are also reconstructed 
similar to the leptons in a cone of $\Delta R = 0.2$, with the maximum energy deposit 
from  other objects in the cone being at most 10$\%$ of the $p_{T}$ of the photon. 
%All the benchmarks are checked using the public software \texttt{CheckMATE} 
%\cite{Drees:2013wra,Dercks:2016npn} to ensure that the 8 TeV and existing 13 TeV 
%constraints from LHC are respected. Amongst the results not implemented in \texttt{CheckMATE}, we have recast them in \texttt{MadAnalysis-v5} to check the benchmarks pass the current limits.

SM backgrounds have also been generated using \texttt{MadGraph-v5}, 
\texttt{Pythia-v6} \cite{Sjostrand:2006za} and visible objects reconstructed at the 
detector level using \texttt{Delphes-v3} \cite{Mertens:2015kba,Selvaggi:2014mya,
deFavereau:2013fsa}. Dominant SM backgrounds such as $l \nu + 0,1 j$ and Drell Yan 
($l^+ l^- + 0,1$ j) with large production cross-sections 
have been generated upto 1 extra parton. The matching between shower jets and jets produced at parton level is done using MLM
matching with \textit{showerKT} algorithm using $p_T$ ordered showers
and a matching scale \texttt{QCUT} = 20 GeV. Signal and background analysis has been performed using 
\texttt{MadAnalysis-v5} \cite{Conte:2012fm,Conte:2014zja,Dumont:2014tja}.

\subsection*{Primary Selection Criteria }
We choose the following basic criteria for leptons (only $e^{\pm}$ and $\mu^{\pm}$), jets 
and photons for both signal and background: 
\begin{itemize}
 \item We select leptons (e, $\mu$)  satisfying $p_T > $ 10 GeV and $|\eta| <$ 2.5. 
 \item We choose photons with $p_T > $ 10 GeV and $|\eta| <$ 2.5. 
 \item Reconstructed jets are identified as signal jets if they have $p_T>$ 40 GeV and 
 $|\eta| <$ 2.5.
  \item Reconstructed b-tagged jets are identified with $p_T>$40 GeV and $|\eta| <$ 2.5.
 \item Jets and leptons are isolated such that $\Delta R_{lj}>$ 0.4 and $\Delta R_{ll}>$ 0.2.
 \end{itemize}
 
\begin{figure}[H]
 \includegraphics[height=2.55in,width=2.9in]{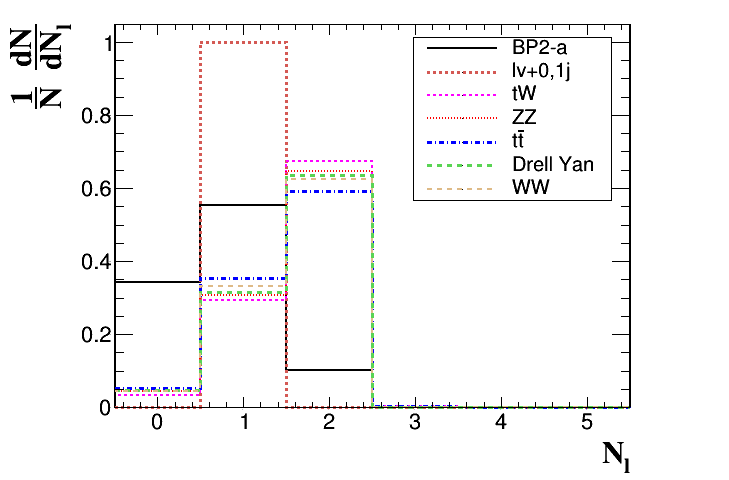}
  \includegraphics[height=2.6in,width=3.2in]{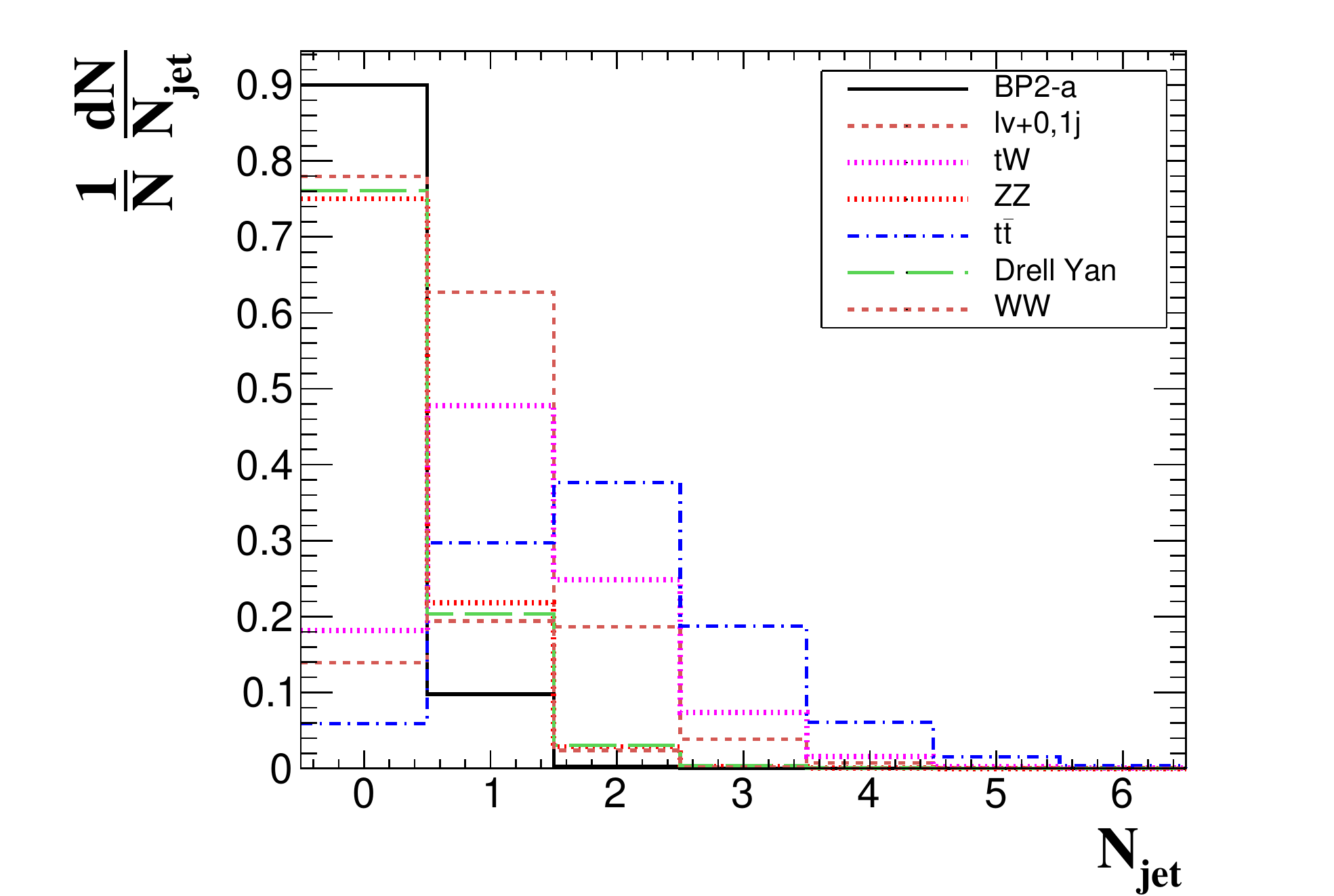}
    \caption{Normalized distributions for lepton and jet multiplicity for benchmark \textbf{BP2} and dominant SM backgrounds channels. 
    respectively.}
   \label{njetsplot}
\end{figure}
\subsection{Mono-lepton + $\leq 1 $ jet $+$ $\slashed{E}_T$ signal}
The presence of a sneutrino LSP opens up decay channels of the lightest chargino 
(neutralino) to a lepton (neutrino) and sneutrino. In such cases, mono-lepton signals 
with missing energy and few jets (mainly from ISR) arise dominantly from 
$\widetilde{\chi}^{\pm}_1\widetilde{\chi}^0_1$, 
$\widetilde{\chi}^{\pm}_1 \widetilde{\chi}^0_2$ with $\widetilde{\chi}^{\pm}_1 \rightarrow 
l \, \widetilde{\nu}$ and $\widetilde{\chi}^{0}_{1/2} \rightarrow \nu \widetilde{\nu}$. 
Sub-dominant contributions to the signal may also arise from $\widetilde{\chi}^{\pm}_1 
\widetilde{\chi}^{\pm}_1$ pair production when one of the chargino decays to a 
soft lepton (via the three body decay to the neutralino) and the other one 
decays to a hard lepton and the LSP. Smaller contributions to the signal also come from 
$\widetilde{\chi}^{0}_1\widetilde{\chi}^{0}_2$ production with $\widetilde{\chi}^{0}_2$ 
decaying to a chargino and soft decay products while $\widetilde{\chi}^{0}_1$ decays 
invisibly or vice versa if the chargino is the lightest among the higgsinos. 
 
Dominant background to this signal come from SM processes: 
\begin{itemize}
 \item $l^{\pm} \nu +$0,1 jets
 (including contributions from both on-shell and off-shell $W$ boson), 
 \item $t \bar{t}$ (where one of the top quark decays hadronically and the other semi-leptonically). 
 \item Single top quark production ($t (\bar{t})\, j, \,\, tW$).
\item $W^+ W^- + $ jets ($W \rightarrow l \nu$, $ W \rightarrow j j $).
\item $t \bar{t} W +$ jets (when both top quarks decay hadronically and $W \rightarrow l \nu$) and 
\item $W Z$ (with $W \rightarrow l \nu$, $Z \rightarrow \nu \bar{\nu} / j j )$. 
\end{itemize}
Other subdominant contributions come from $t \bar{t} $  (where both top quarks decaying 
semi-leptonically), Drell Yan process ($l^+ l^- + 0,1j)$ and $ZZ, (Z \rightarrow l^+ l^-, Z 
\rightarrow \nu \bar{\nu}/ j j)$ from 
misidentification of one of the leptons fail to meet the isolation cuts required to 
identify signal leptons or even hadronic energy mismeasurements leading to jets faking 
leptons. Smaller contributions may also arise from triple gauge boson production with 
one of the gauge boson decaying leptonically and the others hadronically. However, these 
are negligible compared to the $l \nu + jets$ contribution. Other indirect contributions 
may arise from energy mismeasurements of jets as missing energy.

\begin{figure}[ht]
\begin{center}  
 \includegraphics[width=4.2in]{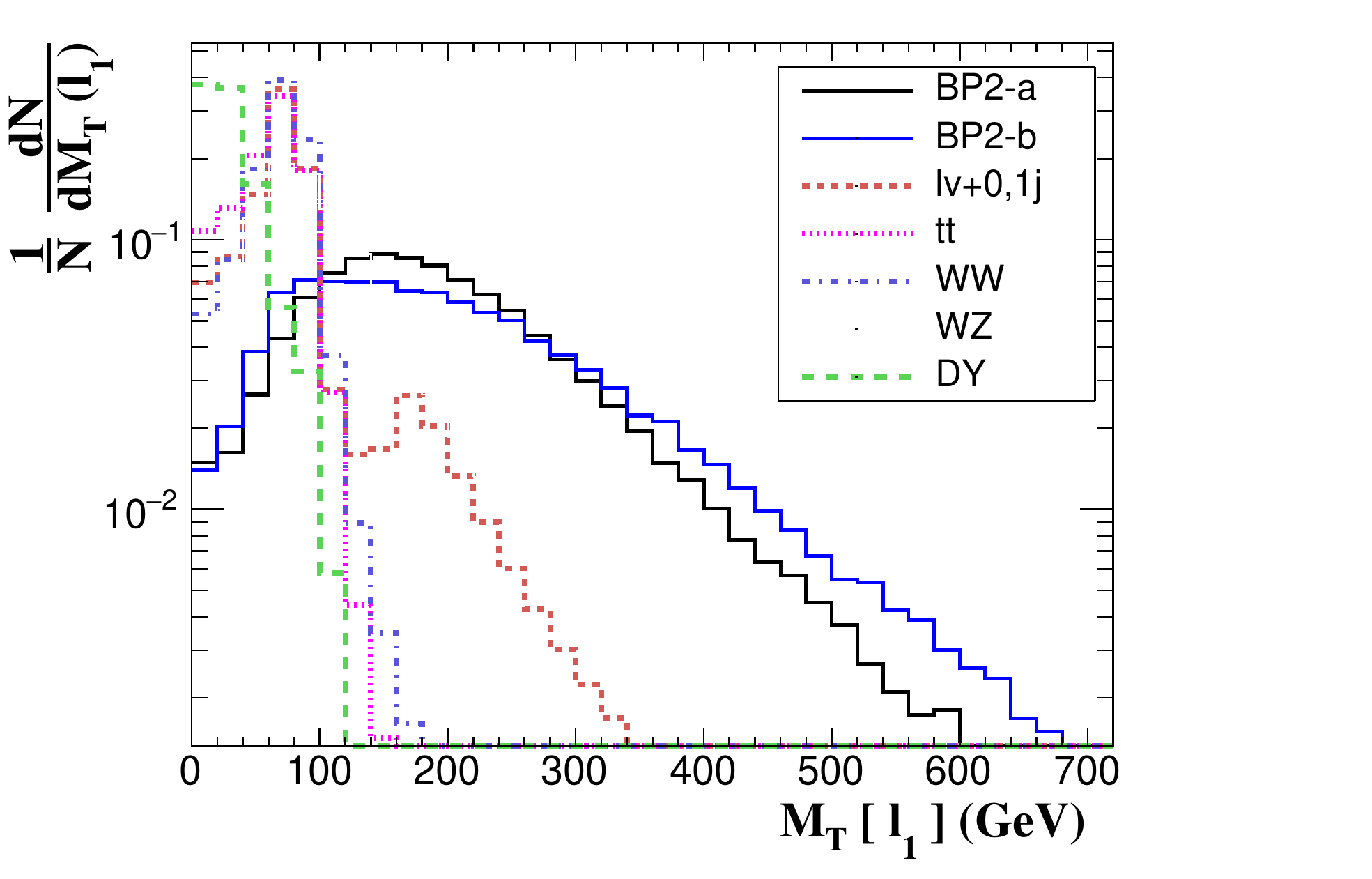}
 \caption{Normalized distribution for $M_T(l_1)$, the transverse mass of the leading 
 lepton for SUSY signal $\bf{BP2-a}$ and $\bf{BP2-b} $ against the  dominant SM backgrounds 
 after preselection cut \bf{M1}.}
\label{fig:mtl1} 
 \end{center}
\end{figure}
In order to select one lepton + missing energy signal, we implement the following criteria 
for both signal and backgrounds:
\begin{itemize}
 \item \textbf{M1}: The final state consists of a single lepton with $p_T>$ 25 GeV and 
 no photons. \item \textbf{M2}: Since the dominant background contributions arise from W 
 bosons, a large cut on the transverse mass, $M_T ( l,\slashed{E}_T) > $ 150 GeV, where 
 \begin{equation}
  M_T (l,\slashed{E}_T) = \sqrt{2 p_T (l) \slashed{E}_T ( 1 - \cos(\Delta \phi))}.
 \end{equation}
  
$\Delta \phi$ denotes the azimuthal angle separation between the charged lepton 
$\vec{p_{T}}$ and $\vec{\slashed{E}_T}$. A large cut on $M_T$  reduces SM background 
contributions from $l \nu + 0,1$ jet, $WZ$, $WW$ and $t \bar{t}$ substantially as 
compared to the signal as seen in cut flow Table ~\ref{signalmono1} and ~\ref{bkgmono1} .
 
 \begin{table}[hb]
\begin{center}
\begin{tabular}{|c|c|c|c|c|c|c|c|}
\hline
\multicolumn{1}{|c|}{Signal} & 
\multicolumn{6}{|c|}{Number of events after cut } \\
\cline{2-7} 
 & Preselection(\bf{M1}) &  \bf{M2}  & \bf{M3}  & \bf{M4}  &\bf{M5}  &  \bf{M6} \\
\hline
BP1 &2543& 1987 &1946&1936 &1601&1429 \\
BP2-a&1495 &944&922 &916&706& 611\\
BP2-b& 6252&3194& 3128& 3118&2462& 2215\\
BP3 & 1.06$\times10^4$ &1664&1614&1601&1138&919\\
%BP4 &1700 &756 &736 & 731 &530 & 448\\
BP4 & 3919&1793 &1751 &1740 &1258 & 1074 \\
\hline
   \end{tabular}
\end{center}
 \caption{Mono-lepton $+$ missing energy signal final state  number of events at 
 100 fb$^{-1}$  for SUSY signals. Note that the events have been rounded-off to the 
 nearest integer. Cross-sections have been scaled using NLO K-factors obtained 
 from \texttt{Prospino}.}
 \label{signalmono1}
\end{table}
 \item \textbf{M3}: Events with atleast one b-tagged jet with $p_T>$ 40 GeV are rejected 
 in order to reduce contribution from channels involving top quarks while leaving SUSY 
 signals mostly unaffected.  
 
 \item \textbf{M4}: As seen from Figure~\ref{njetsplot} the weakly produced SUSY signals 
 have a comparatively lower jet multiplicity compared to SM background processes involving 
 strong production such as $t \bar{t}$ or single top. Thus a  cut  on the jet multiplicity 
 in the signal events help to suppress the large SM background from these 
 sources. Thus, we demand jet multiplicity, $N_{jet} \leq$ 3.
 
 \item \textbf{M5}: Since SUSY signals have a large missing energy compared to the SM 
 background, $\slashed{E}_T > 100$ GeV helps to reduce contributions from background.
 
 \item \textbf{M6}: In addition the events are made  quiet from hadronic activity by 
 demanding at most 1 jet in the final state. This helps to further reduce backgrounds 
 events from $t \bar{t}$ and single top production.
 
\end{itemize}

\begin{table}[H]

  \footnotesize
   \begin{tabular}{|c|c|c|c|c|c|c|}
\hline
\multicolumn{1}{|c|}{SM } &
\multicolumn{6}{|c|}{Number of events after cut} \\
\cline{2-7} 
Backgrounds& Preselection(\bf{M1}) &  \bf{M2}  &   \bf{M3}  & \bf{M4}  &\bf{M5}  &  \bf{M6} \\

\hline
 $l \nu +0,1j $&1.07$\times10^7$&1.09$\times10^6$  &1.08$\times10^6$ &1.08$\times10^5$&5.77$\times10^5$&5.49$\times10^5$\\ 
  $\text{Drell Yan}$ &3.52$\times10^7$&3.47$\times10^4$ &3.34$\times10^4$&5991&5272 &3674 \\
  
  $W W $ &8.04$\times10^{5}$ &5696&5485& 5446&1329 &1130 \\
  $WZ $& 1.56$\times10^5$&2.54$\times10^4$&2.48$\times10^4$& 2.20$\times10^4$&1.55$\times10^4$ &11523\\
  $ZZ $&4938 &912 &900 &899  &551&492 \\
  %  $tj $& & &   & & & & &\\
  $t \bar{t}$ &2.04$\times10^6$ &5.97$\times10^4$&1.79$\times10^4$ &1.68$\times10^4$&1.17$\times10^4$ &6399\\
  Single top  & 3.68$\times10^6$&2.05$\times10^4$  &8517 &8088& 2659 & 1603\\
 \hline
Total&\multicolumn{5}{|c|}{}&5.74$\times10^5$\\
 \hline

 \end{tabular}
\label{bkgmono1}
\caption{Mono-lepton $+$ missing energy signal final state  number of events at 
100 fb$^{-1}$ for SM background. Note that the events have been rounded-off to the 
nearest integer. Cross-sections scaled with K-factors at NLO \cite{Alwall:2014hca} 
and wherever available, NNLO \cite{Grazzini:2016swo,Cascioli:2014yka,Giammanco:2017xyn,
Czakon:2013goa,Czakon:2011xx,Kidonakis:2016sjf} have been used.}
  \end{table}
In Table~\ref{signalmono1} and ~\ref{bkgmono1} we list the number of events observable 
at 13 TeV LHC at 100 fb$^{-1}$, for the signal and SM background respectively. Although 
most of the SM background events could be suppressed, the continuum background from 
$l \, \nu \, + 0,1 \,j$ survives most of the cuts. The required luminosities for 
observing a 3$\sigma$ and $5\sigma$ excess for the mono-lepton $+$ $\slashed{E}_T$ 
channel are given in Table~\ref{significancemono}. The statistical significance is 
computed using:
\begin{equation}
 \mathcal{S} = \sqrt{2[(s+b)\text{ln}(1+\frac{s}{b})-s]}
\end{equation}
where $s$ and $b$ refer to the number of signal and background events after implementing 
the cuts $\textbf{M1}-\textbf{M6}$ respectively. 

\begin{table}[H]
\begin{center}
 \begin{tabular}{|c|c|c|}
 \hline
  Signal & $\mathcal{L}_{3\sigma}$ (fb$^{-1}$)&$\mathcal{L}_{5\sigma}$(fb$^{-1}$)\\
  \hline
   BP1 & 254 & 704  \\
  %BP2-a & 46 & 126 \\
   BP2-a &1384 & 3485  \\
  BP2-b & 106 &293\\
  BP3 &613 & 1701 \\
  BP4 &448 & 1245 \\
  \hline
 \end{tabular}
\caption{Required luminosities for discovery of mono lepton final states with 
missing energy at $\sqrt{s}=13$ TeV LHC.}
\label{significancemono}
\end{center}
\end{table}
We find that the best signal significance is obtained by retaining at least one jet 
in the signal for all the benchmarks since the dominant background $l \nu+0,1 j$ and 
signal both have only ISR jet contributions. We note that requiring large $M_T$, 
$\slashed{E}_T$ and one jet in the final state helps to improve the signal significance. 
Among all the benchmarks, \textbf{BP2-a} and \textbf{BP4} have highest leptonic branching fraction for 
the chargino ($100\%$) as well as a large mass gap $\Delta M$ between the chargino and 
LSP. This leads to a relatively high cut efficiency for the signal. However since \textbf{BP2-a}  corresponds to $|\mu| = 500 $ GeV, 
the overall required luminosity for 3$\sigma$ excess is $\sim 1400 fb^{-1}$. For \textbf{BP4} with $\mu=300$ GeV and
thus a higher production cross-section, the required luminosity is $\sim 500 fb^{-1}$.  \textbf{BP1}, having a 
large $\Delta M$ but lower chargino leptonic branching fraction, i.e., $\sim 12\%$ 
would require $254$ fb$^{-1}$ of data for observing a 3$\sigma$ excess at LHC. The 
relatively compressed spectra \textbf{BP2-b} and \textbf{BP3} although with large 
leptonic branching fractions of the chargino, i.e. $\sim 37\%$ and $100\%$ respectively,
have a lower cut efficiency owing to a smaller $\Delta M \sim 40$ GeV. Thus the corresponding 
leptons would be soft compared to \textbf{BP1} and \textbf{BP2-a}. Therefore \textbf{BP2-b} and  \textbf{BP3} 
require higher luminosity $106$ fb$^{-1}$ and $613$ fb$^{-1}$ respectively for observation.

\subsection{Di-lepton + $0$ jet $+ \slashed{E}_T$ signal}
\label{sec:dilep}
The challenge in having a multi-lepton signal from the production of compressed 
higgsino-like electroweakinos comes from the fact that the decay products usually 
lead to soft final states. However, with a sneutrino LSP and
the possibility of the decay of the chargino to a hard lepton and the LSP leads to a 
healthy di-lepton signal with large missing energy (from $\widetilde{\chi}^{+}_1 \,\, 
\widetilde{\chi}^{-}_1$ as well as $\widetilde{\chi}^{\pm}_1 \,\, \widetilde{\chi}^{0}_{2}$ 
pair production, provided the next-to-lightest neutralino decay yields a lepton 
via the chargino). A sub-dominant contribution also arises from $\widetilde{\chi}^{0}_1 \,\, 
\widetilde{\chi}^{0}_2$ with each of the neutralino decaying to a chargino and an 
off-shell $W$ boson which gives soft decay products. The chargino then decays to a 
charged lepton and sneutrino LSP. This happens most favorably when  chargino is the 
lightest of the higgsinos. Owing to the Majorana nature of $\widetilde{\chi}^{0}_i$ 
we can have signals for opposite-sign and same-sign di-lepton final states with large 
missing transverse energy. Hence we look into both the possibilities:
\begin{itemize}
 \item Opposite sign di-lepton  + $0$ jet $+$ $\slashed{E}_T$
\item Same sign di-lepton  + $0$ jet $+$ $\slashed{E}_T$
 \end{itemize}
 
\subsubsection{Opposite Sign di-lepton  + $0$ jet + $\slashed{E}_T$ signal}
\label{sec:osdl}

Opposite sign di-lepton signal arises mainly from $\widetilde{\chi}^{+}_1\widetilde{\chi}^{-}_1$ 
production process. Sub-dominant contributions arise from $\widetilde{\chi}^{\pm}_1 \,\, \widetilde{\chi}^{0}_1$, $\widetilde{\chi}^{\pm}_1 \,\, \widetilde{\chi}^{0}_2$ and
$\widetilde{\chi}^{0}_1 \,\, \widetilde{\chi}^{0}_2$ as discussed before. The dominant SM 
contributions to the opposite sign di-lepton signal with missing energy come from 
$t \bar{t} $, $t W $ and Drell-Yan production. Among the di-boson processes, 
$W^+ W^-$ $(W^+ \rightarrow l^+ \nu, W^- \rightarrow l^- \bar{\nu})$ , $Z Z$ ( $Z \rightarrow 
l^+ l^-, Z \rightarrow j j/ \nu \bar{\nu}$ ) and $ W Z$+jets ($W \rightarrow j j, 
Z \rightarrow l^+ l^-)$ also contribute substantially to the opposite sign di-lepton 
channel. The triple gauge boson processes may also contribute. However, these have a small 
production cross-section and are expected to be subdominant. There could also be fake 
contributions to missing energy from hadronic energy mismeasurements. 

\begin{figure}
\includegraphics[height=2.2in,width=3.0in]{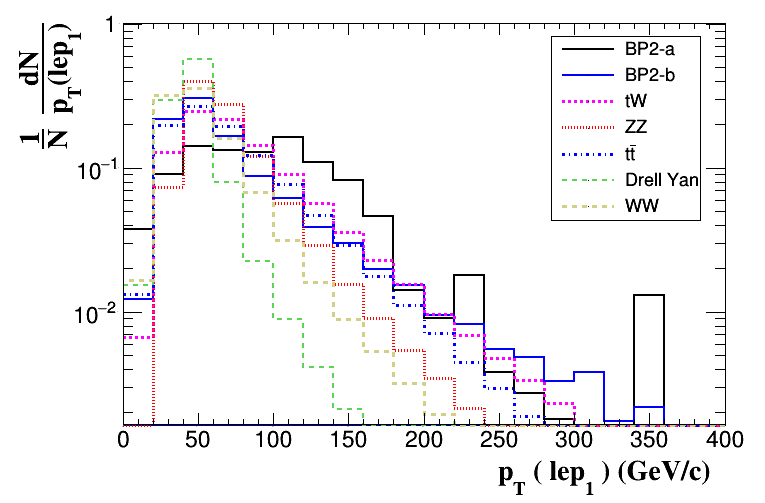}
\includegraphics[height=2.2in,width=3.0in]{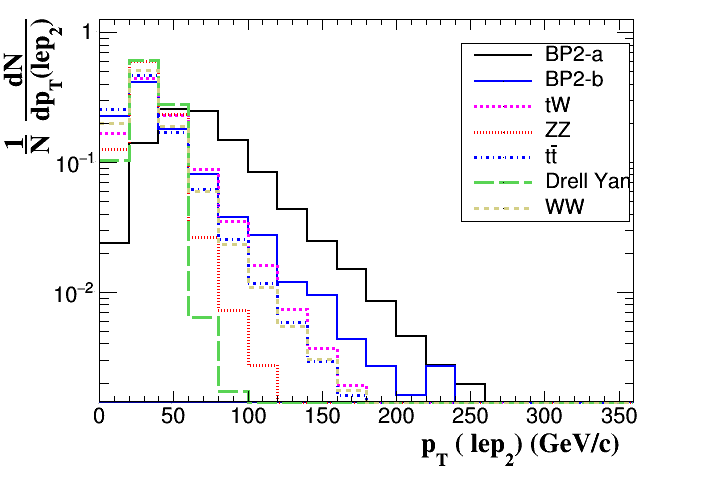}
\includegraphics[height=2.2in,width=3.0in]{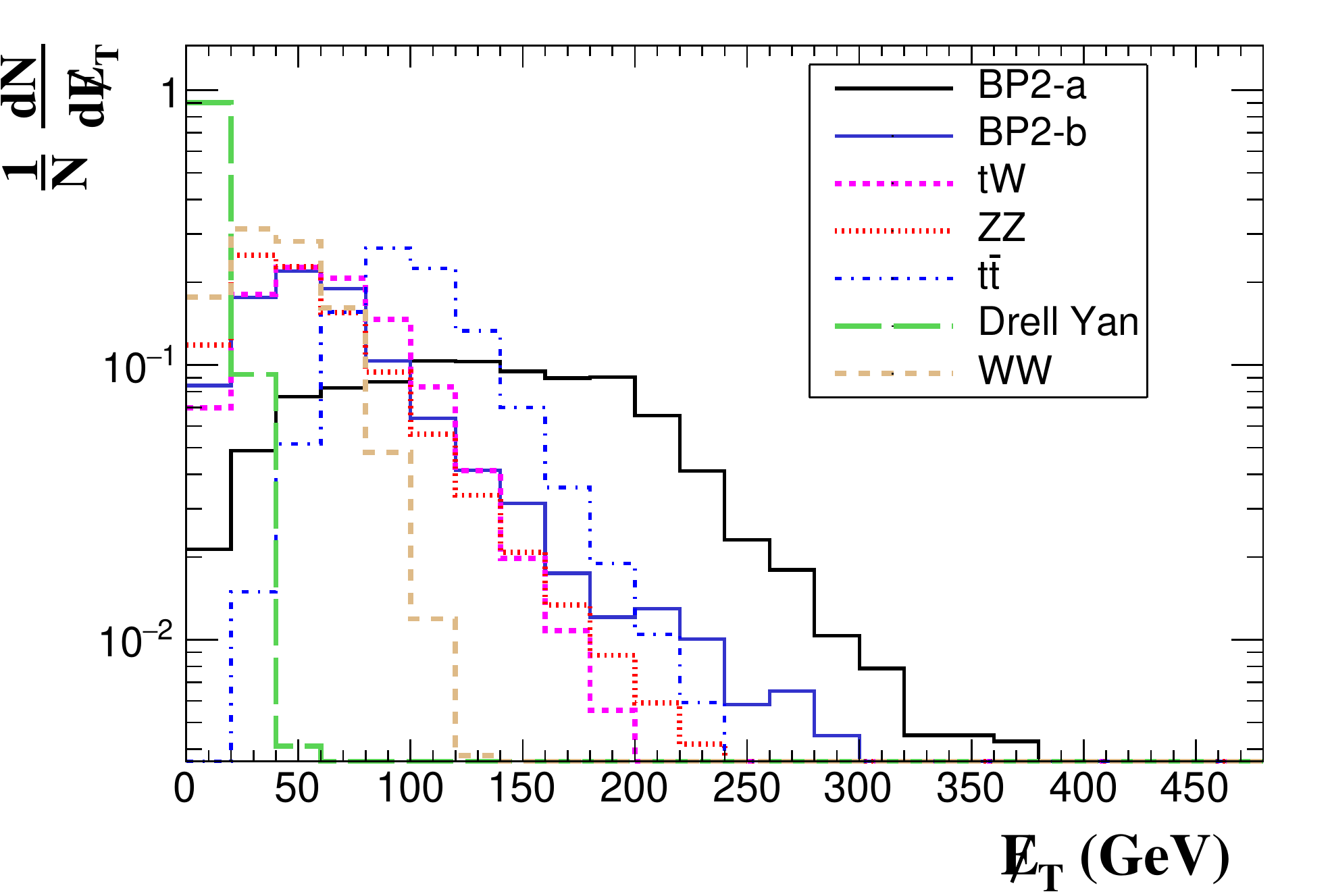}
\includegraphics[height=2.2in,width=3.0in]{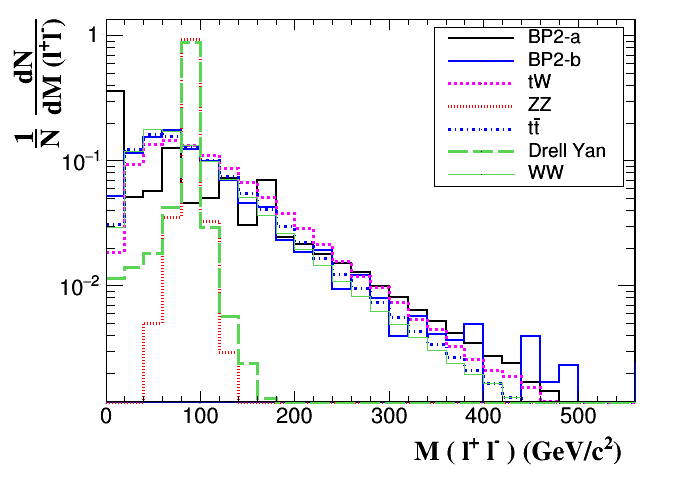}
\includegraphics[height=2.3in,width=3.0in]{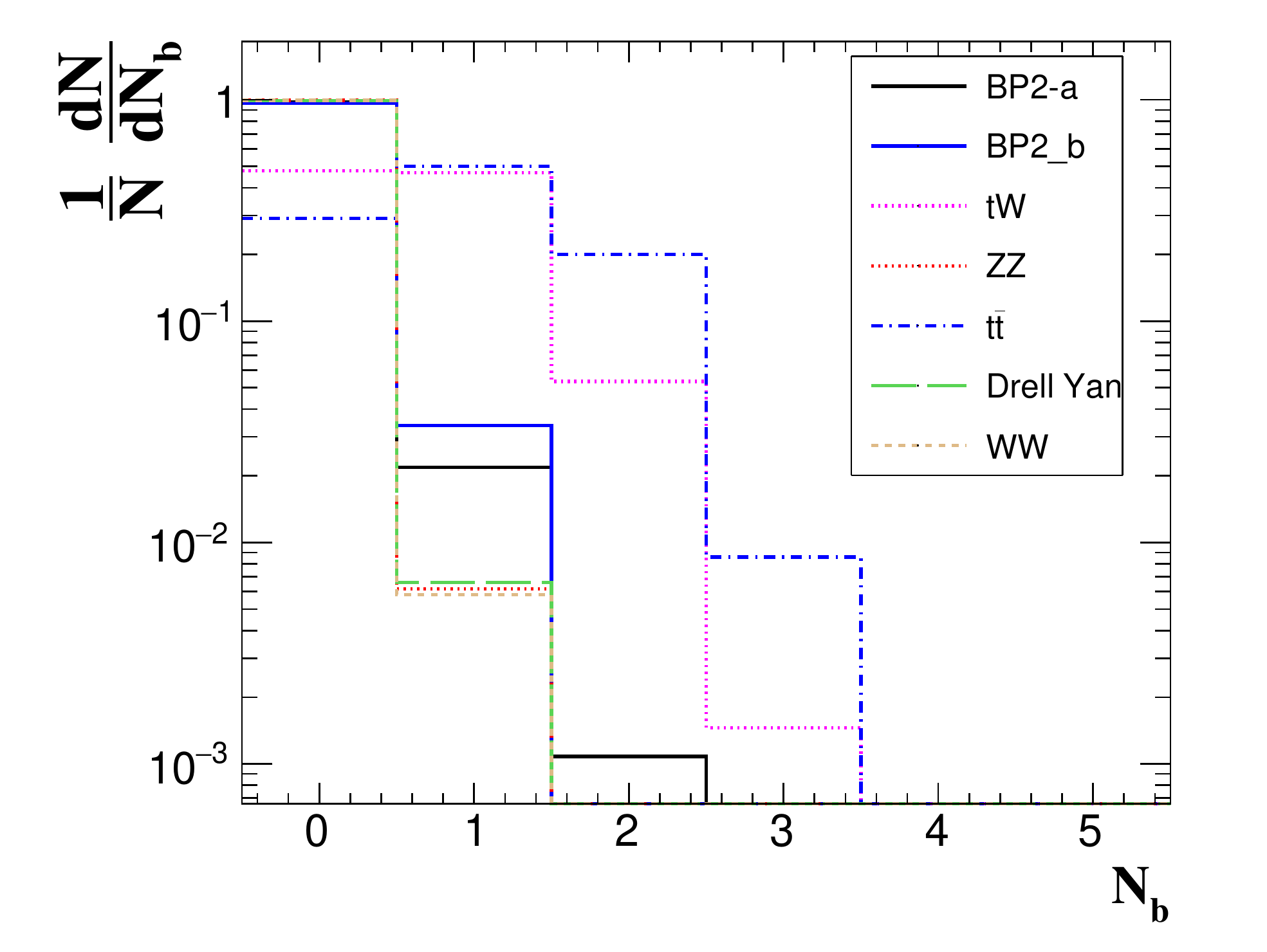}
\includegraphics[height=2.3in,width=2.9in]{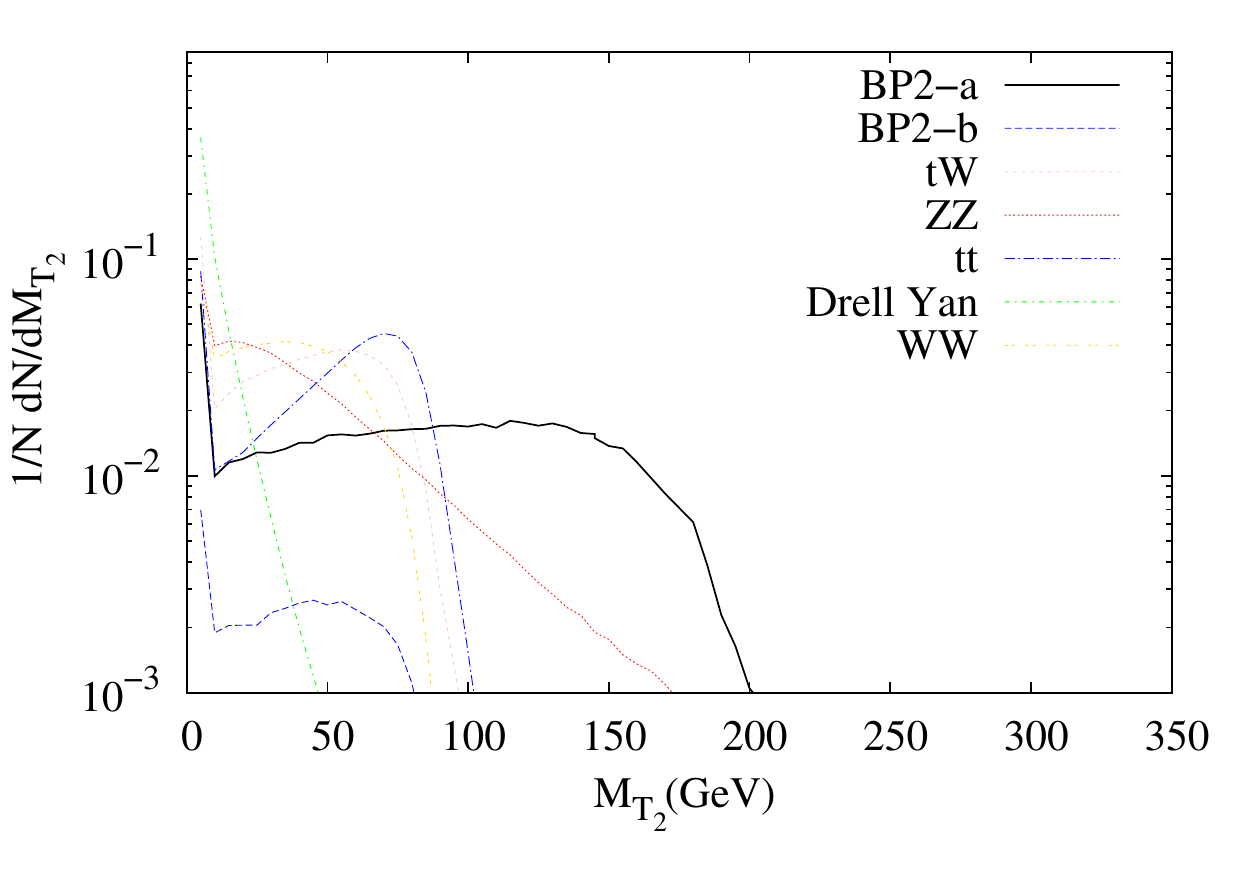}
\caption{Normalized distributions of several kinematic variables after cut \bf{D1}.}
\label{KE1dilep}
\end{figure}

In Figure~\ref{KE1dilep} we show the normalized distributions for important kinematic 
variables for two benchmarks \textbf{BP2-a} and \textbf{BP2-b} with $\Delta M = 100,~ 40$ 
GeV respectively along with the dominant SM backgrounds after selecting the
opposite sign-di-lepton state (\textbf{D1}). We find that as expected the lepton $p_T$ 
distribution for \textbf{BP2-a} is much harder than the SM backgrounds processes 
whereas for \textbf{BP2-b} with a lower mass gap between the chargino and LSP, the 
leptons are much softer and the distributions have substantial overlap with the 
backgrounds. We further use the other kinematic variables, 
\begin{equation}
 \slashed{E}_T  = |\Sigma_{i}  \vec{ p}_{T_i}| \text{ and } M^2_{l^+l^-} = ({p}_{l_1}+{p}_{l_2})^2
\end{equation}
(where $i$ runs over all visible particles in the final state) represent the transverse 
missing energy and invariant mass-squared of the di-lepton final state respectively which 
peak at higher values for SUSY signals over backgrounds in \textbf{BP2-a} whereas 
\textbf{BP2-b} still retains a large overlap with the SM backgrounds. However, the 
largest source of background for the di-lepton background coming from Drell-Yan process
can be removed safely by excluding the $Z$ boson mass window for $M_{l^+l^-}$. 
Since the SUSY signals do not arise from a resonance the exclusion of the $Z$ mass window 
is expected to have very little effect on the signal events. We further note that removing 
b-tagged jets would also be helpful in removing SM background contributions from the strongly 
produced top quark channels which have huge cross sections at the LHC. 

Another kinematic variable of interest to discriminate between SUSY signals and SM backgrounds 
is the $M_{T_2}$ variable \cite{Cheng:2008hk} constructed using the leading and sub-leading 
lepton $\vec{p_T}$ and $\vec{\slashed{E}_T}$. For processes with genuine source of 
$\slashed{E}_T$ there is a kinematic end point of $M_{T_2}$ which terminates near the 
mass of the parent particle producing the leptons and the invisible particle. In SM, 
channels such as $t \bar{t}, t W, W^+W^-$ involving a $W$ boson finally giving the massless 
invisible neutrino in the event, the end-point would be around 80 GeV. For SUSY events 
the invisible particle is not massless and therefore the visible lepton $p_T$ will depend 
on the mass difference. Thus the end-point in the signal distribution would not have a 
cut-off at the parent particle mass anymore. For \textbf{BP2-a} which has a large 
$\Delta M$ the end point is expected at larger values ($\sim 200$) GeV. However for 
\textbf{BP2-b}, where the available phase space is small for the charged lepton due to 
smaller $\Delta M$ the $M_{T_2}$ distribution is not very wide and has an end-point 
at a much lower value. Thus a strong cut on this variable is not favorable when the 
sneutrino LSP mass lies close to the electroweakino's mass.

Following the features of the kinematic distributions, we implement the following optimal 
selection criteria as follows for both signal and backgrounds:
\begin{itemize}
 \item \textbf{D1}: The final state consists of two opposite sign leptons and no photons. 
 \item \textbf{D2}: The leading lepton has $p_T> 20$ GeV and the sub-leading lepton has 
 $p_T> 10$ GeV. 
 \item \textbf{D3}: $M_{l^+l^-}>10 $ GeV helps remove contributions from photon mediated 
 processes while the $Z$ mass window is also removed by demanding that the opposite-sign 
 same flavor di-lepton invariant mass satisfies  $76<M_{l^+l^-}<106$ GeV. This helps to 
 reduce a large resonant contribution form the $Z$ exchange in Drell-Yan process.  
   \item \textbf{D4}: We reject any $b$-jet by putting a $b$-jet veto (for $p_T>40$ GeV). 
   This helps in suppressing background events coming from top quark production. 
 \item \textbf{D5}: We demand a completely hadronically quiet event by choosing zero jet 
 multiplicity ($N_{jet}=0$) in the signal events. This is effective in suppressing  
 contributions from background processes produced via strong interactions.
 \item \textbf{D6}: We demand $\slashed{E}_T > $ 80 GeV to suppress the large Drell-Yan 
 contribution. 
 \item \textbf{D7}: We demand $M_T{_{2}} > $ 90 GeV which helps reduce a majority of the 
 other SM backgrounds.
 \item \textbf{D8}: $\slashed{E}_T > $ 100 GeV is implemented to further reduce the SM 
 backgrounds.
\end{itemize}

 \begin{table}
\begin{center}
\begin{tabular}{|c|c|c|c|c|c|c|c|c|}
\hline
\multicolumn{1}{|c|}{Signal} &
\multicolumn{7}{|c|}{Number of events after cut } \\
\cline{2-8} 
\scriptsize
& Preselection (\bf{D1}) &  \bf{D2}  &   \bf{D3}  & \bf{D4}  &\bf{D5}  &  \bf{D7}  & 
 \bf{D8}    \\
 \hline
BP1 &130  &129  & 112&109 &68&22  &21 \\
BP2-a &306 & 271& 265&161 &108 &76 &72 \\
BP4 & 209&298 &246& 241& 153&81&40 \\

\hline \hline 
\multicolumn{1}{|c|}{Signal} &
\multicolumn{7}{|c|}{Number of events after cut } \\
\cline{2-8} 
\scriptsize
& Preselection (\bf{D1}) &  \bf{D2}  &   \bf{D3}  & \bf{D4}  &\bf{D5}  &  \bf{D6}  & \\
\hline
BP2-b & 455 & 452& 351 & 345 & 230 & 45 & \\
BP3 &2424&2394& 1840&1805&1186& 189 & \\
%BP4 & 654 & 654 & 538  &527  & 331 & 173 & 98\\
\hline
\end{tabular}
\end{center}
\caption{Opposite Sign di-lepton $+$ $\slashed{E}_T$ final state  number of events at 
100 fb$^{-1}$ for SUSY signals. Note that the events have been rounded-off to the nearest 
integer. Cross-sections have been scaled using NLO K-factors obtained from 
\texttt{Prospino}.}
  \label{dilepOSS}
\end{table}
In Table~\ref{dilepOSS} we show the signal events that survive the above listed kinematic 
selections (cut-flow).  We find that among all benchmarks, \textbf{BP2-a} is the most robust followed by \textbf{BP4}.  
Note that we avoid using the $M_{T2}$ cut on the benchmarks where the mass splitting between 
the chargino and the sneutrino LSP is small as {\bf D7} cut makes the signal events 
negligible. As pointed out earlier, the end-point analysis in $M_{T2}$ is not favorable 
for small $\Delta M$ as seen in the signal and background distributions in 
fig. \ref{KE1dilep}. Thus \textbf{BP2-b} and \textbf{BP3} have cuts \textbf{D1}-\textbf{D6}.  
In Table \ref{dilepOSB} we plot the SM background events after each kinematic cuts. 
Quite clearly up to cut \textbf{D6} the SM background numbers are quite large, and
then drastically reduce after the $M_{T2}$ cut (\textbf{D7}) is imposed.

In Table \ref{dilepOSSig} we give the required integrated luminosities to achieve a 
$3 \sigma$ and $5 \sigma$ statistical significance for the signal events of the 
benchmark points. Just like for the mono-lepton case, \textbf{BP2-a} requires the least 
integrated luminosity and is in fact gives a $3\sigma$ significance for much lower 
luminosity compared to mono-lepton signal.  However for the rest of the benchmarks 
mono-lepton channel is more favorable while the opposite-sign di-lepton can act as a 
complementary channel for \textbf{BP1} with higher luminosity and \textbf{BP3} with the 
very-high luminosity option of LHC.  \textbf{BP2-b} type of spectrum for the model 
is strongly suppressed in the di-lepton channel. The signal rates can be attributed 
to the fact that the leptonic branching of the chargino is much larger for 
\textbf{BP2-a}  ($\sim 100\%$) than \textbf{BP1} ($\sim 12\%$) and hence the signal 
is much more suppressed for \textbf{BP1} than in \textbf{BP2-a}. For \textbf{BP4} where the NLSP is the chargino, the opposite sign dilepton signal is a robust channel for discovery.

\begin{table}[h]
\small
\begin{tabular}{|c|c|c|c|c|c|c|c|c|}
\hline
\multicolumn{1}{|c|}{Signal} &
\multicolumn{8}{|c|}{Number of events after cut } \\
\cline{2-9} 
\scriptsize
& Preselection (\bf{D1}) &  \bf{D2}  &   \bf{D3}  & \bf{D4}  &\bf{D5}  &  \bf{D6}  & 
 \bf{D7}  & \bf{D8}  \\
\hline
Drell\text{ }Yan&1.16$\times10^8$&1.14$\times10^8$ &8.89$\times10^6$&8.80$\times10^6$&6.89$\times10^6$&506 & 293&53 \\
$W^+W^-$ &1.44$\times10^5$&1.43$\times10^5$&1.13$\times10^5$ &1.13$\times10^5$&1.00$\times10^5$& 5813& 24& 12 \\
$ZZ$ &1.71$\times10^4$ & 1.71$\times10^4$& 656& 651&504 & 117& 50&45  \\
$WZ$ &6.0$\times10^4$&6.0$\times10^4$& 5399  &5208 &1554& 92& 6 &4 \\
$t \bar{t} $ & 6.19$\times10^5$&6.16$\times10^5$&4.96$\times10^5$ &1.48$\times10^{5}$  &3.5$\times10^4$ &2.63$\times10^4$ &132&106 \\
$t W$  &1.77$\times10^5$ &1.74$\times 10^5$ &1.40$\times10^5$&6.76$\times10^4$&2.99$\times10^4$& 8181& 114&51\\
\hline\hline
Total & \multicolumn{5}{|c|}{}&41009& &271 \\
\hline
\end{tabular}
\caption{Opposite Sign di-lepton $+$ $\slashed{E}_T$ final state  number of events at 
100 fb$^{-1}$  for Standard Model backgrounds. Note that the events have been rounded-off 
to the nearest integer. Cross-sections scaled with K-factors at NLO \cite{Alwall:2014hca} 
and wherever available, NNLO \cite{Grazzini:2016swo,Cascioli:2014yka,Giammanco:2017xyn,
Czakon:2013goa,Czakon:2011xx,Kidonakis:2016sjf} have been used.}
\label{dilepOSB}
\end{table}

Thus this channel is not a likely probe for benchmarks with a smaller phase space, like 
\textbf{BP2-b} and \textbf{BP3} in which cases, as seen in the previous section, mono-lepton 
signals fare better over di-lepton signals. Whereas for spectra like $\bf{BP2-a}$ and \textbf{BP4}, with a large 
phase space available, opposite sign di-lepton signals are much more sensitive than 
mono-lepton signals. In contrast spectra like $\bf{BP1}$ with a lower leptonic 
branching of the chargino, mono-lepton $+$ missing energy signal is still a better 
channel to look for than opposite sign di-lepton channel.
\begin{table}[H]
\begin{center}
 \begin{tabular}{|c|c|c|}
 \hline
  Signal & $\mathcal{L}_{3\sigma}$ (fb$^{-1}$)&$\mathcal{L}_{5\sigma}$(fb$^{-1}$)\\
  \hline
  BP1 & 568 & 1576\\
  BP2-a & 51 & 142\\
  BP2-b & $ 1.83\times10^4$ & 5.07$\times10^4$ \\
  BP3 & 1035  & 2875 \\
  %BP4 & 28 & 79 \\
  BP4 & 160& 444 \\
  \hline
 \end{tabular}
\caption{Required luminosities for discovery of opposite sign di-lepton $+$ $\slashed{E}_T$  
final states at $\sqrt{s} =13$ TeV LHC.}
\label{dilepOSSig}
\end{center}
\end{table}

%%%%%%%%%%%%%%%%%
\subsubsection{Same Sign di-leptons  + $0$ jet + $\slashed{E}_T$ signal }
A more interesting and unique new physics signal at LHC in the di-lepton channel is the 
same-sign di-lepton mode. The same-sign di-lepton in the absence of missing transverse 
energy is a clear signal for lepton number violation and forms the backbone for most studies 
of models with heavy right-handed Majorana neutrinos.  Even with missing energy, the 
same-sign di-lepton is a difficult final state to find within the SM and therefore a 
signal with very little SM background. Thus finding signal events in this channel would give 
very clear hints of physics beyond the SM.  
 
In our framework of SUSY model the same-sign di-lepton signal with missing energy and 
few jets come from the production modes $\widetilde{\chi}^{\pm}_1 \widetilde{\chi}^{0}_2$ 
and$/$or $\widetilde{\chi}^{\pm}_1 \widetilde{\chi}^{0}_1$ where the lepton number violating 
contribution comes from the decay of the Majorana-like neutralinos given by 
$\widetilde{\chi}^{0}_2 \rightarrow W^{\mp*} \widetilde{\chi}^{\pm}_1$ with $\widetilde{\chi}^{\pm}_1 \rightarrow l \widetilde{\nu}$. We note that same-sign di-lepton backgrounds are 
rare in SM, with some small contributions coming from processes such as $p\,p \to WZ, ZZ, 
W^+W^+/W^-W^- +$jets,  $t\bar{t}W$ and  $t\bar{t}Z$ as well as from triple gauge boson 
productions such as $WWW $ where with two of the $W$ bosons being of same sign and the 
other decaying hadronically. Other indirect backgrounds can arise from energy 
mismeasurements, i.e, when jets or photons or opposite sign leptons fake a same sign 
di-lepton signal.\footnote{There may be additional contributions for same-sign di-lepton 
coming from non-prompt and conversions which we have not considered \cite{CMS-PAS-SUS-16-039}.}

  \begin{table}
\begin{center}
\begin{tabular}{|c|c|c|c|c|c|}
\hline
%\scriptsize
\multicolumn{1}{|c|}{Signal} &
\multicolumn{5}{|c|}{Number of events after cut: }\\
\cline{2-6} 
 & Preselection (\bf{S1}) &  \bf{S2}  &   \bf{S3}  & \bf{S4}  &\bf{S5} \\
 \hline
BP1 &5 & 5 & 5 & 4 &  2 \\
BP2-a &  3 & 2.5& 2.3 & 1.6& 1  \\ 
BP2-b & 32 & 23 & 22 & 15 &  8 \\ 
BP3 & 2 & 0.5 & 0.5 & 0.2 &  0.1 \\
BP4 & 30 &23 & 23 & 13& 7 \\ 
 \hline
\end{tabular}
 \caption{Same sign di-lepton $+$ $\slashed{E}_T$ final state number of events at 
 100 fb$^{-1}$ for SUSY  signals. Note that the events have been rounded-off to the 
 nearest integer where relevant. Cross-sections have been scaled using NLO K-factors 
 obtained from \texttt{Prospino}. }
 \label{dilepSSS}
\end{center}
\end{table} 

For our analysis we select the same-sign di-lepton events using optimal cuts for 
both signal and background using the following kinematic criteria:
\begin{itemize}
 \item \textbf{S1}: The final state consists of two charged leptons with same-sign 
and the leading lepton in $p_T$ must satisfy $p_T> 20$ GeV with the sub-leading lepton 
having $p_T>15$ GeV. Additionally we ensure that there are no isolated photon and $b$-jets 
in the final state. 
 \item \textbf{S2}: A minimal cut on the transverse mass constructed with the leading 
 charged lepton ($l_1$),  $M_T(l_1,\vec{\slashed{E}_T})> 100$ GeV is chosen to reject 
 background contributions coming from $W$ boson.
 \item \textbf{S3}: To suppress background from $W^{\pm}W^{\pm}$jj as well as those from 
 $t\bar{t}W$, $t\bar{t}Z$ with higher jet multiplicities than the SUSY signal, we keep 
 events with only up to 2 jets.
 \item \textbf{S4}: A large missing energy cut, $\slashed{E}_T > $ 100 GeV is implemented 
 to reduce SM backgrounds.
 \item \textbf{S5}: Finally we choose the events to be completely hadronically quiet 
 and demand zero jets in the event.
\end{itemize}

 \begin{table}
 \small
\begin{center} 
\begin{tabular}{|c|c|c|c|c|c|c|}
\hline
%\scriptsize

\multicolumn{1}{|c|}{SM Backgrounds} &  
\multicolumn{5}{|c|}{Number of events after cut }\\
\cline{2-6} 
 & Preselection (\bf{S1}) &  \bf{S2}  &   \bf{S3}  & \bf{S4}  &\bf{S5}  \\
\hline
$WZ$  & 3856 &1053 &930 &194 & 39\\
$ZZ$  &94  &6 &5 & 0.5& 0.2\\
$WWW$& 60 & 29&21&7 & 0.6\\
$W^+ W^+ j j $ &416 &175&116&56&2\\
$W^- W^- j j $  &188 &82&57 &18 &0.5\\
$tW$  &40 &20&19&7& 4\\
$t \bar{t} W$  &128&60&30&13& 1\\
$t \bar{t}$  &90 &65& 50&28&8 \\
\hline
Total background &\multicolumn{4}{|c|}{}&55\\

\hline
\end{tabular}
 \caption{Same sign di-lepton $+$ $\slashed{E}_T$ final state  number of events at 100 fb$^{-1}$ for SM 
 background. Note that the events have been rounded-off to the nearest integer where relevant.
Cross-sections scaled with K-factors at NLO \cite{Alwall:2014hca} and wherever available, 
 NNLO \cite{Grazzini:2016swo,Cascioli:2014yka,Giammanco:2017xyn,Czakon:2013goa,Czakon:2011xx,Kidonakis:2016sjf} have been used.}
   \label{dilepSSB}
 \end{center}
\end{table}

In Tables \ref{dilepSSS} and \ref{dilepSSB} we show the signal and backgrounds events after 
each selection cuts are imposed. As the same-sign signal is strongly constrained by existing 
LHC data, our benchmarks have been chosen to comply with the existing limits. Thus we find that 
our benchmark choices do not seem too robust in terms of signal rates, especially 
\textbf{BP3} and \textbf{BP4} which has the chargino as the NLSP. It is therefore important to point out that 
\textbf{BP1} and  \textbf{BP2-a} like spectra is naturally not favored to give a same-sign di-lepton signal while \textbf{BP3} and \textbf{BP4}
are the most probable to give the same-sign signal but have been chosen to suppress the signal to respect 
existing constraints (by choosing very small branching for the neutralinos to decay to 
chargino) for two different $\mu$ values. However the spectra as reflected by \textbf{BP2-b} and \textbf{BP4} satisfying 
existing constraints do present us with a significant number of event rates when compared to the background
after cuts. 

From the above cuts, we find that a large $M_T(l_1)$ cut coupled with a large $\slashed{E}_T$ and 
the requirement of jet veto removes a large fraction of the dominant $WZ$ background as well as other 
fake contributions coming from $t \bar{t}$.  
Other genuine contributions to this channel from $W^{\pm}W^{\pm}\, jj$, $t \bar{t} W$ and 
$WWW$  having a lower production cross-section and are efficiently suppressed  by cuts on 
$\slashed{E}_T$, $M_T$ and applying a jet veto. 
Amongst all benchmarks, the most sensitive to the same sign di-lepton analysis are 
\textbf{BP2-b} where BR($\widetilde{\chi}^{0}_2 \rightarrow \widetilde{\chi}^{\pm}_{1} 
W^{\mp} \rightarrow l \widetilde{\nu})\sim 3.3\%$ and \textbf{BP4} with BR($\widetilde{\chi}^{0}_2 \rightarrow \widetilde{\chi}^{\pm}_{1} 
W^{\mp} \rightarrow l \widetilde{\nu})\sim10\%$.  Note that \textbf{BP2-b}, with a smaller 
$\Delta M$ gives soft leptons and is therefore slightly suppressed and
requires larger integrated luminosity $\sim 810 $ fb$^{-1}$ of data.
Although \textbf{BP4} has a larger branching fraction, it requires $1052 $ fb$^{-1}$ of data at LHC for observing a 
$3 \sigma$ excess owing to a higher $\mu$ value compared to \textbf{BP2-b}. 
Thus the same-sign di-lepton can be a complementary channel to observe 
for benchmarks of \textbf{BP2-b} and \textbf{BP4}. 
\begin{table}[H]
\begin{center}
 \begin{tabular}{|c|c|c|}
 \hline
  Signal & $\mathcal{L}_{3\sigma}$ (fb$^{-1}$)&$\mathcal{L}_{5\sigma}$(fb$^{-1}$)\\
  \hline
 BP2-b & 811 & 2251 \\
 BP4 & 1052 & 3845 \\
 \hline
 \end{tabular}
\caption{Required luminosities for discovery of same sign di-lepton final states 
with missing energy at $\sqrt{s}=13$ TeV LHC.}
\label{dilepSSSig}
\end{center}
\end{table}
We must again point out here that for \textbf{BP3}-like spectra with
$\widetilde{\chi}^{\pm}_1$ NLSP the same-sign di-lepton would be the most sensitive 
channel of discovery, for large $|\mu|$ and small $M_2$, where the neutralino decay 
to chargino NLSP becomes large (see figs. \ref{fig:brpmu} and \ref{fig:brnmu}) because 
of the small SM background. In such a case both $\widetilde{\chi}^{0}_1$ and 
$\widetilde{\chi}^{0}_2$ will decay to the NLSP along with soft jets or leptons. 
Thus both $\widetilde{\chi}^{\pm}_1  \widetilde{\chi}^{0}_2$ and $\widetilde{\chi}^{\pm}_1  \widetilde{\chi}^{0}_1$ production channels would have contributed to the signal 
leading to a two-fold increase of the number of signal events and would be more 
sensitive to detect a sneutrino LSP scenario. 
 
We conclude that conventional channels such as mono-lepton or opposite sign di-lepton 
channels however with low hadronic activity, i.e, with at most 1 jet or no jet would 
be extremely useful channels to look for cases of a sneutrino LSP.  Detecting  same 
sign di-lepton signals at higher luminosities would further serve as a 
strong confirmatory channel for a sneutrino LSP scenario over a $\widetilde{\chi}^{0}_1$ 
LSP scenario as in the MSSM from the compressed higgsino sector and can exclude large 
portions of the regions with $M_1<0$. Our analyses also shows better signal 
significance for a given integrated luminosity, when compared to the forecast shown 
in table~\ref{LHCforecast}. Note that our estimates do not include any systematic 
uncertainties that may be present and would be dependent on the specific analysis of event 
topologies. However it is worthwhile to ascertain how our results fare in presence of such 
systematic uncertainties. To highlight this we assume a conservative 10$\%$ systematic uncertainty 
in  each case. We find that the required integrated luminosities follow a similar scaling 
and our results for the luminosity vary by atmost 10$\%$ in most cases.

\subsubsection*{Dependence on flavor of $\widetilde{\nu}_R$ LSP}
LHC searches explore different search channels involving the flavor of the leptons 
owing to their high reconstruction efficiency at the detector,
for instance, e+$\slashed{E}_T$, $\mu+\slashed{E}_T$ \cite{Aaboud:2017efa,Khachatryan:2016jww}, 
$ee /\mu\mu/ e\mu$ final states associated with $\slashed{E}_T$ \cite{CMS-PAS-SUS-16-048,
ATLAS-CONF-2017-039}. As we have considered both first and second generation sneutrinos 
to be light in this study, we qualitatively analyze the prospects of the signals studied by 
tagging the flavor of the leptons as well as consequences of a single light generation of 
sneutrino LSP assuming the net leptonic branching to be the same in both cases.\footnote{This
may not correspond to the same parameter point since the presence of the other decay modes of 
$\widetilde{\chi}^{\pm}_1$ affect the leptonic branching for the single light sneutrino 
LSP case. However, when $\widetilde{\chi}^{\pm}_1$ is the NLSP, the net leptonic branching 
is the same in both cases.}  
Hence, for a single light sneutrino LSP, the observed events in the mono-lepton and 
di-lepton signals contribute to only a single choice of lepton flavor and vanishes for 
the rest. We compare the signal and background in this case for the same luminosity as 
before and comment on the results obtained for our benchmarks. 

For \textit{mono-lepton} signals with degenerate sneutrino LSP (first two generations), 
say, we look at only an electron in the final state. This would lead to reduction of 
both signal and background in Table~\ref{signalmono1} and \ref{bkgmono1} by half such 
that the significance falls by a factor of $\sqrt{2}$. If a single generation of 
right-sneutrino was light, say $\widetilde{\nu}_e$, then only the background would reduce 
by a factor 1/2. Since the signal remains unchanged as the chargino now decays completely 
to an electron and the lightest sneutrino the signal significance increases by a factor 
of $\sqrt{2}$. Consequently no signal is observed for the other flavor lepton channel, 
in this case $\mu$, where although the background decreases by half, no signal events 
are present. 

For the di-lepton signal there are three possible channels $ee, \text{ } \mu\mu $ and 
$e \mu$ with net branching fraction of around 1/4, 1/4 and 1/2 respectively. We consider 
first the opposite-sign di-lepton channel. For $e\mu$ final states, only different flavor 
lepton backgrounds such as from $WW,  \,\, t \bar{t}$ or $tW$ contribute with a BR 
$\simeq 2/3$. However contributions from same flavor di-lepton sources such as involving 
$Z$ boson fall. The total background  thus reduces to nearly 70$\%$. Since signal in 
this channel also reduces to half thereby the significance falls. For channels with 
same flavor (SF) leptons, i.e, $ee/\mu\mu$, dominant SF contributions are from $Z$ 
boson whereas sub-dominant contributions from top quark production channel reduce. 
Although SM background reduces so does the signal statistics and hence the significance.
However, in presence of a single generation of light sneutrino we find that the signal 
significance improves by a factor of about $\sqrt{2}$. Note that if the LSP is 
$\widetilde{\nu}_e$ then the chargino  decays to an electron and the LSP. Therefore 
$ee + \slashed{E}_T$ channel significance improves whereas $\mu\mu$ and $e\mu$ channels 
vanish. Similarly, for an $\widetilde{\nu}_{\mu}$ LSP, $\mu\mu+ \slashed{E}_T$ channels 
improve whereas the rest vanish. Similar conclusions may be drawn for \textit{same sign 
di-lepton} channel, where the dominant backgrounds are $WZ$ and $W^{\pm}W^{\pm}$, the 
significance is expected to improve only for a single light generation of sneutrinos. 

\subsubsection*{Some comments on the prospect of $\tau$ flavor searches and other channels}
In this context, we also explore the discovery prospects of a natural higgsino sector and 
a single light $\widetilde{\nu}_{\tau}$ as the LSP. LHC has looked at final states with 
tau leptons, decaying hadronically, in the context of electroweakino searches. 
The electroweakino mass limits considerably reduce for tau lepton searches owing to the 
reduced reconstruction efficiency of hadronically decaying $\tau$ leptons
($\sim$ $60\%$) \cite{CMS-PAS-TAU-16-002} compared to that of the light leptons 
(e,$\mu$) ($\sim 95\%$). From searches with one or two hadronically decaying tau leptons 
associated with light leptons lead to stronger limits from $\widetilde{\chi}^{\pm}_1$,$
\widetilde{\chi}^{0}_2$ production on $m_{\widetilde{\chi}^{\pm}_1},m_{\widetilde{\chi}^{0}_2} 
> 800$ GeV for a bino-like $m_{\widetilde{\chi}^{0}_1} < 200 $ GeV for stau mass midway between the $\widetilde{\chi}^{\pm}_1$ and
$\widetilde{\chi}^{0}_1$. For stau closer to the $\widetilde{\chi}^{\pm}_1$, the limit \cite{CMS-PAS-SUS-16-039}, $m_{\widetilde{\chi}^{\pm}_1},m_{\widetilde{\chi}^{0}_2} \sim 1000$ GeV for $m_{\widetilde{\chi}^{0}_1} \sim 200$ GeV. Limits on electroweakino 
searches from three tau lepton searches exclude wino-like degenerate $m_{\widetilde{\chi}^{0}_2},
m_{\widetilde{\chi}^{\pm}_1} > 600$ GeV for a bino-like $m_{\widetilde{\chi}^{0}_1} < 200 $ 
GeV \cite{CMS-PAS-SUS-16-039} for stau mass midway between the $\widetilde{\chi}^{\pm}_1$ and
$\widetilde{\chi}^{0}_1$. Limits from opposite sign tau lepton searches 
\cite{Aaboud:2017nhr} reinterpreted from $\widetilde{\chi}^{\pm}_1\widetilde{\chi}^{0}_2 $ 
production and decaying via intermediate sleptons lead to $m_{\widetilde{\chi}^{0}_2},m_{\widetilde{\chi}^{\pm}_1}> 760 $ GeV for $m_{\widetilde{\chi}^{0}_1}< 200$ GeV.
From opposite sign di-tau searches reinterpreted in context of chargino pair production 
leads to a bound close to 650 GeV on chargino for LSP masses up to 
100 GeV \cite{Aaboud:2017nhr} .

For the current scenario of a compressed electroweakino sector in presence of a light 
$\widetilde{\nu}_{\tau}$ LSP,
the signals from the low-lying compressed higgsino sector would be:
\begin{itemize}
 \item Mono-$\tau$ jet + $\slashed{E}_T$
 \item Di $\tau$ jets  + $\slashed{E}_T$
% \item Same-sign $\tau$ leptons + $\slashed{E}_T$
 \end{itemize}

For the mono-tau channel, both signal and background scale by the tau reconstruction 
efficiency, $\epsilon_R = 0.6$ is the tau reconstruction efficiency. Further a factor of 
$\frac{1}{2}$ comes in for the background since the branching of $W$ or $Z$ boson to light 
leptons is roughly twice that to the tau lepton as for a $\widetilde{\nu}_{e/\mu}$ LSP. 
However, owing to the reduced tau reconstruction efficiency, the signal significance 
falls by $\sim \sqrt{\epsilon_R} \sim 0.78$. Similarly, for the di-tau channels, the 
significance scales by $\epsilon_R \sim 0.6$.  Hence, the estimated reach of the higgsino mass 
parameter, $\mu$ is expected to weaken for a $\widetilde{\nu}_{\tau}$ LSP compared to 
$\widetilde{\nu}_{e/\mu}$ LSP. 

Note that, the pionic decay modes of the $\tilde{\chi}_1^{\pm}$ and $\tilde{\chi}_2^0$ 
can dominate among the hadronic modes as the respective mass differences become less 
than about a GeV.  While we have used form factors to estimate the pionic branching 
fractions, we have not considered the possibility of late decay into pions in this work. 
This is because we have ensured that in the parameter space of our interest the two body 
mode to the lightest sneutrino(s) always remain prompt. Further, the potential of the 
loop-induced channel $\tilde{\chi}_2^0 \rightarrow \tilde{\chi}_1^0 \gamma$ in deciphering 
the scenario has not been explored in the present work. While the photons, thus produced 
in the cascade, would be soft in the rest frame of $\tilde{\chi}_2^0$, it may be possible 
to tag hard photons in the lab frame. Note that the choice of light higgsinos are 
motivated by ``naturalness" at the electroweak scale and we do not discuss the discovery 
potential for stop squarks and gluino in the present work which we plan to do in 
a subsequent extension. 

\section{Conclusion}
\label{sec:conclusion}
To summarize, motivated by ``naturalness" criteria at the electroweak scale, we 
have studied a simplified scenario with low $\mu$ parameter in the presence of a 
right-sneutrino LSP. For simplicity, we have assumed the gaugino mass parameters 
to be quite heavy $\gtrsim 1$ TeV. In such a scenario, with $\mathcal{O}(100)$ 
GeV Majorana mass parameter the neutrino Yukawa coupling can be as large as $ 
10^{-6} - 10^{-7}$. In contrast with the MSSM with light-higgsinos, in the present 
context, the higgsino-like states can decay to the sneutrino LSP. While the 
neutral higgsinos can decay into neutrino and sneutrino, the lightest chargino
can decay into a lepton and sneutrino. We have demonstrated that the latter
decay channel can lead to various leptonic final states  with up to two leptons 
(i.e. mono-lepton, same-sign di-lepton and opposite-sign di-lepton) and missing 
transverse energy  at the LHC, which can be important in searching for or 
constraining this scenario. We have only considered prompt decay into leptons, 
which require $y_{\nu} > 10^{-7}$ and/or small $\mathcal{O}(10^{-5}-10^{-1})$ left-right 
mixing in the sneutrino sector. For smaller values of $y_{\nu}$, contribution from 
the latter dominates and the leptonic  partial width on small gaugino-higgsino 
mixing ($\lesssim \mathcal{O}(10^{-2})$). Further, the mass split between the 
three states, the lightest chargino and the two lightest neutralinos depend 
on the choice of the gaugino mass parameters, as well as on one-loop 
contributions. We have shown how these mass differences significantly affect the 
three-body partial widths, thus affecting the branching ratios to the sneutrino. 
Therefore, even assuming the gaugino-like states to be above a TeV, as in our 
benchmark scenarios, the viability of a low $\mu$ parameter depends crucially on 
the choice of $M_1, \, M_2$. This has been emphasized in great detail. 
{Consequently, there are regions of the parameter space where 
$BR(\widetilde{\chi}^{\pm}_1 \rightarrow l \, \widetilde{\nu}) \sim 100\%$
especially in the negative $M_1$ parameter space. Such regions of parameter
space would lead to enhanced leptonic rates, thereby a large fraction of 
negative $M_1$ parameter space can be excluded from current leptonic searches 
at LHC. For a given $|\mu|$, we check the existing constraints by recasting our 
signal in \texttt{CheckMATE} against existing LHC analysis relevant for our model 
parameters to search for a viable parameter region of the model. We then choose 
some representative benchmarks and observe that mono-lepton signals with large 
$\slashed{E}_T$ and little hadronic activity could successfully probe $\mu$ as 
low as 300 GeV at the ongoing run of LHC with 106 fb$^{-1}$ of data at 3$\sigma$. 
Additional confirmatory channels for the $\widetilde{\nu}$ LSP scenario are 
opposite-sign di-lepton and same-sign di-lepton signal which require $\sim 50$ fb$^{-1}$ 
and $\sim800$ fb$^{-1}$  for observing 3$\sigma$ excess at LHC. While our benchmarks assume the 
first two generations of sneutrinos to be degenerate and consider only $e,\mu$ for 
the charged leptons which can be detected efficiently at the LHC, the reach may be 
substantially reduced if only tau-sneutrino appears as the lightest flavor due to 
the low tau reconstruction efficiency.
\section*{Acknowledgement}
AC acknowledges financial support from the Department of Science and Technology,
Government of India through the INSPIRE Faculty Award: /2016/DST/INSPIRE/04 /2015/000110.
The work of JD and SKR is partially supported by funding available from the Department 
of Atomic Energy, Government of India, for the Regional Centre for Accelerator-based 
Particle Physics (RECAPP), Harish-Chandra Research Institute. The authors would like to thank 
W. Porod for his help in providing several clarifications on our model implementation in {\tt SPheno} 
and S. Choubey, S. Mondal and J. Beuria for useful discussions. Computational work for 
this work was carried out at the cluster computing facility (http://www.hri.res.in/cluster)
and the RECAPP cluster in the Harish-Chandra Research Institute.

\bibliographystyle{JHEP}
\bibliography{nRNSUSY_ac1}

\providecommand{\href}[2]{#2}\begingroup\raggedright\begin{thebibliography}{100}

\bibitem{Barbieri:1987fn}
R.~Barbieri and G.~F. Giudice, {\it {Upper Bounds on Supersymmetric Particle
  Masses}},  {\em Nucl. Phys.} {\bf B306} (1988) 63--76.

\bibitem{Ellis:1986yg}
J.~R. Ellis, K.~Enqvist, D.~V. Nanopoulos, and F.~Zwirner, {\it {Observables in
  Low-Energy Superstring Models}},  {\em Mod. Phys. Lett.} {\bf A1} (1986) 57.

\bibitem{Feng:2013pwa}
J.~L. Feng, {\it {Naturalness and the Status of Supersymmetry}},  {\em Ann.
  Rev. Nucl. Part. Sci.} {\bf 63} (2013) 351--382,
  [\href{http://arxiv.org//abs/1302.6587}{{\tt arXiv:1302.6587}}].

\bibitem{Giudice:2013nak}
G.~F. Giudice, {\it {Naturalness after LHC8}},  {\em PoS} {\bf EPS-HEP2013}
  (2013) 163, [\href{http://arxiv.org//abs/1307.7879}{{\tt arXiv:1307.7879}}].

\bibitem{Baer:2012cf}
H.~Baer, V.~Barger, P.~Huang, D.~Mickelson, A.~Mustafayev, and X.~Tata, {\it
  {Radiative natural supersymmetry: Reconciling electroweak fine-tuning and the
  Higgs boson mass}},  {\em Phys. Rev.} {\bf D87} (2013), no.~11 115028,
  [\href{http://arxiv.org//abs/1212.2655}{{\tt arXiv:1212.2655}}].

\bibitem{Mustafayev:2014lqa}
A.~Mustafayev and X.~Tata, {\it {Supersymmetry, Naturalness, and Light
  Higgsinos}},  {\em Indian J. Phys.} {\bf 88} (2014) 991--1004,
  [\href{http://arxiv.org//abs/1404.1386}{{\tt arXiv:1404.1386}}].

\bibitem{Baer:2012up}
H.~Baer, V.~Barger, P.~Huang, A.~Mustafayev, and X.~Tata, {\it {Radiative
  natural SUSY with a 125 GeV Higgs boson}},  {\em Phys. Rev. Lett.} {\bf 109}
  (2012) 161802, [\href{http://arxiv.org//abs/1207.3343}{{\tt
  arXiv:1207.3343}}].

\bibitem{Baer:2013ava}
H.~Baer, V.~Barger, P.~Huang, D.~Mickelson, A.~Mustafayev, and X.~Tata, {\it
  {Naturalness, Supersymmetry and Light Higgsinos: A Snowmass Whitepaper}},  in
  {\em {Proceedings, Community Summer Study 2013: Snowmass on the Mississippi
  (CSS2013): Minneapolis, MN, USA, July 29-August 6, 2013}}, 2013.
\newblock \href{http://arxiv.org//abs/1306.2926}{{\tt arXiv:1306.2926}}.

\bibitem{Baer:2015rja}
H.~Baer, V.~Barger, and M.~Savoy, {\it {Upper bounds on sparticle masses from
  naturalness or how to disprove weak scale supersymmetry}},  {\em Phys. Rev.}
  {\bf D93} (2016), no.~3 035016, [\href{http://arxiv.org//abs/1509.02929}{{\tt
  arXiv:1509.02929}}].

\bibitem{Baer2017}
H.~Baer, V.~Barger, J.~S. Gainer, P.~Huang, M.~Savoy, D.~Sengupta, and X.~Tata,
  {\it Gluino reach and mass extraction at the lhc in radiatively-driven
  natural susy},  {\em The European Physical Journal C} {\bf 77} (Jul, 2017)
  499.

\bibitem{MoriondATLAS}
{\bf ATLAS} Collaboration,
  \url{https://atlas.web.cern.ch/Atlas/GROUPS/PHYSICS/CombinedSummaryPlots/SUSY/}.

\bibitem{MoriondCMS}
{\bf CMS} Collaboration,
  \url{https://twiki.cern.ch/twiki/bin/view/CMSPublic/PhysicsResultsSUS}.

\bibitem{Choudhury:2012tc}
A.~Choudhury and A.~Datta, {\it {Many faces of low mass neutralino dark matter
  in the unconstrained MSSM, LHC data and new signals}},  {\em JHEP} {\bf 06}
  (2012) 006, [\href{http://arxiv.org//abs/1203.4106}{{\tt arXiv:1203.4106}}].

\bibitem{Belanger:2013pna}
G.~Bélanger, G.~Drieu La~Rochelle, B.~Dumont, R.~M. Godbole, S.~Kraml, and
  S.~Kulkarni, {\it {LHC constraints on light neutralino dark matter in the
  MSSM}},  {\em Phys. Lett.} {\bf B726} (2013) 773--780,
  [\href{http://arxiv.org//abs/1308.3735}{{\tt arXiv:1308.3735}}].

\bibitem{Choudhury:2013jpa}
A.~Choudhury and A.~Datta, {\it {Neutralino dark matter confronted by the LHC
  constraints on Electroweak SUSY signals}},  {\em JHEP} {\bf 09} (2013) 119,
  [\href{http://arxiv.org//abs/1305.0928}{{\tt arXiv:1305.0928}}].

\bibitem{Boehm:2013gst}
C.~Boehm, P.~S.~B. Dev, A.~Mazumdar, and E.~Pukartas, {\it {Naturalness of
  Light Neutralino Dark Matter in pMSSM after LHC, XENON100 and Planck Data}},
  {\em JHEP} {\bf 06} (2013) 113, [\href{http://arxiv.org//abs/1303.5386}{{\tt
  arXiv:1303.5386}}].

\bibitem{Chakraborti:2014gea}
M.~Chakraborti, U.~Chattopadhyay, A.~Choudhury, A.~Datta, and S.~Poddar, {\it
  {The Electroweak Sector of the pMSSM in the Light of LHC - 8 TeV and Other
  Data}},  {\em JHEP} {\bf 07} (2014) 019,
  [\href{http://arxiv.org//abs/1404.4841}{{\tt arXiv:1404.4841}}].

\bibitem{diCortona:2014yua}
G.~Grilli~di Cortona, {\it {Hunting electroweakinos at future hadron colliders
  and direct detection experiments}},  {\em JHEP} {\bf 05} (2015) 035,
  [\href{http://arxiv.org//abs/1412.5952}{{\tt arXiv:1412.5952}}].

\bibitem{Drees:2015aeo}
M.~Drees and J.~S. Kim, {\it {Minimal natural supersymmetry after the LHC8}},
  {\em Phys. Rev.} {\bf D93} (2016), no.~9 095005,
  [\href{http://arxiv.org//abs/1511.04461}{{\tt arXiv:1511.04461}}].

\bibitem{Chakraborti:2015mra}
M.~Chakraborti, U.~Chattopadhyay, A.~Choudhury, A.~Datta, and S.~Poddar, {\it
  {Reduced LHC constraints for higgsino-like heavier electroweakinos}},  {\em
  JHEP} {\bf 11} (2015) 050, [\href{http://arxiv.org//abs/1507.01395}{{\tt
  arXiv:1507.01395}}].

\bibitem{Badziak:2015qca}
M.~Badziak, A.~Delgado, M.~Olechowski, S.~Pokorski, and K.~Sakurai, {\it
  {Detecting underabundant neutralinos}},  {\em JHEP} {\bf 11} (2015) 053,
  [\href{http://arxiv.org//abs/1506.07177}{{\tt arXiv:1506.07177}}].

\bibitem{Cao:2015efs}
J.~Cao, Y.~He, L.~Shang, W.~Su, and Y.~Zhang, {\it {Testing the light dark
  matter scenario of the MSSM at the LHC}},  {\em JHEP} {\bf 03} (2016) 207,
  [\href{http://arxiv.org//abs/1511.05386}{{\tt arXiv:1511.05386}}].

\bibitem{Beneke:2016jpw}
M.~Beneke, A.~Bharucha, A.~Hryczuk, S.~Recksiegel, and P.~Ruiz-Femenia, {\it
  {The last refuge of mixed wino-Higgsino dark matter}},
  \href{http://arxiv.org//abs/1611.00804}{{\tt arXiv:1611.00804}}.

\bibitem{Baer:2015tva}
H.~Baer, V.~Barger, P.~Huang, D.~Mickelson, M.~Padeffke-Kirkland, and X.~Tata,
  {\it {Natural SUSY with a bino- or wino-like LSP}},  {\em Phys. Rev.} {\bf
  D91} (2015), no.~7 075005, [\href{http://arxiv.org//abs/1501.06357}{{\tt
  arXiv:1501.06357}}].

\bibitem{Abdughani:2017dqs}
M.~Abdughani, L.~Wu, and J.~M. Yang, {\it {The mixed bino-higgsino dark matter
  in natural SUSY confronted with XENON1T/PandaX and LHC data}},
  \href{http://arxiv.org//abs/1705.09164}{{\tt arXiv:1705.09164}}.

\bibitem{Buckley2017}
M.~R. Buckley, D.~Feld, S.~Macaluso, A.~Monteux, and D.~Shih, {\it Cornering
  natural susy at lhc run ii and beyond},  {\em Journal of High Energy Physics}
  {\bf 2017} (Aug, 2017) 115.

\bibitem{Baer:2011ec}
H.~Baer, V.~Barger, and P.~Huang, {\it {Hidden SUSY at the LHC: the light
  higgsino-world scenario and the role of a lepton collider}},  {\em JHEP} {\bf
  11} (2011) 031, [\href{http://arxiv.org//abs/1107.5581}{{\tt
  arXiv:1107.5581}}].

\bibitem{Han:2013usa}
C.~Han, A.~Kobakhidze, N.~Liu, A.~Saavedra, L.~Wu, and J.~M. Yang, {\it
  {Probing Light Higgsinos in Natural SUSY from Monojet Signals at the LHC}},
  {\em JHEP} {\bf 02} (2014) 049, [\href{http://arxiv.org//abs/1310.4274}{{\tt
  arXiv:1310.4274}}].

\bibitem{Han:2014kaa}
Z.~Han, G.~D. Kribs, A.~Martin, and A.~Menon, {\it {Hunting quasidegenerate
  Higgsinos}},  {\em Phys. Rev.} {\bf D89} (2014), no.~7 075007,
  [\href{http://arxiv.org//abs/1401.1235}{{\tt arXiv:1401.1235}}].

\bibitem{Barducci:2015ffa}
D.~Barducci, A.~Belyaev, A.~Bharucha, W.~Porod, and V.~Sanz, {\it {Uncovering
  Natural Supersymmetry via the interplay between the LHC and Direct Dark
  Matter Detection}},  {\em JHEP} {\bf 07} (2015) 066,
  [\href{http://arxiv.org//abs/1504.02472}{{\tt arXiv:1504.02472}}].

\bibitem{Baer:2016usl}
H.~Baer, V.~Barger, M.~Savoy, and X.~Tata, {\it {Multichannel assault on
  natural supersymmetry at the high luminosity LHC}},  {\em Phys. Rev.} {\bf
  D94} (2016), no.~3 035025, [\href{http://arxiv.org//abs/1604.07438}{{\tt
  arXiv:1604.07438}}].

\bibitem{Fukuda:2017jmk}
H.~Fukuda, N.~Nagata, H.~Otono, and S.~Shirai, {\it {Higgsino Dark Matter or
  Not: Role of Disappearing Track Searches at the LHC and Future Colliders}},
  \href{http://arxiv.org//abs/1703.09675}{{\tt arXiv:1703.09675}}.

\bibitem{Mahbubani:2017gjh}
R.~Mahbubani, P.~Schwaller, and J.~Zurita, {\it {Closing the window for
  compressed Dark Sectors with disappearing charged tracks}},  {\em JHEP} {\bf
  06} (2017) 119, [\href{http://arxiv.org//abs/1703.05327}{{\tt
  arXiv:1703.05327}}]. [Erratum: JHEP10,061(2017)].

\bibitem{Ambrosanio:2000ik}
S.~Ambrosanio, B.~Mele, S.~Petrarca, G.~Polesello, and A.~Rimoldi, {\it
  {Measuring the SUSY breaking scale at the LHC in the slepton NLSP scenario of
  GMSB models}},  {\em JHEP} {\bf 01} (2001) 014,
  [\href{http://arxiv.org//abs/hep-ph/0010081}{{\tt hep-ph/0010081}}].

\bibitem{Ellis:2004bx}
J.~R. Ellis, K.~A. Olive, Y.~Santoso, and V.~C. Spanos, {\it {Prospects for
  sparticle discovery in variants of the MSSM}},  {\em Phys. Lett.} {\bf B603}
  (2004) 51, [\href{http://arxiv.org//abs/hep-ph/0408118}{{\tt
  hep-ph/0408118}}].

\bibitem{Roszkowski:2004jd}
L.~Roszkowski, R.~Ruiz~de Austri, and K.-Y. Choi, {\it {Gravitino dark matter
  in the CMSSM and implications for leptogenesis and the LHC}},  {\em JHEP}
  {\bf 08} (2005) 080, [\href{http://arxiv.org//abs/hep-ph/0408227}{{\tt
  hep-ph/0408227}}].

\bibitem{Ellis:2006vu}
J.~R. Ellis, A.~R. Raklev, and O.~K. Oye, {\it {Gravitino dark matter scenarios
  with massive metastable charged sparticles at the LHC}},  {\em JHEP} {\bf 10}
  (2006) 061, [\href{http://arxiv.org//abs/hep-ph/0607261}{{\tt
  hep-ph/0607261}}].

\bibitem{Kribs:2008hq}
G.~D. Kribs, A.~Martin, and T.~S. Roy, {\it {Supersymmetry with a Chargino NLSP
  and Gravitino LSP}},  {\em JHEP} {\bf 01} (2009) 023,
  [\href{http://arxiv.org//abs/0807.4936}{{\tt arXiv:0807.4936}}].

\bibitem{Bailly:2009pe}
S.~Bailly, K.-Y. Choi, K.~Jedamzik, and L.~Roszkowski, {\it {A Re-analysis of
  Gravitino Dark Matter in the Constrained MSSM}},  {\em JHEP} {\bf 05} (2009)
  103, [\href{http://arxiv.org//abs/0903.3974}{{\tt arXiv:0903.3974}}].

\bibitem{Feng:2010ij}
J.~L. Feng, M.~Kamionkowski, and S.~K. Lee, {\it {Light Gravitinos at Colliders
  and Implications for Cosmology}},  {\em Phys. Rev.} {\bf D82} (2010) 015012,
  [\href{http://arxiv.org//abs/1004.4213}{{\tt arXiv:1004.4213}}].

\bibitem{Figy:2010hu}
T.~Figy, K.~Rolbiecki, and Y.~Santoso, {\it {Tau-Sneutrino NLSP and Multilepton
  Signatures at the LHC}},  {\em Phys. Rev.} {\bf D82} (2010) 075016,
  [\href{http://arxiv.org//abs/1005.5136}{{\tt arXiv:1005.5136}}].

\bibitem{deAquino:2012ru}
P.~de~Aquino, F.~Maltoni, K.~Mawatari, and B.~Oexl, {\it {Light Gravitino
  Production in Association with Gluinos at the LHC}},  {\em JHEP} {\bf 10}
  (2012) 008, [\href{http://arxiv.org//abs/1206.7098}{{\tt arXiv:1206.7098}}].

\bibitem{Barnard:2012au}
J.~Barnard, B.~Farmer, T.~Gherghetta, and M.~White, {\it {Natural gauge
  mediation with a bino NLSP at the LHC}},  {\em Phys. Rev. Lett.} {\bf 109}
  (2012) 241801, [\href{http://arxiv.org//abs/1208.6062}{{\tt
  arXiv:1208.6062}}].

\bibitem{Bobrovskyi:2012dc}
S.~Bobrovskyi, J.~Hajer, and S.~Rydbeck, {\it {Long-lived higgsinos as probes
  of gravitino dark matter at the LHC}},  {\em JHEP} {\bf 02} (2013) 133,
  [\href{http://arxiv.org//abs/1211.5584}{{\tt arXiv:1211.5584}}].

\bibitem{Roszkowski:2012nq}
L.~Roszkowski, S.~Trojanowski, K.~Turzynski, and K.~Jedamzik, {\it {Gravitino
  dark matter with constraints from Higgs boson mass and sneutrino decays}},
  {\em JHEP} {\bf 03} (2013) 013, [\href{http://arxiv.org//abs/1212.5587}{{\tt
  arXiv:1212.5587}}].

\bibitem{Cyburt:2013fda}
R.~H. Cyburt, J.~Ellis, B.~D. Fields, F.~Luo, K.~A. Olive, and V.~C. Spanos,
  {\it {Gravitino Decays and the Cosmological Lithium Problem in Light of the
  LHC Higgs and Supersymmetry Searches}},  {\em JCAP} {\bf 1305} (2013) 014,
  [\href{http://arxiv.org//abs/1303.0574}{{\tt arXiv:1303.0574}}].

\bibitem{DHondt:2013cwd}
J.~D'Hondt, K.~De~Causmaecker, B.~Fuks, A.~Mariotti, K.~Mawatari, C.~Petersson,
  and D.~Redigolo, {\it {Multilepton signals of gauge mediated supersymmetry
  breaking at the LHC}},  {\em Phys. Lett.} {\bf B731} (2014) 7--12,
  [\href{http://arxiv.org//abs/1310.0018}{{\tt arXiv:1310.0018}}].

\bibitem{Heisig:2013sva}
J.~Heisig, {\it {Gravitino LSP and leptogenesis after the first LHC results}},
  {\em JCAP} {\bf 1404} (2014) 023,
  [\href{http://arxiv.org//abs/1310.6352}{{\tt arXiv:1310.6352}}].

\bibitem{Covi:2014fba}
L.~Covi and F.~Dradi, {\it {Long-Lived stop at the LHC with or without
  R-parity}},  {\em JCAP} {\bf 1410} (2014), no.~10 039,
  [\href{http://arxiv.org//abs/1403.4923}{{\tt arXiv:1403.4923}}].

\bibitem{Maltoni:2015twa}
F.~Maltoni, A.~Martini, K.~Mawatari, and B.~Oexl, {\it {Signals of a superlight
  gravitino at the LHC}},  {\em JHEP} {\bf 04} (2015) 021,
  [\href{http://arxiv.org//abs/1502.01637}{{\tt arXiv:1502.01637}}].

\bibitem{Arvey:2015nra}
A.~Arbey, M.~Battaglia, L.~Covi, J.~Hasenkamp, and F.~Mahmoudi, {\it {LHC
  constraints on Gravitino Dark Matter}},  {\em Phys. Rev.} {\bf D92} (2015),
  no.~11 115008, [\href{http://arxiv.org//abs/1505.04595}{{\tt
  arXiv:1505.04595}}].

\bibitem{Kim:2017pvm}
J.~S. Kim, M.~E. Krauss, and V.~Martin-Lozano, {\it {Probing the Electroweakino
  Sector of General Gauge Mediation at the LHC}},
  \href{http://arxiv.org//abs/1705.06497}{{\tt arXiv:1705.06497}}.

\bibitem{Dutta:2017jpe}
J.~Dutta, P.~Konar, S.~Mondal, B.~Mukhopadhyaya, and S.~K. Rai, {\it {Search
  for a compressed supersymmetric spectrum with a light Gravitino}},  {\em
  JHEP} {\bf 09} (2017) 026, [\href{http://arxiv.org//abs/1704.04617}{{\tt
  arXiv:1704.04617}}].

\bibitem{Baer:2010kd}
H.~Baer, R.~Dermisek, S.~Rajagopalan, and H.~Summy, {\it {Neutralino, axion and
  axino cold dark matter in minimal, hypercharged and gaugino AMSB}},  {\em
  JCAP} {\bf 1007} (2010) 014, [\href{http://arxiv.org//abs/1004.3297}{{\tt
  arXiv:1004.3297}}].

\bibitem{Baer:2011hx}
H.~Baer, A.~Lessa, S.~Rajagopalan, and W.~Sreethawong, {\it {Mixed
  axion/neutralino cold dark matter in supersymmetric models}},  {\em JCAP}
  {\bf 1106} (2011) 031, [\href{http://arxiv.org//abs/1103.5413}{{\tt
  arXiv:1103.5413}}].

\bibitem{Bae:2013hma}
K.~J. Bae, H.~Baer, and E.~J. Chun, {\it {Mixed axion/neutralino dark matter in
  the SUSY DFSZ axion model}},  {\em JCAP} {\bf 1312} (2013) 028,
  [\href{http://arxiv.org//abs/1309.5365}{{\tt arXiv:1309.5365}}].

\bibitem{Bae:2014efa}
K.~J. Bae, H.~Baer, E.~J. Chun, and C.~S. Shin, {\it {Mixed axion/gravitino
  dark matter from SUSY models with heavy axinos}},  {\em Phys. Rev.} {\bf D91}
  (2015), no.~7 075011, [\href{http://arxiv.org//abs/1410.3857}{{\tt
  arXiv:1410.3857}}].

\bibitem{Bae:2015rra}
K.~J. Bae, H.~Baer, A.~Lessa, and H.~Serce, {\it {Mixed axion-wino dark
  matter}},  {\em Front.in Phys.} {\bf 3} (2015) 49,
  [\href{http://arxiv.org//abs/1502.07198}{{\tt arXiv:1502.07198}}].

\bibitem{ArkaniHamed:2000bq}
N.~Arkani-Hamed, L.~J. Hall, H.~Murayama, D.~Tucker-Smith, and N.~Weiner, {\it
  {Small neutrino masses from supersymmetry breaking}},  {\em Phys. Rev.} {\bf
  D64} (2001) 115011, [\href{http://arxiv.org//abs/hep-ph/0006312}{{\tt
  hep-ph/0006312}}].

\bibitem{Gopalakrishna:2006kr}
S.~Gopalakrishna, A.~de~Gouvea, and W.~Porod, {\it {Right-handed sneutrinos as
  nonthermal dark matter}},  {\em JCAP} {\bf 0605} (2006) 005,
  [\href{http://arxiv.org//abs/hep-ph/0602027}{{\tt hep-ph/0602027}}].

\bibitem{Asaka:2005cn}
T.~Asaka, K.~Ishiwata, and T.~Moroi, {\it {Right-handed sneutrino as cold dark
  matter}},  {\em Phys. Rev.} {\bf D73} (2006) 051301,
  [\href{http://arxiv.org//abs/hep-ph/0512118}{{\tt hep-ph/0512118}}].

\bibitem{Asaka:2006fs}
T.~Asaka, K.~Ishiwata, and T.~Moroi, {\it {Right-handed sneutrino as cold dark
  matter of the universe}},  {\em Phys. Rev.} {\bf D75} (2007) 065001,
  [\href{http://arxiv.org//abs/hep-ph/0612211}{{\tt hep-ph/0612211}}].

\bibitem{Page:2007sh}
V.~Page, {\it {Non-thermal right-handed sneutrino dark matter and the
  Omega(DM)/Omega(b) problem}},  {\em JHEP} {\bf 04} (2007) 021,
  [\href{http://arxiv.org//abs/hep-ph/0701266}{{\tt hep-ph/0701266}}].

\bibitem{Kumar:2009sf}
A.~Kumar, D.~Tucker-Smith, and N.~Weiner, {\it {Neutrino Mass, Sneutrino Dark
  Matter and Signals of Lepton Flavor Violation in the MRSSM}},  {\em JHEP}
  {\bf 09} (2010) 111, [\href{http://arxiv.org//abs/0910.2475}{{\tt
  arXiv:0910.2475}}].

\bibitem{Kadota:2009fg}
K.~Kadota and K.~A. Olive, {\it {Heavy Right-Handed Neutrinos and Dark Matter
  in the nuCMSSM}},  {\em Phys. Rev.} {\bf D80} (2009) 095015,
  [\href{http://arxiv.org//abs/0909.3075}{{\tt arXiv:0909.3075}}].

\bibitem{Belanger:2010cd}
G.~Belanger, M.~Kakizaki, E.~K. Park, S.~Kraml, and A.~Pukhov, {\it {Light
  mixed sneutrinos as thermal dark matter}},  {\em JCAP} {\bf 1011} (2010) 017,
  [\href{http://arxiv.org//abs/1008.0580}{{\tt arXiv:1008.0580}}].

\bibitem{Dumont:2012ee}
B.~Dumont, G.~Belanger, S.~Fichet, S.~Kraml, and T.~Schwetz, {\it {Mixed
  sneutrino dark matter in light of the 2011 XENON and LHC results}},  {\em
  JCAP} {\bf 1209} (2012) 013, [\href{http://arxiv.org//abs/1206.1521}{{\tt
  arXiv:1206.1521}}].

\bibitem{Kakizaki:2015nua}
M.~Kakizaki, E.-K. Park, J.-h. Park, and A.~Santa, {\it {Phenomenological
  constraints on light mixed sneutrino dark matter scenarios}},  {\em Phys.
  Lett.} {\bf B749} (2015) 44--49,
  [\href{http://arxiv.org//abs/1503.06783}{{\tt arXiv:1503.06783}}].

\bibitem{Chatterjee:2014bva}
A.~Chatterjee, D.~Das, B.~Mukhopadhyaya, and S.~K. Rai, {\it {Right Sneutrino
  Dark Matter and a Monochromatic Photon Line}},  {\em JCAP} {\bf 1407} (2014)
  023, [\href{http://arxiv.org//abs/1401.2527}{{\tt arXiv:1401.2527}}].

\bibitem{Cerna-Velazco:2017cmn}
N.~Cerna-Velazco, T.~Faber, J.~Jones-Perez, and W.~Porod, {\it {Constraining
  sleptons at the LHC in a supersymmetric low-scale seesaw scenario}},  {\em
  Eur. Phys. J.} {\bf C77} (2017), no.~10 661,
  [\href{http://arxiv.org//abs/1705.06583}{{\tt arXiv:1705.06583}}].

\bibitem{Falk:1994es}
T.~Falk, K.~A. Olive, and M.~Srednicki, {\it {Heavy sneutrinos as dark
  matter}},  {\em Phys. Lett.} {\bf B339} (1994) 248--251,
  [\href{http://arxiv.org//abs/hep-ph/9409270}{{\tt hep-ph/9409270}}].

\bibitem{Arina:2007tm}
C.~Arina and N.~Fornengo, {\it {Sneutrino cold dark matter, a new analysis:
  Relic abundance and detection rates}},  {\em JHEP} {\bf 11} (2007) 029,
  [\href{http://arxiv.org//abs/0709.4477}{{\tt arXiv:0709.4477}}].

\bibitem{deGouvea:2006wd}
A.~de~Gouvea, S.~Gopalakrishna, and W.~Porod, {\it {Stop Decay into
  Right-handed Sneutrino LSP at Hadron Colliders}},  {\em JHEP} {\bf 11} (2006)
  050, [\href{http://arxiv.org//abs/hep-ph/0606296}{{\tt hep-ph/0606296}}].

\bibitem{Gupta:2007ui}
S.~K. Gupta, B.~Mukhopadhyaya, and S.~K. Rai, {\it {Right-chiral sneutrinos and
  long-lived staus: Event characteristics at the large hadron collider}},  {\em
  Phys. Rev.} {\bf D75} (2007) 075007,
  [\href{http://arxiv.org//abs/hep-ph/0701063}{{\tt hep-ph/0701063}}].

\bibitem{Choudhury:2008gb}
D.~Choudhury, S.~K. Gupta, and B.~Mukhopadhyaya, {\it {Right sneutrinos in a
  supergravity model and the signals of a stable stop at the Large Hadron
  Collider}},  {\em Phys. Rev.} {\bf D78} (2008) 015023,
  [\href{http://arxiv.org//abs/0804.3560}{{\tt arXiv:0804.3560}}].

\bibitem{Guo:2013asa}
J.~Guo, Z.~Kang, J.~Li, T.~Li, and Y.~Liu, {\it {Simplified Supersymmetry with
  Sneutrino LSP at 8 TeV LHC}},  {\em JHEP} {\bf 10} (2014) 164,
  [\href{http://arxiv.org//abs/1312.2821}{{\tt arXiv:1312.2821}}].

\bibitem{Arina2014}
C.~Arina and M.~E. Cabrera, {\it Multi-lepton signatures at lhc from sneutrino
  dark matter},  {\em Journal of High Energy Physics} {\bf 2014} (Apr, 2014)
  100.

\bibitem{Arina:2015uea}
C.~Arina, M.~E.~C. Catalan, S.~Kraml, S.~Kulkarni, and U.~Laa, {\it
  {Constraints on sneutrino dark matter from LHC Run 1}},  {\em JHEP} {\bf 05}
  (2015) 142, [\href{http://arxiv.org//abs/1503.02960}{{\tt
  arXiv:1503.02960}}].

\bibitem{Banerjee:2016uyt}
S.~Banerjee, G.~Bélanger, B.~Mukhopadhyaya, and P.~D. Serpico, {\it
  {Signatures of sneutrino dark matter in an extension of the CMSSM}},  {\em
  JHEP} {\bf 07} (2016) 095, [\href{http://arxiv.org//abs/1603.08834}{{\tt
  arXiv:1603.08834}}].

\bibitem{Grossman:1997is}
Y.~Grossman and H.~E. Haber, {\it {Sneutrino mixing phenomena}},  {\em Phys.
  Rev. Lett.} {\bf 78} (1997) 3438--3441,
  [\href{http://arxiv.org//abs/hep-ph/9702421}{{\tt hep-ph/9702421}}].

\bibitem{MINKOWSKI1977421}
P.~Minkowski, {\it μ→eγ at a rate of one out of 109 muon decays?},  {\em
  Physics Letters B} {\bf 67} (1977), no.~4 421 -- 428.

\bibitem{Yanagida:1979as}
T.~Yanagida, {\it {Horizontal Symmetry and masses of neutrinos}},  {\em Conf.
  Proc.} {\bf C7902131} (1979) 95--99.

\bibitem{Mohapatra}
R.~N. Mohapatra and G.~Senjanovi\ifmmode~\acute{c}\else \'{c}\fi{}, {\it
  Neutrino mass and spontaneous parity nonconservation},  {\em Phys. Rev.
  Lett.} {\bf 44} (Apr, 1980) 912--915.

\bibitem{Casas:2001sr}
J.~A. Casas and A.~Ibarra, {\it {Oscillating neutrinos and muon ---> e,
  gamma}},  {\em Nucl. Phys.} {\bf B618} (2001) 171--204,
  [\href{http://arxiv.org//abs/hep-ph/0103065}{{\tt hep-ph/0103065}}].

\bibitem{Drewes:2013gca}
M.~Drewes, {\it {The Phenomenology of Right Handed Neutrinos}},  {\em Int. J.
  Mod. Phys.} {\bf E22} (2013) 1330019,
  [\href{http://arxiv.org//abs/1303.6912}{{\tt arXiv:1303.6912}}].

\bibitem{Rasmussen:2016njh}
R.~W. Rasmussen and W.~Winter, {\it {Perspectives for tests of neutrino mass
  generation at the GeV scale: Experimental reach versus theoretical
  predictions}},  {\em Phys. Rev.} {\bf D94} (2016), no.~7 073004,
  [\href{http://arxiv.org//abs/1607.07880}{{\tt arXiv:1607.07880}}].

\bibitem{Dedes:2007ef}
A.~Dedes, H.~E. Haber, and J.~Rosiek, {\it {Seesaw mechanism in the sneutrino
  sector and its consequences}},  {\em JHEP} {\bf 11} (2007) 059,
  [\href{http://arxiv.org//abs/0707.3718}{{\tt arXiv:0707.3718}}].

\bibitem{Drees:2004jm}
M.~Drees, R.~Godbole, and P.~Roy, {\em {Theory and phenomenology of sparticles:
  An account of four-dimensional N=1 supersymmetry in high energy physics}}.
\newblock 2004.

\bibitem{Drees:1996pk}
M.~Drees, M.~M. Nojiri, D.~P. Roy, and Y.~Yamada, {\it {Light Higgsino dark
  matter}},  {\em Phys. Rev.} {\bf D56} (1997) 276--290,
  [\href{http://arxiv.org//abs/hep-ph/9701219}{{\tt hep-ph/9701219}}].
  [Erratum: Phys. Rev.D64,039901(2001)].

\bibitem{Giudice:1995np}
G.~F. Giudice and A.~Pomarol, {\it {Mass degeneracy of the Higgsinos}},  {\em
  Phys. Lett.} {\bf B372} (1996) 253--258,
  [\href{http://arxiv.org//abs/hep-ph/9512337}{{\tt hep-ph/9512337}}].

\bibitem{Pierce:1993gj}
D.~Pierce and A.~Papadopoulos, {\it {Radiative corrections to neutralino and
  chargino masses in the minimal supersymmetric model}},  {\em Phys. Rev.} {\bf
  D50} (1994) 565--570, [\href{http://arxiv.org//abs/hep-ph/9312248}{{\tt
  hep-ph/9312248}}].

\bibitem{Pierce:1994ew}
D.~Pierce and A.~Papadopoulos, {\it {The Complete radiative corrections to the
  gaugino and Higgsino masses in the minimal supersymmetric model}},  {\em
  Nucl. Phys.} {\bf B430} (1994) 278--294,
  [\href{http://arxiv.org//abs/hep-ph/9403240}{{\tt hep-ph/9403240}}].

\bibitem{Lahanas:1993ib}
A.~B. Lahanas, K.~Tamvakis, and N.~D. Tracas, {\it {One loop corrections to the
  neutralino sector and radiative electroweak breaking in the MSSM}},  {\em
  Phys. Lett.} {\bf B324} (1994) 387--396,
  [\href{http://arxiv.org//abs/hep-ph/9312251}{{\tt hep-ph/9312251}}].

\bibitem{Staub:2008uz}
F.~Staub, {\it {SARAH}},  \href{http://arxiv.org//abs/0806.0538}{{\tt
  arXiv:0806.0538}}.

\bibitem{Staub:2013tta}
F.~Staub, {\it {SARAH 4 : A tool for (not only SUSY) model builders}},  {\em
  Comput. Phys. Commun.} {\bf 185} (2014) 1773--1790,
  [\href{http://arxiv.org//abs/1309.7223}{{\tt arXiv:1309.7223}}].

\bibitem{Porod:2003um}
W.~Porod, {\it {SPheno, a program for calculating supersymmetric spectra, SUSY
  particle decays and SUSY particle production at e+ e- colliders}},  {\em
  Comput. Phys. Commun.} {\bf 153} (2003) 275--315,
  [\href{http://arxiv.org//abs/hep-ph/0301101}{{\tt hep-ph/0301101}}].

\bibitem{Porod:2011nf}
W.~Porod and F.~Staub, {\it {SPheno 3.1: Extensions including flavour,
  CP-phases and models beyond the MSSM}},  {\em Comput. Phys. Commun.} {\bf
  183} (2012) 2458--2469, [\href{http://arxiv.org//abs/1104.1573}{{\tt
  arXiv:1104.1573}}].

\bibitem{Skands:2003cj}
P.~Z. Skands et~al., {\it {SUSY Les Houches accord: Interfacing SUSY spectrum
  calculators, decay packages, and event generators}},  {\em JHEP} {\bf 07}
  (2004) 036, [\href{http://arxiv.org//abs/hep-ph/0311123}{{\tt
  hep-ph/0311123}}].

\bibitem{Djouadi:2001fa}
A.~Djouadi, Y.~Mambrini, and M.~Muhlleitner, {\it {Chargino and neutralino
  decays revisited}},  {\em Eur. Phys. J.} {\bf C20} (2001) 563--584,
  [\href{http://arxiv.org//abs/hep-ph/0104115}{{\tt hep-ph/0104115}}].

\bibitem{Chen:1995yu}
C.~H. Chen, M.~Drees, and J.~F. Gunion, {\it {Searching for invisible and
  almost invisible particles at e+ e- colliders}},  {\em Phys. Rev. Lett.} {\bf
  76} (1996) 2002--2005, [\href{http://arxiv.org//abs/hep-ph/9512230}{{\tt
  hep-ph/9512230}}].

\bibitem{Chen:1996ap}
C.~H. Chen, M.~Drees, and J.~F. Gunion, {\it {A Nonstandard string / SUSY
  scenario and its phenomenological implications}},  {\em Phys. Rev.} {\bf D55}
  (1997) 330--347, [\href{http://arxiv.org//abs/hep-ph/9607421}{{\tt
  hep-ph/9607421}}]. [Erratum: Phys. Rev.D60,039901(1999)].

\bibitem{Chen:1999yf}
C.~H. Chen, M.~Drees, and J.~F. Gunion, {\it {Addendum/erratum for 'searching
  for invisible and almost invisible particles at e+ e- colliders'
  [hep-ph/9512230] and 'a nonstandard string/SUSY scenario and its
  phenomenological implications' [hep-ph/9607421]}},
  \href{http://arxiv.org//abs/hep-ph/9902309}{{\tt hep-ph/9902309}}.

\bibitem{Haber:1988px}
H.~E. Haber and D.~Wyler, {\it {Radiative Neutralino Decay}},  {\em Nucl.
  Phys.} {\bf B323} (1989) 267--310.

\bibitem{Ambrosanio:1995az}
S.~Ambrosanio and B.~Mele, {\it {Neutralino decays in the minimal
  supersymmetric Standard Model}},  {\em Phys. Rev.} {\bf D53} (1996)
  2541--2562, [\href{http://arxiv.org//abs/hep-ph/9508237}{{\tt
  hep-ph/9508237}}].

\bibitem{Ambrosanio:1996gz}
S.~Ambrosanio and B.~Mele, {\it {Supersymmetric scenarios with dominant
  radiative neutralino decay}},  {\em Phys. Rev.} {\bf D55} (1997) 1399--1417,
  [\href{http://arxiv.org//abs/hep-ph/9609212}{{\tt hep-ph/9609212}}].
  [Erratum: Phys. Rev.D56,3157(1997)].

\bibitem{Baer:2002kv}
H.~Baer and T.~Krupovnickas, {\it {Radiative neutralino decay in supersymmetric
  models}},  {\em JHEP} {\bf 09} (2002) 038,
  [\href{http://arxiv.org//abs/hep-ph/0208277}{{\tt hep-ph/0208277}}].

\bibitem{Aad:2015zhl}
{\bf ATLAS, CMS} Collaboration, G.~Aad et~al., {\it {Combined Measurement of
  the Higgs Boson Mass in $pp$ Collisions at $\sqrt{s}=7$ and 8 TeV with the
  ATLAS and CMS Experiments}},  {\em Phys. Rev. Lett.} {\bf 114} (2015) 191803,
  [\href{http://arxiv.org//abs/1503.07589}{{\tt arXiv:1503.07589}}].

\bibitem{Aad:2012tfa}
{\bf ATLAS} Collaboration, G.~Aad et~al., {\it {Observation of a new particle
  in the search for the Standard Model Higgs boson with the ATLAS detector at
  the LHC}},  {\em Phys. Lett.} {\bf B716} (2012) 1--29,
  [\href{http://arxiv.org//abs/1207.7214}{{\tt arXiv:1207.7214}}].

\bibitem{Chatrchyan:2012xdj}
{\bf CMS} Collaboration, S.~Chatrchyan et~al., {\it {Observation of a new boson
  at a mass of 125 GeV with the CMS experiment at the LHC}},  {\em Phys. Lett.}
  {\bf B716} (2012) 30--61, [\href{http://arxiv.org//abs/1207.7235}{{\tt
  arXiv:1207.7235}}].

\bibitem{Carena:2013ytb}
M.~Carena, S.~Heinemeyer, O.~Stål, C.~E.~M. Wagner, and G.~Weiglein, {\it
  {MSSM Higgs Boson Searches at the LHC: Benchmark Scenarios after the
  Discovery of a Higgs-like Particle}},  {\em Eur. Phys. J.} {\bf C73} (2013),
  no.~9 2552, [\href{http://arxiv.org//abs/1302.7033}{{\tt arXiv:1302.7033}}].

\bibitem{LEPSUSYWG}
LEP2SUSYWG et~al., {\it Lep2 joint susy working group}, .
  \url{http://lepsusy.web.cern.ch/lepsusy/,
  http://lepsusy.web.cern.ch/lepsusy/www/inoslowdmsummer02/charginolowdm_pub.html,
  http://lepsusy.web.cern.ch/lepsusy/www/inos_moriond01/charginos_pub.html}.

\bibitem{Khachatryan:2016whc}
{\bf CMS} Collaboration, V.~Khachatryan et~al., {\it {Searches for invisible
  decays of the Higgs boson in pp collisions at sqrt(s) = 7, 8, and 13 TeV}},
  {\em JHEP} {\bf 02} (2017) 135, [\href{http://arxiv.org//abs/1610.09218}{{\tt
  arXiv:1610.09218}}].

\bibitem{PDG}
C.~Patrignani and P.~D. Group, {\it Review of particle physics},  {\em Chinese
  Physics C} {\bf 40} (2016), no.~10 100001.

\bibitem{Aaij:2012nna}
{\bf LHCb} Collaboration, R.~Aaij et~al., {\it {First Evidence for the Decay
  $B_s^0 \to \mu^+ \mu^-$}},  {\em Phys. Rev. Lett.} {\bf 110} (2013), no.~2
  021801, [\href{http://arxiv.org//abs/1211.2674}{{\tt arXiv:1211.2674}}].

\bibitem{Amhis:2012bh}
{\bf Heavy Flavor Averaging Group} Collaboration, Y.~Amhis et~al., {\it
  {Averages of B-Hadron, C-Hadron, and tau-lepton properties as of early
  2012}},  \href{http://arxiv.org//abs/1207.1158}{{\tt arXiv:1207.1158}}.

\bibitem{Ade:2015xua}
{\bf Planck} Collaboration, P.~A.~R. Ade et~al., {\it {Planck 2015 results.
  XIII. Cosmological parameters}},  {\em Astron. Astrophys.} {\bf 594} (2016)
  A13, [\href{http://arxiv.org//abs/1502.01589}{{\tt arXiv:1502.01589}}].

\bibitem{Akerib:2016vxi}
D.~S. Akerib et~al., {\it {Results from a search for dark matter in the
  complete LUX exposure}},  \href{http://arxiv.org//abs/1608.07648}{{\tt
  arXiv:1608.07648}}.

\bibitem{Tan:2016zwf}
{\bf PandaX-II} Collaboration, A.~Tan et~al., {\it {Dark Matter Results from
  First 98.7 Days of Data from the PandaX-II Experiment}},  {\em Phys. Rev.
  Lett.} {\bf 117} (2016), no.~12 121303,
  [\href{http://arxiv.org//abs/1607.07400}{{\tt arXiv:1607.07400}}].

\bibitem{Aprile:2017iyp}
{\bf XENON} Collaboration, E.~Aprile et~al., {\it {First Dark Matter Search
  Results from the XENON1T Experiment}},
  \href{http://arxiv.org//abs/1705.06655}{{\tt arXiv:1705.06655}}.

\bibitem{Belanger:2013oya}
G.~Belanger, F.~Boudjema, A.~Pukhov, and A.~Semenov, {\it {micrOMEGAs-3: A
  program for calculating dark matter observables}},  {\em Comput. Phys.
  Commun.} {\bf 185} (2014) 960--985,
  [\href{http://arxiv.org//abs/1305.0237}{{\tt arXiv:1305.0237}}].

\bibitem{Hall:1997ah}
L.~J. Hall, T.~Moroi, and H.~Murayama, {\it {Sneutrino cold dark matter with
  lepton number violation}},  {\em Phys. Lett.} {\bf B424} (1998) 305--312,
  [\href{http://arxiv.org//abs/hep-ph/9712515}{{\tt hep-ph/9712515}}].

\bibitem{TuckerSmith:2001hy}
D.~Tucker-Smith and N.~Weiner, {\it {Inelastic dark matter}},  {\em Phys. Rev.}
  {\bf D64} (2001) 043502, [\href{http://arxiv.org//abs/hep-ph/0101138}{{\tt
  hep-ph/0101138}}].

\bibitem{Baer:2014eja}
H.~Baer, K.-Y. Choi, J.~E. Kim, and L.~Roszkowski, {\it {Dark matter production
  in the early Universe: beyond the thermal WIMP paradigm}},  {\em Phys. Rept.}
  {\bf 555} (2015) 1--60, [\href{http://arxiv.org//abs/1407.0017}{{\tt
  arXiv:1407.0017}}].

\bibitem{Acharya:2008bk}
B.~S. Acharya, P.~Kumar, K.~Bobkov, G.~Kane, J.~Shao, and S.~Watson, {\it
  {Non-thermal Dark Matter and the Moduli Problem in String Frameworks}},  {\em
  JHEP} {\bf 06} (2008) 064, [\href{http://arxiv.org//abs/0804.0863}{{\tt
  arXiv:0804.0863}}].

\bibitem{Aaboud:2017iio}
{\bf ATLAS} Collaboration, M.~Aaboud et~al., {\it {Search for long-lived,
  massive particles in events with displaced vertices and missing transverse
  momentum in $\sqrt{s}$ = 13 TeV $pp$ collisions with the ATLAS detector}},
  \href{http://arxiv.org//abs/1710.04901}{{\tt arXiv:1710.04901}}.

\bibitem{Beenakker:1996ed}
W.~Beenakker, R.~Hopker, and M.~Spira, {\it {PROSPINO: A Program for the
  production of supersymmetric particles in next-to-leading order QCD}},
  \href{http://arxiv.org//abs/hep-ph/9611232}{{\tt hep-ph/9611232}}.

\bibitem{Plehn:2004rp}
T.~Plehn, {\it {Measuring the MSSM Lagrangean}},  {\em Czech. J. Phys.} {\bf
  55} (2005) B213--B220, [\href{http://arxiv.org//abs/hep-ph/0410063}{{\tt
  hep-ph/0410063}}].

\bibitem{Spira:2002rd}
M.~Spira, {\it {Higgs and SUSY particle production at hadron colliders}},  in
  {\em {Supersymmetry and unification of fundamental interactions. Proceedings,
  10th International Conference, SUSY'02, Hamburg, Germany, June 17-23, 2002}},
  pp.~217--226, 2002.
\newblock \href{http://arxiv.org//abs/hep-ph/0211145}{{\tt hep-ph/0211145}}.

\bibitem{CMS-PAS-SUS-16-039}
{\bf CMS Collaboration} Collaboration, {\it {Search for electroweak production
  of charginos and neutralinos in multilepton final states in pp collision data
  at $\sqrt{s}=13~\mathrm{TeV}$}},  Tech. Rep. CMS-PAS-SUS-16-039, CERN,
  Geneva, 2017.

\bibitem{CMS-PAS-SUS-16-048}
{\bf CMS Collaboration} Collaboration, {\it {Search for new physics in events
  with two low momentum opposite-sign leptons and missing transverse energy at
  $\sqrt{s}=13~\mathrm{TeV}$}},  Tech. Rep. CMS-PAS-SUS-16-048, CERN, Geneva,
  2017.

\bibitem{Aaboud:2017leg}
{\bf ATLAS} Collaboration, M.~Aaboud et~al., {\it {Search for electroweak
  production of supersymmetric states in scenarios with compressed mass spectra
  at $\sqrt{s}=13$ TeV with the ATLAS detector}},
  \href{http://arxiv.org//abs/1712.08119}{{\tt arXiv:1712.08119}}.

\bibitem{ATLAS-CONF-2017-039}
{\bf ATLAS Collaboration} Collaboration, {\it {Search for electroweak
  production of supersymmetric particles in the two and three lepton final
  state at $\boldmath{\sqrt{s}=13\,}$TeV with the ATLAS detector}},  Tech. Rep.
  ATLAS-CONF-2017-039, CERN, Geneva, Jun, 2017.

\bibitem{Aad:2014nua}
{\bf ATLAS} Collaboration, G.~Aad et~al., {\it {Search for direct production of
  charginos and neutralinos in events with three leptons and missing transverse
  momentum in $\sqrt{s} =$ 8TeV $pp$ collisions with the ATLAS detector}},
  {\em JHEP} {\bf 04} (2014) 169, [\href{http://arxiv.org//abs/1402.7029}{{\tt
  arXiv:1402.7029}}].

\bibitem{Aad:2014qaa}
{\bf ATLAS} Collaboration, G.~Aad et~al., {\it {Search for direct top-squark
  pair production in final states with two leptons in pp collisions at
  $\sqrt{s} =$ 8TeV with the ATLAS detector}},  {\em JHEP} {\bf 06} (2014) 124,
  [\href{http://arxiv.org//abs/1403.4853}{{\tt arXiv:1403.4853}}].

\bibitem{Aad:2014mha}
{\bf ATLAS} Collaboration, G.~Aad et~al., {\it {Search for direct top squark
  pair production in events with a Z boson, b-jets and missing transverse
  momentum in sqrt(s)=8 TeV pp collisions with the ATLAS detector}},  {\em Eur.
  Phys. J.} {\bf C74} (2014), no.~6 2883,
  [\href{http://arxiv.org//abs/1403.5222}{{\tt arXiv:1403.5222}}].

\bibitem{Aad:2014pda}
{\bf ATLAS} Collaboration, G.~Aad et~al., {\it {Search for supersymmetry at
  $\sqrt{s}$=8 TeV in final states with jets and two same-sign leptons or three
  leptons with the ATLAS detector}},  {\em JHEP} {\bf 06} (2014) 035,
  [\href{http://arxiv.org//abs/1404.2500}{{\tt arXiv:1404.2500}}].

\bibitem{Aad:2014kra}
{\bf ATLAS} Collaboration, G.~Aad et~al., {\it {Search for top squark pair
  production in final states with one isolated lepton, jets, and missing
  transverse momentum in $\sqrt s =$8 TeV $pp$ collisions with the ATLAS
  detector}},  {\em JHEP} {\bf 11} (2014) 118,
  [\href{http://arxiv.org//abs/1407.0583}{{\tt arXiv:1407.0583}}].

\bibitem{Aad:2015wqa}
{\bf ATLAS} Collaboration, G.~Aad et~al., {\it {Search for supersymmetry in
  events containing a same-flavour opposite-sign dilepton pair, jets, and large
  missing transverse momentum in $\sqrt{s}=8$ TeV pp collisions with the ATLAS
  detector}},  {\em Eur. Phys. J.} {\bf C75} (2015), no.~7 318,
  [\href{http://arxiv.org//abs/1503.03290}{{\tt arXiv:1503.03290}}]. [Erratum:
  Eur. Phys. J.C75,no.10,463(2015)].

\bibitem{Aad2015}
G.~Aad and {etal}, {\it Search for new phenomena in final states with an
  energetic jet and large missing transverse momentum in pp collisions at
  $\sqrt{s}$=8 tev with the atlas detector},  {\em The European Physical
  Journal C} {\bf 75} (Jul, 2015) 299.

\bibitem{Khachatryan2015}
V.~Khachatryan and {etal}, {\it Search for dark matter, extra dimensions, and
  unparticles in monojet events in proton--proton collisions at $\sqrt{s} = 8 $
  tev}, .

\bibitem{Aad:2015pfx}
{\bf ATLAS} Collaboration, G.~Aad et~al., {\it {ATLAS Run 1 searches for direct
  pair production of third-generation squarks at the Large Hadron Collider}},
  {\em Eur. Phys. J.} {\bf C75} (2015), no.~10 510,
  [\href{http://arxiv.org//abs/1506.08616}{{\tt arXiv:1506.08616}}]. [Erratum:
  Eur. Phys. J.C76,no.3,153(2016)].

\bibitem{ATLAS:2012tna}
{\bf ATLAS} Collaboration, {\it {Search for supersymmetry at $\sqrt{s} = 8$ TeV
  in final states with jets, missing transverse momentum and one isolated
  lepton}}, .

\bibitem{ATLAS:2013rla}
{\bf ATLAS} Collaboration, {\it {Search for direct production of charginos and
  neutralinos in events with three leptons and missing transverse momentum in
  21$\,$fb$^{-1}$ of pp collisions at $\sqrt{s}=8\,$TeV with the ATLAS
  detector}}, .

\bibitem{TheATLAScollaboration:2013hha}
{\bf ATLAS} Collaboration, T.~A. collaboration, {\it {Search for direct-slepton
  and direct-chargino production in final states with two opposite-sign
  leptons, missing transverse momentum and no jets in 20/fb of pp collisions at
  sqrt(s) = 8 TeV with the ATLAS detector}}, .

\bibitem{TheATLAScollaboration:2013tha}
{\bf ATLAS} Collaboration, T.~A. collaboration, {\it {Search for strong
  production of supersymmetric particles in final states with missing
  transverse momentum and at least three b-jets using 20.1 fb−1 of pp
  collisions at sqrt(s) = 8 TeV with the ATLAS Detector.}}, .

\bibitem{s}
{\it {Search for strongly produced supersymmetric particles in decays with two
  leptons at $\sqrt{s}$ = 8 TeV}},  Tech. Rep. ATLAS-CONF-2013-089, CERN,
  Geneva, Aug, 2013.

\bibitem{Khachatryan:2015lwa}
{\bf CMS} Collaboration, V.~Khachatryan et~al., {\it {Search for Physics Beyond
  the Standard Model in Events with Two Leptons, Jets, and Missing Transverse
  Momentum in pp Collisions at sqrt(s) = 8 TeV}},  {\em JHEP} {\bf 04} (2015)
  124, [\href{http://arxiv.org//abs/1502.06031}{{\tt arXiv:1502.06031}}].

\bibitem{Khachatryan:2015nua}
{\bf CMS} Collaboration, V.~Khachatryan et~al., {\it {Search for the production
  of dark matter in association with top-quark pairs in the single-lepton final
  state in proton-proton collisions at sqrt(s) = 8 TeV}},  {\em JHEP} {\bf 06}
  (2015) 121, [\href{http://arxiv.org//abs/1504.03198}{{\tt
  arXiv:1504.03198}}].

\bibitem{CMS-PAS-SUS-13-016}
{\bf CMS Collaboration} Collaboration, {\it {Search for supersymmetry in pp
  collisions at sqrt(s) = 8 Tev in events with two opposite sign leptons, large
  number of jets, b-tagged jets, and large missing transverse energy.}},  Tech.
  Rep. CMS-PAS-SUS-13-016, CERN, Geneva, 2013.

\bibitem{Aad:2016tuk}
{\bf ATLAS} Collaboration, G.~Aad et~al., {\it {Search for supersymmetry at
  $\sqrt{s}=13$ TeV in final states with jets and two same-sign leptons or
  three leptons with the ATLAS detector}},  {\em Eur. Phys. J.} {\bf C76}
  (2016), no.~5 259, [\href{http://arxiv.org//abs/1602.09058}{{\tt
  arXiv:1602.09058}}].

\bibitem{Aaboud:2016tnv}
{\bf ATLAS} Collaboration, M.~Aaboud et~al., {\it {Search for new phenomena in
  final states with an energetic jet and large missing transverse momentum in
  $pp$ collisions at $\sqrt{s}=13$  TeV using the ATLAS detector}},  {\em
  Phys. Rev.} {\bf D94} (2016), no.~3 032005,
  [\href{http://arxiv.org//abs/1604.07773}{{\tt arXiv:1604.07773}}].

\bibitem{Aad:2016qqk}
{\bf ATLAS} Collaboration, G.~Aad et~al., {\it {Search for gluinos in events
  with an isolated lepton, jets and missing transverse momentum at $\sqrt{s}$ =
  13 Te V with the ATLAS detector}},  {\em Eur. Phys. J.} {\bf C76} (2016),
  no.~10 565, [\href{http://arxiv.org//abs/1605.04285}{{\tt
  arXiv:1605.04285}}].

\bibitem{Aad:2016eki}
{\bf ATLAS} Collaboration, G.~Aad et~al., {\it {Search for pair production of
  gluinos decaying via stop and sbottom in events with $b$-jets and large
  missing transverse momentum in $pp$ collisions at $\sqrt{s} = 13$ TeV with
  the ATLAS detector}},  {\em Phys. Rev.} {\bf D94} (2016), no.~3 032003,
  [\href{http://arxiv.org//abs/1605.09318}{{\tt arXiv:1605.09318}}].

\bibitem{Aaboud:2016lwz}
{\bf ATLAS} Collaboration, M.~Aaboud et~al., {\it {Search for top squarks in
  final states with one isolated lepton, jets, and missing transverse momentum
  in $\sqrt{s}=13$ TeV $pp$ collisions with the ATLAS detector}},  {\em Phys.
  Rev.} {\bf D94} (2016), no.~5 052009,
  [\href{http://arxiv.org//abs/1606.03903}{{\tt arXiv:1606.03903}}].

\bibitem{TheATLAScollaboration:2015nxu}
T.~A. collaboration, {\it {A search for Supersymmetry in events containing a
  leptonically decaying $Z$ boson, jets and missing transverse momentum in
  $\sqrt{s}=13~$TeV $pp$ collisions with the ATLAS detector}}, .

\bibitem{TheATLAScollaboration:2016gxs}
T.~A. collaboration, {\it {Search for production of vector-like top quark pairs
  and of four top quarks in the lepton-plus-jets final state in $pp$ collisions
  at $\sqrt{s}=13$ TeV with the ATLAS detector}}, .

\bibitem{ATLAS:2016xcm}
{\bf ATLAS} Collaboration, T.~A. collaboration, {\it {Search for direct top
  squark pair production and dark matter production in final states with two
  leptons in $\sqrt{s} = 13$ TeV $pp$ collisions using 13.3 fb$^{-1}$ of ATLAS
  data}}, .

\bibitem{ATLAS-CONF-2016-050}
{\bf ATLAS Collaboration} Collaboration, {\it {Search for top squarks in final
  states with one isolated lepton, jets, and missing transverse momentum in
  $\sqrt{s}$ = 13 TeV pp collisions with the ATLAS detector}},  Tech. Rep.
  ATLAS-CONF-2016-050, CERN, Geneva, Aug, 2016.

\bibitem{CMS:2015bsf}
{\bf CMS} Collaboration, C.~Collaboration, {\it {Search for new physics in
  final states with two opposite-sign same-flavor leptons, jets and missing
  transverse momentum in pp collisions at sqrt(s)=13 TeV}}, .

\bibitem{Aaboud:2017phn}
{\bf ATLAS} Collaboration, M.~Aaboud et~al., {\it {Search for dark matter and
  other new phenomena in events with an energetic jet and large missing
  transverse momentum using the ATLAS detector}},
  \href{http://arxiv.org//abs/1711.03301}{{\tt arXiv:1711.03301}}.

\bibitem{Aaboud:2017efa}
{\bf ATLAS} Collaboration, M.~Aaboud et~al., {\it {Search for a new heavy gauge
  boson resonance decaying into a lepton and missing transverse momentum in 36
  fb$^{-1}$ of $pp$ collisions at $\sqrt{s} =$ 13 TeV with the ATLAS
  experiment}},  \href{http://arxiv.org//abs/1706.04786}{{\tt
  arXiv:1706.04786}}.

\bibitem{Khachatryan:2016jww}
{\bf CMS} Collaboration, V.~Khachatryan et~al., {\it {Search for heavy gauge W'
  boson in events with an energetic lepton and large missing transverse
  momentum at $ \sqrt{s} = $ 13 TeV}},  {\em Phys. Lett.} {\bf B770} (2017)
  278--301, [\href{http://arxiv.org//abs/1612.09274}{{\tt arXiv:1612.09274}}].

\bibitem{CMS-PAS-SUS-16-052}
{\bf CMS Collaboration} Collaboration, {\it {Search for supersymmetry in events
  with at least one soft lepton, low jet multiplicity, and missing transverse
  momentum in proton-proton collisions at $\sqrt{s}=13~\mathrm{TeV}$}},  Tech.
  Rep. CMS-PAS-SUS-16-052, CERN, Geneva, 2017.

\bibitem{201757}
{\it Search for narrow resonances in dilepton mass spectra in proton–proton
  collisions at s=13 tev and combination with 8 tev data},  {\em Physics
  Letters B} {\bf 768} (2017) 57 -- 80.

\bibitem{CMS-PAS-SUS-17-009}
{\bf CMS Collaboration} Collaboration, {\it {Search for selectrons and smuons
  at $\sqrt{s}=13$ TeV}},  Tech. Rep. CMS-PAS-SUS-17-009, CERN, Geneva, 2017.

\bibitem{CMS-PAS-SUS-17-002}
{\bf CMS Collaboration} Collaboration, {\it {Search for supersymmetry in events
  with tau leptons and missing transverse momentum in proton-proton collisions
  at sqrt(s)=13 TeV}},  Tech. Rep. CMS-PAS-SUS-17-002, CERN, Geneva, 2017.

\bibitem{Sirunyan:2017lae}
{\bf CMS} Collaboration, A.~M. Sirunyan et~al., {\it {Search for electroweak
  production of charginos and neutralinos in multilepton final states in
  proton-proton collisions at $\sqrt{s}=$ 13 TeV}},
  \href{http://arxiv.org//abs/1709.05406}{{\tt arXiv:1709.05406}}.

\bibitem{Aaboud:2017dmy}
{\bf ATLAS} Collaboration, M.~Aaboud et~al., {\it {Search for supersymmetry in
  final states with two same-sign or three leptons and jets using 36 fb$^{-1}$
  of $\sqrt{s}=13$ TeV $pp$ collision data with the ATLAS detector}},
  \href{http://arxiv.org//abs/1706.03731}{{\tt arXiv:1706.03731}}.

\bibitem{ATLAS-CONF-2016-075}
{\bf ATLAS Collaboration} Collaboration, {\it {Search for supersymmetry in
  events with four or more leptons in $\sqrt{s}=13\,{\rm TeV}$ pp collisions
  using $13.3\,{\rm fb}^{−1}$ of ATLAS data.}},  Tech. Rep.
  ATLAS-CONF-2016-075, CERN, Geneva, Aug, 2016.

\bibitem{Sirunyan:2018ubx}
{\bf CMS} Collaboration, A.~M. Sirunyan et~al., {\it {Combined search for
  electroweak production of charginos and neutralinos in proton-proton
  collisions at $\sqrt{s} =$ 13 TeV}},
  \href{http://arxiv.org//abs/1801.03957}{{\tt arXiv:1801.03957}}.

\bibitem{Khachatryan2017}
V.~Khachatryan and {etal}.

\bibitem{20169}
{\it Search for supersymmetry in events with soft leptons, low jet
  multiplicity, and missing transverse energy in proton–proton collisions at
  s=8 tev},  {\em Physics Letters B} {\bf 759} (2016), no.~Supplement C 9 --
  35.

\bibitem{Beauchesne:2017yhh}
H.~Beauchesne, E.~Bertuzzo, G.~Grilli~di Cortona, and Z.~Tabrizi, {\it
  {Collider phenomenology of Hidden Valley mediators of spin 0 or 1/2 with
  semivisible jets}},  \href{http://arxiv.org//abs/1712.07160}{{\tt
  arXiv:1712.07160}}.

\bibitem{Sirunyan:2017qaj}
{\bf CMS} Collaboration, A.~M. Sirunyan et~al., {\it {Search for new phenomena
  in final states with two opposite-charge, same-flavor leptons, jets, and
  missing transverse momentum in pp collisions at $\sqrt{s} = $ 13 TeV}},
  \href{http://arxiv.org//abs/1709.08908}{{\tt arXiv:1709.08908}}.

\bibitem{Drees:2013wra}
M.~Drees, H.~Dreiner, D.~Schmeier, J.~Tattersall, and J.~S. Kim, {\it
  {CheckMATE: Confronting your Favourite New Physics Model with LHC Data}},
  {\em Comput. Phys. Commun.} {\bf 187} (2015) 227--265,
  [\href{http://arxiv.org//abs/1312.2591}{{\tt arXiv:1312.2591}}].

\bibitem{Dercks:2016npn}
D.~Dercks, N.~Desai, J.~S. Kim, K.~Rolbiecki, J.~Tattersall, and T.~Weber, {\it
  {CheckMATE 2: From the model to the limit}},
  \href{http://arxiv.org//abs/1611.09856}{{\tt arXiv:1611.09856}}.

\bibitem{Alwall:2011uj}
J.~Alwall, M.~Herquet, F.~Maltoni, O.~Mattelaer, and T.~Stelzer, {\it {MadGraph
  5 : Going Beyond}},  {\em JHEP} {\bf 06} (2011) 128,
  [\href{http://arxiv.org//abs/1106.0522}{{\tt arXiv:1106.0522}}].

\bibitem{Alwall:2014hca}
J.~Alwall, R.~Frederix, S.~Frixione, V.~Hirschi, F.~Maltoni, O.~Mattelaer,
  H.~S. Shao, T.~Stelzer, P.~Torrielli, and M.~Zaro, {\it {The automated
  computation of tree-level and next-to-leading order differential cross
  sections, and their matching to parton shower simulations}},  {\em JHEP} {\bf
  07} (2014) 079, [\href{http://arxiv.org//abs/1405.0301}{{\tt
  arXiv:1405.0301}}].

\bibitem{Sjostrand:2006za}
T.~Sjostrand, S.~Mrenna, and P.~Z. Skands, {\it {PYTHIA 6.4 Physics and
  Manual}},  {\em JHEP} {\bf 05} (2006) 026,
  [\href{http://arxiv.org//abs/hep-ph/0603175}{{\tt hep-ph/0603175}}].

\bibitem{Sjostrand:2007gs}
T.~Sjostrand, S.~Mrenna, and P.~Z. Skands, {\it {A Brief Introduction to PYTHIA
  8.1}},  {\em Comput. Phys. Commun.} {\bf 178} (2008) 852--867,
  [\href{http://arxiv.org//abs/0710.3820}{{\tt arXiv:0710.3820}}].

\bibitem{Sjostrand:2014zea}
T.~Sjöstrand, S.~Ask, J.~R. Christiansen, R.~Corke, N.~Desai, P.~Ilten,
  S.~Mrenna, S.~Prestel, C.~O. Rasmussen, and P.~Z. Skands, {\it {An
  Introduction to PYTHIA 8.2}},  {\em Comput. Phys. Commun.} {\bf 191} (2015)
  159--177, [\href{http://arxiv.org//abs/1410.3012}{{\tt arXiv:1410.3012}}].

\bibitem{deFavereau:2013fsa}
{\bf DELPHES 3} Collaboration, J.~de~Favereau, C.~Delaere, P.~Demin,
  A.~Giammanco, V.~Lemaître, A.~Mertens, and M.~Selvaggi, {\it {DELPHES 3, A
  modular framework for fast simulation of a generic collider experiment}},
  {\em JHEP} {\bf 02} (2014) 057, [\href{http://arxiv.org//abs/1307.6346}{{\tt
  arXiv:1307.6346}}].

\bibitem{Selvaggi:2014mya}
M.~Selvaggi, {\it {DELPHES 3: A modular framework for fast-simulation of
  generic collider experiments}},  {\em J. Phys. Conf. Ser.} {\bf 523} (2014)
  012033.

\bibitem{Mertens:2015kba}
A.~Mertens, {\it {New features in Delphes 3}},  {\em J. Phys. Conf. Ser.} {\bf
  608} (2015), no.~1 012045.

\bibitem{Pumplin:2002vw}
J.~Pumplin, D.~R. Stump, J.~Huston, H.~L. Lai, P.~M. Nadolsky, and W.~K. Tung,
  {\it {New generation of parton distributions with uncertainties from global
  QCD analysis}},  {\em JHEP} {\bf 07} (2002) 012,
  [\href{http://arxiv.org//abs/hep-ph/0201195}{{\tt hep-ph/0201195}}].

\bibitem{Cacciari:2011ma}
M.~Cacciari, G.~P. Salam, and G.~Soyez, {\it {FastJet User Manual}},  {\em Eur.
  Phys. J.} {\bf C72} (2012) 1896, [\href{http://arxiv.org//abs/1111.6097}{{\tt
  arXiv:1111.6097}}].

\bibitem{Cacciari:2008gp}
M.~Cacciari, G.~P. Salam, and G.~Soyez, {\it {The Anti-k(t) jet clustering
  algorithm}},  {\em JHEP} {\bf 04} (2008) 063,
  [\href{http://arxiv.org//abs/0802.1189}{{\tt arXiv:0802.1189}}].

\bibitem{Conte:2012fm}
E.~Conte, B.~Fuks, and G.~Serret, {\it {MadAnalysis 5, A User-Friendly
  Framework for Collider Phenomenology}},  {\em Comput. Phys. Commun.} {\bf
  184} (2013) 222--256, [\href{http://arxiv.org//abs/1206.1599}{{\tt
  arXiv:1206.1599}}].

\bibitem{Conte:2014zja}
E.~Conte, B.~Dumont, B.~Fuks, and C.~Wymant, {\it {Designing and recasting LHC
  analyses with MadAnalysis 5}},  {\em Eur. Phys. J.} {\bf C74} (2014), no.~10
  3103, [\href{http://arxiv.org//abs/1405.3982}{{\tt arXiv:1405.3982}}].

\bibitem{Dumont:2014tja}
B.~Dumont, B.~Fuks, S.~Kraml, S.~Bein, G.~Chalons, E.~Conte, S.~Kulkarni,
  D.~Sengupta, and C.~Wymant, {\it {Toward a public analysis database for LHC
  new physics searches using MADANALYSIS 5}},  {\em Eur. Phys. J.} {\bf C75}
  (2015), no.~2 56, [\href{http://arxiv.org//abs/1407.3278}{{\tt
  arXiv:1407.3278}}].

\bibitem{Grazzini:2016swo}
M.~Grazzini, S.~Kallweit, D.~Rathlev, and M.~Wiesemann, {\it {$W^{\pm}Z$
  production at hadron colliders in NNLO QCD}},  {\em Phys. Lett.} {\bf B761}
  (2016) 179--183, [\href{http://arxiv.org//abs/1604.08576}{{\tt
  arXiv:1604.08576}}].

\bibitem{Cascioli:2014yka}
F.~Cascioli, T.~Gehrmann, M.~Grazzini, S.~Kallweit, P.~Maierhöfer, A.~von
  Manteuffel, S.~Pozzorini, D.~Rathlev, L.~Tancredi, and E.~Weihs, {\it {ZZ
  production at hadron colliders in NNLO QCD}},  {\em Phys. Lett.} {\bf B735}
  (2014) 311--313, [\href{http://arxiv.org//abs/1405.2219}{{\tt
  arXiv:1405.2219}}].

\bibitem{Giammanco:2017xyn}
A.~Giammanco and R.~Schwienhorst, {\it {Single top-quark production at the
  Tevatron and the LHC}},  \href{http://arxiv.org//abs/1710.10699}{{\tt
  arXiv:1710.10699}}.

\bibitem{Czakon:2013goa}
M.~Czakon, P.~Fiedler, and A.~Mitov, {\it {Total Top-Quark Pair-Production
  Cross Section at Hadron Colliders Through $O(α\frac{4}{S})$}},  {\em Phys.
  Rev. Lett.} {\bf 110} (2013) 252004,
  [\href{http://arxiv.org//abs/1303.6254}{{\tt arXiv:1303.6254}}].

\bibitem{Czakon:2011xx}
M.~Czakon and A.~Mitov, {\it {Top++: A Program for the Calculation of the
  Top-Pair Cross-Section at Hadron Colliders}},  {\em Comput. Phys. Commun.}
  {\bf 185} (2014) 2930, [\href{http://arxiv.org//abs/1112.5675}{{\tt
  arXiv:1112.5675}}].

\bibitem{Kidonakis:2016sjf}
N.~Kidonakis, {\it {Soft-gluon corrections for $tW$ production at N$^3$LO}},
  {\em Phys. Rev.} {\bf D96} (2017), no.~3 034014,
  [\href{http://arxiv.org//abs/1612.06426}{{\tt arXiv:1612.06426}}].

\bibitem{Cheng:2008hk}
H.-C. Cheng and Z.~Han, {\it {Minimal Kinematic Constraints and m(T2)}},  {\em
  JHEP} {\bf 12} (2008) 063, [\href{http://arxiv.org//abs/0810.5178}{{\tt
  arXiv:0810.5178}}].

\bibitem{CMS-PAS-TAU-16-002}
{\bf CMS Collaboration} Collaboration, {\it {Performance of reconstruction and
  identification of tau leptons in their decays to hadrons and tau neutrino in
  LHC Run-2}},  Tech. Rep. CMS-PAS-TAU-16-002, CERN, Geneva, 2016.

\bibitem{Aaboud:2017nhr}
{\bf ATLAS} Collaboration, M.~Aaboud et~al., {\it {Search for the direct
  production of charginos and neutralinos in $\sqrt{s} = $ 13 TeV $pp$
  collisions with the ATLAS detector}},
  \href{http://arxiv.org//abs/1708.07875}{{\tt arXiv:1708.07875}}.

\end{thebibliography}\endgroup

\end{document}